\documentclass[preprint,3p,12pt]{elsarticle}

\usepackage{mathrsfs}
\usepackage{amsmath}
\usepackage{stmaryrd}
\usepackage{bbding}
\usepackage{dcolumn}
\usepackage{graphicx}
\usepackage{amsfonts}
\usepackage{amssymb}
\usepackage{psfrag}
\usepackage{wrapfig}
\usepackage{makeidx}
\usepackage{bm}
\usepackage{epsf}
\usepackage{epsfig}
\usepackage{setspace}
\usepackage{epstopdf}
\usepackage{color}
\usepackage{comment}
\usepackage{algorithm}
\usepackage{algorithmic}
\usepackage{enumitem}
\usepackage{subcaption}
\usepackage{placeins}
\usepackage{booktabs}
\usepackage{xcolor}
\newcommand \td {\mathrm{~d}}

\journal{}

\begin{document}

\begin{frontmatter}



\title{Kinetic Non-Equilibrium Receptivity Analysis of Hypersonic Flows via a Complex Linearized Unified Gas-Kinetic Scheme}

\author[HKUST1]{Yue Zhang}
\ead{yzhangnl@connect.ust.hk}	
\author[TJU]{Caihong Su} 
\ead{su_ch@tju.edu.cn}
\author[TJU]{Chao Li} 
\author[HKUST1,HKUST2,HKUST3]{Kun Xu\corref{cor1}}
\ead{makxu@ust.hk}

\address[HKUST1]{Department of Mathematics, Hong Kong University of Science and Technology, Clear Water Bay, Kowloon, Hong Kong}
\address[HKUST2]{Department of Mechanical and Aerospace Engineering, Hong Kong University of Science and Technology, Clear Water Bay, Kowloon, Hong Kong}
\address[HKUST3]{Shenzhen Research Institute, Hong Kong University of Science and Technology, Shenzhen, China}
\address[TJU]{Laboratory of High-Speed Aerodynamics, Tianjin University, 300072 Tianjin, People’s Republic of China}
\cortext[cor1]{Corresponding author}

\begin{abstract}

In the near-space flight regime, local rarefaction near the leading-edge shock and the wall degrades the accuracy of Navier--Stokes-based stability and receptivity analyses due to velocity slip, temperature jumps, and non-equilibrium transport.
The unified gas-kinetic scheme (UGKS) provides a multiscale kinetic description spanning the continuum to rarefied regimes.
Building on this capability, a complex linearized UGKS (CLUGKS) is developed for linear stability and receptivity analysis: first-order perturbations of the discrete UGKS residual are formulated under a time-harmonic assumption with complex amplitudes, converting the disturbance equations into a steady elliptic system solved via pseudo-time marching with LU-SGS acceleration.
Consequently, kinetic shock structures, wall slip, temperature jumps, and non-Fourier heat fluxes derived from distribution-function moments are inherently captured without requiring hybrid continuum--kinetic coupling.
Three benchmarks verify the framework: a standing acoustic wave at an isothermal diffuse-reflection wall, a Mach\,2 normal shock interacting with an obliquely incident fast acoustic wave (validated against spatial linear interaction analysis), and the freestream acoustic receptivity of a hypersonic blunt wedge.
The scheme is then applied to argon blunt-wedge flows at equivalent flight altitudes of \(40\,\mathrm{km}\) and \(50\,\mathrm{km}\).
At \(40\,\mathrm{km}\), the UGKS and Navier--Stokes wall responses show close agreement in period and phase, while near-wall slip sustains a weakly positive spatial growth rate in UGKS further downstream.
At \(50\,\mathrm{km}\), intensified rarefaction produces a kinetic temperature precursor and a thickened, merged shock layer; while the wall-pressure responses from UGKS and Navier--Stokes remain comparatively close, significant deviations emerge in the temperature and heat-flux disturbances.
These findings demonstrate that while continuum-based models can capture dominant wall-pressure signatures under moderate rarefaction, accounting for kinetic non-equilibrium across the leading-edge shock---particularly its thermal footprint---becomes essential at higher altitudes.

\end{abstract}


\end{frontmatter}


\section{Introduction}

Near-space hypersonic vehicles operate at high speeds across altitudes typically ranging from $20$ to $100\,\mathrm{km}$.
At relatively low altitudes (below approximately $40\,\mathrm{km}$), the vehicle boundary layer frequently undergoes laminar-to-turbulent transition, whereas at high altitudes (above roughly $70\,\mathrm{km}$), strong rarefaction effects tend to suppress transition altogether.
Consequently, boundary-layer instability and transition at intermediate altitudes ($40\text{--}60\,\mathrm{km}$) are of critical interest~\cite{ou2024rarefaction, klothakis2022linear}.
Accurate transition prediction is vital for aerothermodynamic design and thermal protection systems, as the onset of turbulence substantially amplifies skin friction and surface heat flux~\cite{lee2019flow, tu2021progress}.

In low-disturbance freestream environments, natural transition conventionally progresses through three canonical stages: receptivity, linear modal growth, and nonlinear breakdown to turbulence~\cite{zhong2012direct}.
The receptivity stage governs how external perturbations---such as freestream acoustic waves, entropy spots, and wall disturbances---penetrate the boundary layer to excite instability modes.
Because it dictates the initial amplitudes, frequencies, and phases that seed subsequent modal growth, receptivity is central to reliable transition prediction~\cite{goldstein1989boundary, saric2002boundary, fedorov2011transition}.

For continuum flows, hypersonic boundary-layer stability and transition have been studied extensively using the compressible Navier--Stokes equations under no-slip wall conditions~\cite{zhong2012direct,fedorov2011transition}.
Linear stability theory (LST) and the parabolized stability equations (PSE) serve as standard frameworks for modal analysis~\cite{lees1946investigation,mack1984boundary,bertolotti1991analysis}, while direct numerical simulation (DNS) has been widely employed to examine freestream acoustic receptivity over flat plates, wedges, and cones~\cite{zhong2000receptivity,balakumar2015receptivity}.
Mack~\cite{mack1975linear,mack1990inviscid} demonstrated that multiple instability modes can emerge in high-speed shear flows, with the second and higher acoustic modes dominating at sufficiently high Mach numbers.
The influences of wall cooling, real-gas non-equilibrium, and nose bluntness have also been thoroughly investigated within the continuum framework~\cite{liang2010effects,malik1991real,wan2020receptivity,chen2021receptivity,wan2023effects}.
Nevertheless, the vast majority of these studies rely on the continuum hypothesis and no-slip wall boundaries.
To address the near-continuum regime, Ou and Chen developed a DSMC data-improved Navier--Stokes approach (DiNS) incorporating nonlinear constitutive relations alongside velocity-slip and temperature-jump boundary conditions~\cite{ou2019dins,ou2020nonlinear}.
This framework was subsequently utilized to evaluate slip effects on freestream acoustic receptivity~\cite{wang2024slip} as well as rarefaction effects on blunt-cone second-mode stability under wall cooling and nose bluntness~\cite{wang2025bluntcone}.
Additionally, rarefaction effects on boundary-layer transition over a straight cone were analyzed using a nonlinear coupled constitutive model with slip conditions coupled to a modified transition model~\cite{zeng2026}.
While these continuum-based extensions successfully capture critical near-wall slip phenomena, a fully kinetic treatment of the leading-edge shock structure and the non-equilibrium Knudsen layer remains largely missing.

This deficiency becomes critical for near-space hypersonic vehicles, where localized rarefaction frequently emerges within the leading-edge shock layer and the near-wall region.
The extent of rarefaction is quantified by the Knudsen number, $\mathit{Kn}$, defined as the ratio of the molecular mean free path to a characteristic macroscopic length scale.
While the Navier--Stokes (NS) equations are well established in the continuum regime ($\mathit{Kn} < 0.001$), they break down in rarefied hypersonic flows due to wall velocity slip, temperature jump, and non-equilibrium transport across shock waves and shear layers.
In such regimes, kinetic formulations based on the Boltzmann equation become indispensable.
Two widely adopted approaches are the stochastic Direct Simulation Monte Carlo (DSMC) method~\cite{bird1994molecular} and the deterministic discrete velocity method (DVM)~\cite{broadwell1964study}. However, both methods are essentially single-scale: relying on operator splitting, their numerical resolution is constrained by the molecular mean free path and mean collision time. Because substantial portions of a hypersonic boundary-layer flow remain near the continuum regime, enforcing global kinetic resolution across the entire domain incurs prohibitive computational costs.
Multiscale methods therefore offer a compelling alternative.
The unified gas-kinetic scheme (UGKS)~\cite{xu2010unified,xu2015direct} couples molecular transport and collision within a single flux evaluation, thereby removing the requirement that spatial cell sizes resolve the mean free path in near-continuum regions.
Although the subsequent unified gas-kinetic wave-particle (UGKWP) method~\cite{liu2020ugkwp,zhu2019ugkwp,chen2020three,long2024nonequilibrium} mitigates the cost of velocity-space discretization, its inherent particle sampling introduces statistical scatter.
For resolving minute disturbance fields in stability and receptivity problems within shear and boundary layers, a deterministic UGKS is distinctly advantageous.
Furthermore, UGKS has been successfully extended to account for rotational and vibrational non-equilibrium~\cite{liu2014unified,wei2024adaptive} as well as chemically reacting flows~\cite{wei2026unified}.
Long et~al.~\cite{long2026applicability} demonstrated that implementing kinetic wall conditions within this framework delivers accurate and reliable predictions for near-continuum hypersonic flows.

Stability analyses based on kinetic theory have also advanced considerably in recent years.
Eigenvalue problems formulated via the BGK and Shakhov--BGK models have been investigated for microchannel flows~\cite{zou2022linear}, rarefied thermal convection~\cite{wu2024novel}, and high-speed shear flows~\cite{zou2024normal}.
These studies demonstrate that rarefaction effects can markedly alter modal growth rates compared with continuum predictions.
Nonetheless, solving the kinetic eigenvalue problem remains computationally demanding due to the high dimensionality of phase space, despite efficiency improvements offered by synthetic iterative schemes~\cite{wu2024efficient}.
Consequently, extending kinetic stability analyses to complex, multi-dimensional hypersonic configurations remains a significant challenge.

In this work, we develop a complex linearized unified gas-kinetic scheme (CLUGKS) for flow stability and receptivity analysis spanning continuum to rarefied regimes.
Building upon the discrete UGKS formulation~\cite{xu2010unified,xu2015direct}, first-order perturbations of the residual are constructed in the spirit of the linearized UGKS previously applied to adjoint sensitivity analysis~\cite{zhang2026asymptoticpreservingadjointunifiedgas}.
Invoking a time-harmonic assumption with complex amplitudes transforms the governing disturbance equations into a steady elliptic system, which is subsequently solved using pseudo-time marching accelerated by the lower--upper symmetric Gauss--Seidel (LU-SGS) method.
By retaining the multiscale capabilities of UGKS, the resulting framework seamlessly captures kinetic shock structures, diffuse-reflection wall slip, temperature jumps, and collisional relaxation across continuum and rarefied regions without resorting to hybrid continuum--kinetic coupling.
Freestream acoustic, entropy, and vorticity disturbances are directly prescribed as harmonic boundary conditions, rendering the scheme uniquely well-suited for hypersonic receptivity problems characterized by strong leading-edge rarefaction.

The remainder of this paper is organized as follows.
Section~2 briefly reviews the UGKS formulation and details the derivation of the complex linearized system along with the pseudo-time marching procedure.
Section~3 verifies the proposed framework against three canonical benchmarks: a standing acoustic wave near an isothermal diffuse-reflection wall, a near-critical fast acoustic wave--shock interaction validated against spatial linear interaction analysis, and freestream acoustic receptivity over a hypersonic blunt wedge.
Sections~4 and~5 apply CLUGKS to hypersonic argon flows over a blunt wedge at equivalent flight altitudes of \(40\,\mathrm{km}\) and \(50\,\mathrm{km}\), focusing on shock-layer receptivity and wall-mode responses under progressive rarefaction.
Finally, key conclusions and concluding remarks are summarized in Section~6.

\section{The linear unified gas-kinetic scheme}

\subsection{The unified gas-kinetic scheme}
In this work, the Shakhov model of the Boltzmann equation is used: a BGK-type relaxation collision operator with a heat-flux correction so that the correct Prandtl number is recovered.
The governing equation is
\begin{equation}
  \label{BGK}
  \frac{\partial f}{\partial t} + \mathbf{v}\cdot\nabla f =\frac{f^+-f}{\tau},
\end{equation}
where \(f=f(\mathbf{x},t,\mathbf{v},\mathbf{\xi})\) is the gas distribution function,
\(\mathbf{x}\) and \(t\) are the physical space and time,
\(\mathbf{v}\) is the microscopic particle translation velocity,
and \(\mathbf{\xi}\) is the internal variable associated with \(K\) rotational and vibrational degrees of freedom, with \(\mathbf{\xi}^2=\xi_1^2+\cdots+\xi_K^2\).
Here \(\tau\) is the particle collision time, and \(f^+\) is the Shakhov equilibrium
\begin{equation*}
 f^+=g\left[ 1+(1-\text{Pr})2\lambda\mathbf{c}\cdot\mathbf{q}\left(2\lambda c^2-5\right)/(5p)\right]=g+g^+,
\end{equation*}
in which the term proportional to the heat flux \(\mathbf{q}\) is the Shakhov correction to the Maxwellian \(g\).
Here \(\mathrm{Pr}\) is the Prandtl number, \(\mathbf{c}=\mathbf{v}-\mathbf{V}\) is the peculiar velocity, \(\mathbf{V}\) is the macroscopic velocity, \(\lambda=m/(2k_B T)\), \(m\) is the molecular mass, \(k_B\) is the Boltzmann constant and \(T\) is the temperature.
The Maxwellian distribution is
\begin{equation*}
 g=\rho {\left(\frac{\lambda}{\pi}\right)}^\frac{K+D}{2}e^{-\lambda ((\mathbf{v}-\mathbf{V})^2+\mathbf{\xi}^2)},
\end{equation*}
where \(D\) is the spatial dimension.
The heat flux is obtained from the distribution function,
\begin{equation*}
  \mathbf{q}=\frac{1}{2}\int (\mathbf{v}-\mathbf{V})(|\mathbf{v}-\mathbf{V}|^2+\mathbf{\xi}^2)f\td\mathbf{v} \td \mathbf{\xi}.
\end{equation*}
The collision term satisfies the compatibility condition
\begin{equation*}
 \int (f^+-f)\mathbf{\Psi}\td \mathbf{v}\td\mathbf{\xi}=0,
\end{equation*}
where \(\mathbf{\Psi}=(1,\mathbf{v},1/2(\mathbf{v}^2+\mathbf{\xi}^2))^T\) denotes the collision invariants
and \(\td\mathbf{\xi}=\td\xi_1\cdots\td\xi_K\).

The finite volume method is used in this paper. The domain is divided into cells $\Omega_i$ and the velocity space is divided into discrete points $\delta \mathcal{V}_k$. For each cell $\Omega_i$, the governing equation of the distribution function is
\begin{equation}
  \label{UGKSmicro}
 \frac{f_{i,k}^{n+1}-f_{i,k}^{n}}{\Delta t} + \frac{1}{|\Omega_i|}\underbrace{  \frac{1}{\Delta t}\sum_{j\in N(i)} S_{ij}\int_{t^n}^{t^{n+1}}v_k\cdot \mathbf{n}_{ij}f_{ij,k}\td t}_{\mathcal{R}_{i,k}^n}=\frac{f_{i,k}^{+,n+1}-f_{i,k}^{n+1}}{\tau},
\end{equation}
where the index $k$ denotes the discrete points in the particle velocity space. $|\Omega_i|$ denotes the volume of cell $i$, $N(i)$ is the set of all interface-adjcent neighboring cells of cell $i$ and $j$ is one of the neighboring cells of $i$.  The interface between cells $i$ and $j$ is labeled $ij$ and has area $S_{ij}$. $v_k$ is the velocity of the $k$-th discrete point. $\mathbf{n}_{ij}$ is the normal vector of interface $ij$. The 
To construct the numerical flux at cell interface $\mathbf{x}_0=(0,0,0)^T$, the integral solution along the characteristic line $\mathbf{x}^\prime =\mathbf{x}_0-\mathbf{v}(t-t^\prime)$ of the BGK equation, Eq.~\eqref{BGK}, gives
\begin{equation*}
 f(\mathbf{x},t,\mathbf{\xi})=\frac{1}{\tau}\int_{t_0}^t f^+(\mathbf{x}^\prime,t^\prime)e^{-(t-t^\prime)/\tau}\td t^\prime + e^{-t/\tau}f_{0}(\mathbf{x}-\mathbf{v} t),
\end{equation*}
where $f_{0}(\mathbf{x})$ is the initial gas distribution function at the beginning of each step $t_n$, and $f^+(\mathbf{x},t)$ is the effective equilibrium state distributed in space and time. The integral solution provides a multiscale model of the evolution from an initial non-equilibrium distribution $f$ to an equilibrium distribution $f^+$ via collisions. To achieve second-order accuracy, the initial distribution function $f_{0}(\mathbf{x})$ is approximated as

\begin{equation*}
 f_{0}(\mathbf{x}) = \begin{cases}
 f^l+\mathbf{x}\cdot \nabla f^l,&v_n>0,\\
 f^r+\mathbf{x}\cdot \nabla  f^r,&v_n<0,\\
 \end{cases}
\end{equation*}
where $f^l$ and $f^r$ are the reconstructed initial distribution functions at the left and right sides of the interface. The equilibrium state is approximated as

\begin{equation*}
 f^+(\mathbf{x},t) \approx g_{0}+ g^+_{0} + \mathbf{x}\cdot\nabla g_{0}+\frac{\partial g_{0}}{\partial t}t,
\end{equation*}
where $g_{0}$ is the Maxwellian distribution function obtained from the conservative flow variables of colliding particles from both sides of the interface.
Substituting these reconstructions into the integral solution yields the interface distribution
\begin{equation*}
 f_k(\mathbf{0},t,\mathbf{\xi}) =
 \begin{cases}
 c_1 f_k^l + c_2 \mathbf{v}_k\cdot\nabla f_k^l
 + c_3(g_{0,k}+g_{0,k}^+)
 + c_4 \mathbf{v}_k\cdot\nabla g_{0,k}
 + c_5\partial_t g_{0,k},
 & v_{k,n}>0,\\[0.4em]
 c_1 f_k^r + c_2 \mathbf{v}_k\cdot\nabla f_k^r
 + c_3(g_{0,k}+g_{0,k}^+)
 + c_4 \mathbf{v}_k\cdot\nabla g_{0,k}
 + c_5\partial_t g_{0,k},
 & v_{k,n}<0,
 \end{cases}
\end{equation*}
with
\begin{equation*}
 \begin{aligned}
 c_1&=e^{-t/\tau},\\
 c_2&=-t\,e^{-t/\tau},\\
 c_3&=1-e^{-t/\tau},\\
 c_4&=t\,e^{-t/\tau}-\tau\bigl(1-e^{-t/\tau}\bigr),\\
 c_5&=t-\tau\bigl(1-e^{-t/\tau}\bigr),
 \end{aligned}
\end{equation*}
where \(v_{k,n}=\mathbf{v}_k\cdot\mathbf{n}\) is the normal component of the discrete particle velocity.

In addition, the governing equation of cell-averaged macroscopic conservative values $\mathbf{W}_i=(\rho_i,\rho_i \mathbf{V}_i,\rho_i E_i)^T$, i.e. density, momentum and energy, is also updated in the UGKS. The governing equation is given by

\begin{equation*}
 \mathbf{W}_i^{n+1}=\mathbf{W}_i^n -\frac{1}{|\Omega_i|}\int_{t^n}^{t^{n+1}}\sum_{j\in N(i)} S_{ij} \mathbf{F}_{ij}  \td t,
\end{equation*}
where $\mathbf{W}_i^n$ and $\mathbf{W}_i^{n+1}$ are the cell-averaged conservative variables at times $t^n$ and $t^{n+1}$.
The macroscopic flux $\mathbf{F}_{ij}$ is obtained by taking moments of the microscopic flux at the cell interface $ij$,
\begin{equation*}
  \mathbf{F}_{ij}=\sum_{k} w_k\mathbf{v}_k\cdot \mathbf{n}_{ij} {f}_{ij,k}\mathbf{\Psi}_k,
\end{equation*}
where \(w_k\) is the quadrature weight of the \(k\)-th discrete velocity point,
\(\mathbf{n}_{ij}\) is the interface unit normal,
\({f}_{ij,k}\) is the interface distribution function,
and \(\mathbf{\Psi}_k\) is \(\mathbf{\Psi}\) evaluated at the discrete velocity \(\mathbf{v}_k\).
Correspondingly, the discrete compatibility condition and the moment relation for the conservative variables are
\begin{equation}
  \label{eq:compat_discrete}
  \sum_k w_k\bigl(f_{i,k}^+-f_{i,k}\bigr)\mathbf{\Psi}_k=0,
\end{equation}
\begin{equation}
  \label{eq:W_moment}
  \mathbf{W}_i=\sum_k w_k f_{i,k}\mathbf{\Psi}_k.
\end{equation}

Far-field and other artificial boundaries are imposed with ghost cells: the macroscopic ghost state follows the corresponding Euler-type prescription, the distribution is taken as the associated Maxwellian, and interface fluxes are selected according to the particle velocity direction.
Solid walls use the fully accommodating isothermal diffuse-reflection condition of UGKS~\cite{xu2010unified,zhu2016implicit}.
On the incident side (\(v_{k,n}\) into the wall), the interface distribution is constructed exactly as in the interior, from the adjacent fluid cell with the same coefficients \(c_1\)--\(c_5\).
On the reflected side, diffuse reflection returns the wall equilibrium \(g_w(\rho_w,T_w,\mathbf{V}_w)\), with the wall density \(\rho_w\) fixed by the no-penetration condition on the mass flux.
In addition, reduced distribution functions are employed to lower the velocity-space dimension; their definitions and update formulae follow Refs.~\cite{xu2010unified,xu2015direct}.

\subsection{The Linearized System of the Unified Gas-Kinetic Scheme}

The linearized unified gas-kinetic scheme (LUGKS) is obtained by freezing a converged base state \((\mathbf{W}_i,f_{i,k})\) and retaining first-order perturbations
\((\delta\mathbf{W}_i,\delta f_{i,k})\).
Linearizing the moment relation, Eq.~\eqref{eq:W_moment}, gives the macroscopic constraint
\begin{equation}
  \label{eq:macro_lin_compact}
  \delta\mathbf{W}_i- \sum_k w_k \delta f_{i,k}\mathbf{\Psi}_k = 0.
\end{equation}
Because the Shakhov equilibrium \(f^+\) and the interface residual \(\mathcal{R}_{i,k}\) depend on the heat flux, the heat-flux perturbation is also needed,
\begin{equation*}
\delta\mathbf{q}_i = \sum_{k'} w_{k'}\,\frac{\partial \mathbf{q}_i}{\partial f_{i,k'}}\,\delta f_{i,k'}+\frac{\partial \mathbf{q}_i}{\partial \mathbf{W}_i}\delta \mathbf{W}_i.
\end{equation*}
With these first-order closures, the discrete microscopic equation, Eq.~\eqref{UGKSmicro}, linearizes to
\begin{equation}
  \label{eq:L_micro_lin}
  \begin{aligned} \frac{\delta f_{i,k}^{n+1}-\delta f_{i,k}^{n}}{\Delta t}
  &+ \frac{1}{|\Omega_i|}\sum_j
  \left(
  \frac{\partial \mathcal{R}_{i,k}}{\partial \mathbf{W}_j}\,\delta\mathbf{W}_j^n
  + \frac{\partial \mathcal{R}_{i,k}}{\partial f_{j,k}}\,\delta f_{j,k}^n
  + \frac{\partial \mathcal{R}_{i,k}}{\partial \mathbf{q}_j^n}\,\delta\mathbf{q}_j
  \right) \\
  &- \frac{1}{\tau_i}\left(
  \frac{\partial f_{i,k}^+}{\partial \mathbf{W}_i}\,\delta\mathbf{W}_i^{n+1}
  + \frac{\partial f_{i,k}^+}{\partial \mathbf{q}_i}\,\delta\mathbf{q}_i^n
  - \delta f_{i,k}^{n+1}
  \right) \\
  &+ \frac{1}{\tau_i}\,\frac{\partial \tau_i}{\partial \mathbf{W}_i} \delta\mathbf{W}_i^{n+1}
  \frac{f_{i,k}^+-f_{i,k}}{\tau_i}\,
  = 0.
  \end{aligned}
\end{equation}
Here the time indices on the base-flow variables are omitted, while the perturbation fields keep the discrete time levels of the update.

Taking the moment of Eq.~\eqref{eq:L_micro_lin} with \(w_k\mathbf{\Psi}_k\) and using
\begin{equation*}
  \sum_{k} w_k \mathbf{\Psi}_k \frac{\partial f_{i,k}^+}{\partial \mathbf{W}_i}
  = \frac{\partial}{\partial \mathbf{W}_i}\sum_{k} w_k \mathbf{\Psi}_k f_{i,k}^+
  = \mathbf{I},
\end{equation*}
together with Eq.~\eqref{eq:compat_discrete}, one obtains the linearized macroscopic evolution
\begin{equation}
  \label{eq:ap_linear_macro}
  \begin{aligned}
  \frac{\delta \mathbf{W}_i^{n+1}-\delta\mathbf{W}_i^{n}}{\Delta t}
  &+ \frac{1}{|\Omega_i|}\sum_k w_k\mathbf{\Psi}_k^T\sum_j
  \left(
  \frac{\partial \mathcal{R}_{i,k}}{\partial \mathbf{W}_j}\,\delta\mathbf{W}_j^n
  + \frac{\partial \mathcal{R}_{i,k}}{\partial f_{j,k}}\,\delta f_{j,k}^n
  + \frac{\partial \mathcal{R}_{i,k}}{\partial \mathbf{q}_j^n}\,\delta\mathbf{q}_j
\right)  = 0.
  \end{aligned}
\end{equation}
The LUGKS system therefore consists of the microscopic equation~\eqref{eq:L_micro_lin} and the macroscopic equation~\eqref{eq:ap_linear_macro}, linked by the linearized moment constraint~\eqref{eq:macro_lin_compact}.
Detailed expressions for the spatial Jacobians in the linearized residual are given in the appendix of Zhang et~al.~\cite{zhang2026asymptoticpreservingadjointunifiedgas}.

\subsection{The complex linear unified gas-kinetic scheme}

For time-harmonic small disturbances about a steady base flow, every first-order perturbation field is written as
\begin{equation}
  \label{eq:complex_ansatz}
  \boldsymbol{\varphi}(\mathbf{x},t)
  = \widehat{\boldsymbol{\varphi}}(\mathbf{x})\,e^{-i\omega t},
\end{equation}
where \(\omega\) is the angular frequency. The complex spatial amplitude is split into real and imaginary parts,
\begin{equation*}
  \widehat{\boldsymbol{\varphi}}
  = \widehat{\boldsymbol{\varphi}}^{\mathrm{Re}} + i\widehat{\boldsymbol{\varphi}}^{\mathrm{Im}}.
\end{equation*}
Since
\begin{equation*}
  -i\omega\,\widehat{\boldsymbol{\varphi}}
  = -i\omega\bigl(\widehat{\boldsymbol{\varphi}}^{\mathrm{Re}} + i\widehat{\boldsymbol{\varphi}}^{\mathrm{Im}}\bigr)
  = \omega\widehat{\boldsymbol{\varphi}}^{\mathrm{Im}} - i\omega\widehat{\boldsymbol{\varphi}}^{\mathrm{Re}},
\end{equation*}
the physical time derivative acting on the real--imaginary pair is represented by the operator \(D_t\),
\begin{equation}
  \label{eq:time_deriv_matrix}
  D_t
  =
  \omega
  \begin{bmatrix}
  \mathbf{0} & \mathbf{I}\\
  -\mathbf{I} & \mathbf{0}
  \end{bmatrix},
  \qquad
  D_t
  \begin{bmatrix}
  \widehat{\boldsymbol{\varphi}}^{\mathrm{Re}}\\
  \widehat{\boldsymbol{\varphi}}^{\mathrm{Im}}
  \end{bmatrix}
  =
  \omega
  \begin{bmatrix}
  \widehat{\boldsymbol{\varphi}}^{\mathrm{Im}}\\
  -\widehat{\boldsymbol{\varphi}}^{\mathrm{Re}}
  \end{bmatrix},
\end{equation}
which replaces \(\partial/\partial t\) under the harmonic ansatz, Eq.~\eqref{eq:complex_ansatz}.
Accordingly, the macroscopic and microscopic perturbations take the form
\begin{equation*}
  \begin{aligned}
  \delta\mathbf{W}_i(t)
  &= \bigl(\delta\widehat{\mathbf{W}}_i^{\mathrm{Re}} + i\delta\widehat{\mathbf{W}}_i^{\mathrm{Im}}\bigr)\,e^{-i\omega t},\\
  \delta f_{i,k}(t)
  &= \bigl(\delta\widehat{f}_{i,k}^{\mathrm{Re}} + i\delta\widehat{f}_{i,k}^{\mathrm{Im}}\bigr)\,e^{-i\omega t},\\
  \delta\mathbf{q}_i(t)
  &= \bigl(\delta\widehat{\mathbf{q}}_i^{\mathrm{Re}} + i\delta\widehat{\mathbf{q}}_i^{\mathrm{Im}}\bigr)\,e^{-i\omega t}.
  \end{aligned}
\end{equation*}
For brevity, the hats on the spatial amplitudes \(\delta\widehat{\mathbf{W}}\), \(\delta\widehat{f}\) and \(\delta\widehat{\mathbf{q}}\) are omitted below, and \(\delta\mathbf{W}^{\mathrm{Re}/\mathrm{Im}}\), \(\delta f^{\mathrm{Re}/\mathrm{Im}}\) and \(\delta\mathbf{q}^{\mathrm{Re}/\mathrm{Im}}\) denote these complex amplitudes.
The linearized compatibility condition, Eq.~\eqref{eq:macro_lin_compact}, and the heat-flux relation hold for both components,
\begin{equation}
  \label{eq:complex_compat}
  \delta\mathbf{W}_i^{\mathrm{Re}/\mathrm{Im}}
  - \sum_k w_k\,\delta f_{i,k}^{\mathrm{Re}/\mathrm{Im}}\,\mathbf{\Psi}_k
  = 0,
\end{equation}
\begin{equation}
  \label{eq:complex_q}
  \delta\mathbf{q}_i^{\mathrm{Re}/\mathrm{Im}}
  = \sum_{k'} w_{k'}\,\frac{\partial \mathbf{q}_i}{\partial f_{i,k'}}\,\delta f_{i,k'}^{\mathrm{Re}/\mathrm{Im}}
  + \frac{\partial \mathbf{q}_i}{\partial \mathbf{W}_i}\,\delta\mathbf{W}_i^{\mathrm{Re}/\mathrm{Im}}.
\end{equation}

Replacing the physical time derivative in the linearized system by the operator \(D_t\) in Eq.~\eqref{eq:time_deriv_matrix} yields a steady equation for the complex amplitudes.
Denote by \(\mathbf{R}_i^{\mathrm{W,Re}/\mathrm{Im}}\) (respectively \(\mathcal{R}_{i,k}^{f,\mathrm{Re}/\mathrm{Im}}\)) the linearized spatial residual of the macroscopic (respectively microscopic) equation, evaluated on the frozen base state with the indicated real or imaginary amplitudes.
The harmonic problem for the macroscopic amplitudes is
\begin{equation}
  \label{eq:C_macro}
  \begin{aligned}
  \mathbf{R}_i^{\mathrm{W,Re}} - \omega\,\delta\mathbf{W}_i^{\mathrm{Im}}
  &= \mathbf{0},\\
  \mathbf{R}_i^{\mathrm{W,Im}} + \omega\,\delta\mathbf{W}_i^{\mathrm{Re}}
  &= \mathbf{0},
  \end{aligned}
\end{equation}
and likewise for the microscopic amplitudes,
\begin{equation}
  \label{eq:C_micro}
  \begin{aligned}
  \mathcal{R}_{i,k}^{f,\mathrm{Re}} - \omega\,\delta f_{i,k}^{\mathrm{Im}}
  &= 0,\\
  \mathcal{R}_{i,k}^{f,\mathrm{Im}} + \omega\,\delta f_{i,k}^{\mathrm{Re}}
  &= 0.
  \end{aligned}
\end{equation}
Here \(\mathbf{R}_i^{\mathrm{W}}\) collects the linearized macroscopic flux and collision contributions obtained by taking moments of the microscopic residual, and
\begin{equation*}
  \begin{aligned}
  \mathcal{R}_{i,k}^{f,\mathrm{Re}/\mathrm{Im}}
  &= \frac{1}{|\Omega_i|}\sum_j
  \Biggl(
  \frac{\partial \mathcal{R}_{i,k}}{\partial \mathbf{W}_j}\,\delta\mathbf{W}_j^{\mathrm{Re}/\mathrm{Im}}
  + \frac{\partial \mathcal{R}_{i,k}}{\partial f_{j,k}}\,\delta f_{j,k}^{\mathrm{Re}/\mathrm{Im}}
  + \frac{\partial \mathcal{R}_{i,k}}{\partial \mathbf{q}_j}\,\delta\mathbf{q}_j^{\mathrm{Re}/\mathrm{Im}}
  \Biggr) \\
  &\quad - \frac{1}{\tau_i}\Biggl(
  \frac{\partial f_{i,k}^+}{\partial \mathbf{W}_i}\,\delta\mathbf{W}_i^{\mathrm{Re}/\mathrm{Im}}
  + \frac{\partial f_{i,k}^+}{\partial \mathbf{q}_i}\,\delta\mathbf{q}_i^{\mathrm{Re}/\mathrm{Im}}
  - \delta f_{i,k}^{\mathrm{Re}/\mathrm{Im}}
  \Biggr)
  + \frac{1}{\tau_i}\,\frac{\partial \tau_i}{\partial \mathbf{W}_i}\,\delta\mathbf{W}_i^{\mathrm{Re}/\mathrm{Im}}\,
  \frac{f_{i,k}^+-f_{i,k}}{\tau_i}.
  \end{aligned}
\end{equation*}
All spatial Jacobians are evaluated on the steady base state and are therefore real; the frequency \(\omega\) enters only through the coupling between the real and imaginary parts.
The complex LUGKS (CLUGKS) consists of Eqs.~\eqref{eq:C_macro}--\eqref{eq:C_micro} subject to Eqs.~\eqref{eq:complex_compat}--\eqref{eq:complex_q}.

\subsection{Pseudo-time marching}
Eqs.~\eqref{eq:C_macro}--\eqref{eq:C_micro} are elliptic in character for the unknown amplitudes and are not marched in physical time.
A further difficulty is that the microscopic residual \(\mathcal{R}_{i,k}^{f}\) is not closed in \(\delta f\) alone:
the Shakhov state \(f^+\) (and likewise \(\tau\) and \(\mathbf{q}\)) depends on the macroscopic variables \(\mathbf{W}\), so the collision linearization contains terms such as
\(\partial f^+/\partial\mathbf{W}\,\delta\mathbf{W}\) and \(\partial f^+/\partial\mathbf{q}\,\delta\mathbf{q}\).
The microscopic and macroscopic amplitudes are therefore tightly coupled inside each cell.
Assembling a globally coupled micro--macro Jacobian for all discrete velocities would be prohibitive; the solver instead uses pseudo-time iteration with a macro--micro predictor--corrector that updates \(\delta\mathbf{W}\) and \(\delta f\) in sequence while retaining the \(f^+(\mathbf{W})\) dependence in the residual, following the implicit UGKS strategy of Zhu et~al.~\cite{zhu2016implicit}.
A pseudo-time \(\tilde{t}\) is introduced so that the harmonic residual is driven to zero.
For the macroscopic amplitudes,
\begin{equation}
  \label{eq:C_pseudotime}
  \begin{aligned}
  \frac{\partial\,\delta\mathbf{W}_i^{\mathrm{Re}}}{\partial \tilde{t}}
  + \mathbf{R}_i^{\mathrm{W,Re}} - \omega\,\delta\mathbf{W}_i^{\mathrm{Im}}
  &= \mathbf{0},\\
  \frac{\partial\,\delta\mathbf{W}_i^{\mathrm{Im}}}{\partial \tilde{t}}
  + \mathbf{R}_i^{\mathrm{W,Im}} + \omega\,\delta\mathbf{W}_i^{\mathrm{Re}}
  &= \mathbf{0}.
  \end{aligned}
\end{equation}
With a local pseudo-time step \(\Delta\tilde{t}_i\), a backward-Euler discretization of Eq.~\eqref{eq:C_pseudotime} reads
\begin{equation*}
  \begin{aligned}
  \frac{\delta\mathbf{W}_i^{\mathrm{Re},m+1}-\delta\mathbf{W}_i^{\mathrm{Re},m}}{\Delta\tilde{t}_i}
  + \mathbf{R}_i^{\mathrm{W,Re},m+1} - \omega\,\delta\mathbf{W}_i^{\mathrm{Im},m+1}
  &= \mathbf{0},\\
  \frac{\delta\mathbf{W}_i^{\mathrm{Im},m+1}-\delta\mathbf{W}_i^{\mathrm{Im},m}}{\Delta\tilde{t}_i}
  + \mathbf{R}_i^{\mathrm{W,Im},m+1} + \omega\,\delta\mathbf{W}_i^{\mathrm{Re},m+1}
  &= \mathbf{0}.
  \end{aligned}
\end{equation*}
Writing the update as an increment
\(\delta\mathbf{W}^{m+1}=\delta\mathbf{W}^{m}+\Delta\delta\mathbf{W}\)
and linearizing about the current iterate yields the \(2\times 2\) block system
\begin{equation}
  \label{eq:C_increment_ri}
  \begin{bmatrix}
  D_i & -\omega\,\mathbf{I}\\
  \omega\,\mathbf{I} & D_i
  \end{bmatrix}
  \begin{bmatrix}
  \Delta\delta\mathbf{W}_i^{\mathrm{Re}}\\
  \Delta\delta\mathbf{W}_i^{\mathrm{Im}}
  \end{bmatrix}
  =
  -
  \begin{bmatrix}
  \mathbf{R}_i^{\mathrm{W,Re},m} - \omega\,\delta\mathbf{W}_i^{\mathrm{Im},m}\\
  \mathbf{R}_i^{\mathrm{W,Im},m} + \omega\,\delta\mathbf{W}_i^{\mathrm{Re},m}
  \end{bmatrix},
\end{equation}
where the real diagonal \(D_i\) contains the pseudo-time contribution \(1/\Delta\tilde{t}_i\) together with the spectral radius of the linearized inviscid flux, as in the LU-SGS diagonal of the implicit UGKS~\cite{zhu2016implicit,zhu2019implicit}.

The microscopic amplitudes are discretized by the same procedure.
The pseudo-time equations dual to Eq.~\eqref{eq:C_pseudotime} are
\begin{equation*}
  \begin{aligned}
  \frac{\partial\,\delta f_{i,k}^{\mathrm{Re}}}{\partial \tilde{t}}
  + \mathcal{R}_{i,k}^{f,\mathrm{Re}} - \omega\,\delta f_{i,k}^{\mathrm{Im}}
  &= 0,\\
  \frac{\partial\,\delta f_{i,k}^{\mathrm{Im}}}{\partial \tilde{t}}
  + \mathcal{R}_{i,k}^{f,\mathrm{Im}} + \omega\,\delta f_{i,k}^{\mathrm{Re}}
  &= 0,
  \end{aligned}
\end{equation*}
and the corresponding incremental system takes the form
\begin{equation}
  \label{eq:C_increment_f}
  \begin{bmatrix}
  D_{i,k}^{f} & -\omega\\
  \omega & D_{i,k}^{f}
  \end{bmatrix}
  \begin{bmatrix}
  \Delta\delta f_{i,k}^{\mathrm{Re}}\\
  \Delta\delta f_{i,k}^{\mathrm{Im}}
  \end{bmatrix}
  =
  -
  \begin{bmatrix}
  \mathcal{R}_{i,k}^{f,\mathrm{Re},m} - \omega\,\delta f_{i,k}^{\mathrm{Im},m}\\
  \mathcal{R}_{i,k}^{f,\mathrm{Im},m} + \omega\,\delta f_{i,k}^{\mathrm{Re},m}
  \end{bmatrix},
\end{equation}
where \(D_{i,k}^{f}\) collects the microscopic pseudo-time contribution \(1/\Delta\tilde{t}_i\), the local relaxation linearization and the diagonal part of the migration flux.
Eq.~\eqref{eq:C_increment_f} is solved pointwise for each discrete velocity \(k\).

The macroscopic and microscopic fields are advanced by the same predictor--corrector splitting as in the time-domain LUGKS.
At each pseudo-time step the dependence of \(f^+\) on \(\mathbf{W}\) is treated as follows:
\begin{enumerate}[leftmargin=*]
  \item Evaluate the spatial residuals \(\mathbf{R}^{\mathrm{W,Re}/\mathrm{Im}}\) and \(\mathcal{R}^{f,\mathrm{Re}/\mathrm{Im}}\) (including all \(\partial f^+/\partial\mathbf{W}\) and \(\partial f^+/\partial\mathbf{q}\) contributions) and form the right-hand sides of Eqs.~\eqref{eq:C_increment_ri} and~\eqref{eq:C_increment_f}.
  \item Solve Eq.~\eqref{eq:C_increment_ri} for the macroscopic increments by LU-SGS sweeps with linearized neighbor flux coupling, and update \(\delta\mathbf{W}^{\mathrm{Re}/\mathrm{Im}}\).
  \item Insert the predicted macroscopic amplitudes into the collision linearization of \(f^+(\mathbf{W},\mathbf{q})\), then solve Eq.~\eqref{eq:C_increment_f} pointwise for \(\delta f_{i,k}^{\mathrm{Re}/\mathrm{Im}}\).
  Only the pseudo-time term, the local relaxation contribution and the migration flux are treated implicitly; the macro dependence of \(f^+\) enters through the already updated \(\delta\mathbf{W}\) (and the associated \(\delta\mathbf{q}\)).
  \item Correct \(\delta\mathbf{W}^{\mathrm{Re}/\mathrm{Im}}\) through the linearized compatibility condition, Eq.~\eqref{eq:complex_compat}, so that the discrete moment relation between \(\delta\mathbf{W}\) and \(\delta f\) is restored after the staggered update.
\end{enumerate}
The iteration continues until the residuals of Eqs.~\eqref{eq:C_macro}--\eqref{eq:C_micro} fall below a prescribed tolerance.
The physical time-domain response is recovered from
\begin{equation*}
  \mathrm{Re}\bigl\{\widehat{\boldsymbol{\varphi}}(\mathbf{x})\,e^{-i\omega t}\bigr\}
  = \widehat{\boldsymbol{\varphi}}^{\mathrm{Re}}\cos(\omega t)+\widehat{\boldsymbol{\varphi}}^{\mathrm{Im}}\sin(\omega t).
\end{equation*}

\section{Verification and validation}
\label{sec:vv}

\subsection{Standing acoustic wave at an isothermal diffuse-reflection wall}
\label{sec:sound_wall}

This case examines the linearized isothermal diffuse-reflection wall in the complex LUGKS.
A one-dimensional standing acoustic wave is established in a quiescent gas and is forced to satisfy a fully accommodating diffuse-reflection condition at one end.
The inviscid standing wave is known exactly.
Navier--Stokes (NS) adds bulk attenuation, an imperfect wall reflection and a thin thermal layer.

\subsubsection{Case configuration}
\label{sec:sound_wall_config}

\paragraph{Base state and wave}
The channel occupies \(x\in[0,L]\) with \(L=1\).
The undisturbed gas is monatomic,
\begin{equation*}
  \gamma=\frac{5}{3},\qquad
  R_g=1,\qquad
  \mathrm{Pr}=\frac{2}{3},\qquad
  (\rho,U,p,T)_0=(1,0,1,1),
\end{equation*}
so that the sound speed is \(c_0=\sqrt{\gamma R_g T_0}\).
The collision model is of variable-hard-sphere type with temperature exponent \(1/2\) (hard sphere).
The Knudsen number \(\mathrm{Kn}=\lambda/L\) is varied over
\(10^{-3},\,5\times10^{-3},\,10^{-2},\,2\times10^{-2}\),
with the reference viscosity chosen so that
\begin{equation*}
  \mu=\rho_0 c_0\,\mathrm{Kn}\,L\sqrt{\frac{2}{\pi\gamma}}.
\end{equation*}
A standing acoustic wave of fundamental wave number \(k_0=2\pi/L\) and frequency \(\omega=k_0 c_0\) is formed by superposing equal-amplitude right- and left-going acoustics,
\begin{equation*}
  a_{\pm}=10^{-4},
\end{equation*}
corresponding to a pressure amplitude \(A_p=a\rho_0 c_0^2\) for each travelling wave.
The isentropic temperature perturbation of a single acoustic wave is
\begin{equation*}
  A_T
  =\frac{\gamma-1}{\gamma}\,T_0\frac{A_p}{p_0},
\end{equation*}
and the inviscid standing-wave field (pressure antinodes at both ends) is
\begin{equation*}
  \widehat{p}_{\mathrm{isen}}(x)
  =A_p\bigl(e^{-ik_0 x}+e^{ik_0(x-2L)}\bigr),\qquad
  \widehat{T}_{\mathrm{isen}}
  =\frac{\gamma-1}{\gamma}\,\frac{T_0}{p_0}\,\widehat{p}_{\mathrm{isen}}.
\end{equation*}
Root-mean-square amplitudes plotted below are
\(T_{\mathrm{rms}}=|\widehat{T}|/\sqrt{2}\).

\paragraph{Boundary conditions and mesh}
At \(x=L\) the wall is isothermal and at rest, \(T_w=T_0\), with fully accommodating diffuse reflection, so that the linearized wall temperature vanishes.
The opposite end is a far-field boundary at which both acoustic branches are prescribed, thereby maintaining the standing wave.
The streamwise mesh is stretched toward the wall: \(\Delta x_{\min}=4\times10^{-4}\) in a band of width \(0.06L\) adjacent to \(x=L\), growing geometrically at ratio \(1.08\) to \(\Delta x_{\max}=2\times10^{-3}\) in the core, for \(630\) cells.
The one-dimensional velocity space uses \(80\) uniform points on \([-6,6]\).

\paragraph{Navier--Stokes reference}
The continuum comparison is the one-dimensional linearized NS problem at the same \(\omega\), not a thermal layer pasted onto the inviscid standing wave.
With a quiescent base state and the harmonic factor \(e^{-i\omega t}\), Stokes' hypothesis (\(\mu_b=0\)) gives
\begin{equation}
  \label{eq:ns_lin}
  \begin{aligned}
    -i\omega\,\widehat{\rho}+\rho_0\partial_x\widehat{U}
    &=0,\\
    -i\omega\rho_0\widehat{U}
    &=-\partial_x\widehat{p}+\mu_e\partial_{xx}\widehat{U},\\
    -i\omega\rho_0 c_v\widehat{T}+p_0\partial_x\widehat{U}
    &=\kappa\partial_{xx}\widehat{T},\\
    \widehat{p}
    &=R_g\bigl(T_0\widehat{\rho}+\rho_0\widehat{T}\bigr),
  \end{aligned}
\end{equation}
where \(\mu_e=4\mu/3\), \(c_v=R_g/(\gamma-1)\), \(c_p=\gamma c_v\) and \(\kappa=\mu c_p/\mathrm{Pr}\).
Quadratic viscous heating is absent at this order.

A modal solution \(\propto e^{ikx}\) reduces Eq.~\eqref{eq:ns_lin} to a quartic in \(k\).
The four roots comprise a Stokes--Kirchhoff acoustic pair \(k_{a\pm}\) with
\begin{equation*}
  k_{a+}
  =\frac{\omega}{c_0}+\mathrm{i}\alpha+\cdots,
  \qquad
  \alpha
  =\frac{\omega^2\mu}{2\rho_0 c_0^3}
  \left(\frac{4}{3}+\frac{\gamma-1}{\mathrm{Pr}}\right),
\end{equation*}
\(k_{a-}=-k_{a+}\), and a thermal pair \(k_{t\pm}\) with \(|k_t|\sim 1/\delta_{\mathrm{th}}\) and
\begin{equation}
  \label{eq:delta_th}
  \delta_{\mathrm{th}}
  =\sqrt{\frac{2\nu}{\mathrm{Pr}\,\omega}},
  \qquad
  \nu=\frac{\mu}{\rho_0}.
\end{equation}
(The root with \(\mathrm{Im}\,k_t>0\) decays away from \(x=0\); that with \(\mathrm{Im}\,k_t<0\) decays away from the wall.)
The general harmonic field is therefore
\begin{equation}
  \label{eq:ns_modes}
  \widehat{\boldsymbol{\varphi}}(x)
  =\sum_{j\in\{a\pm,t\pm\}}
  C_j\,\widehat{\boldsymbol{\varphi}}^{(j)}\,e^{ik_j x},
\end{equation}
where \(\widehat{\boldsymbol{\varphi}}=(\widehat{\rho},\widehat{U},\widehat{p},\widehat{T})\) and each eigenvector follows by substituting \(k_j\) into Eq.~\eqref{eq:ns_lin}.

The four coefficients are fixed as follows.
At \(x=0\) the incoming thermal (entropy) wave is set to zero, \(C_{t+}=0\), and the incoming acoustic Riemann invariant is that of a right-going wave,
\begin{equation}
  \label{eq:ns_inlet}
  \Bigl(\frac{\widehat{p}}{\rho_0 c_0}+\widehat{U}\Bigr)\Big|_{x=0}
  =2 a c_0.
\end{equation}
At the isothermal diffuse wall \(x=L\) (\(T_w=T_0\)), the continuum NS solver uses first-order Maxwell slip and temperature-jump conditions, with wall-normal pointing into the gas,
\begin{equation}
  \label{eq:ns_maxwell}
  \widehat{U}+\lambda\partial_x\widehat{U}=0,
  \qquad
  \widehat{T}+\frac{2\gamma}{(\gamma+1)\mathrm{Pr}}\,\lambda\partial_x\widehat{T}=0,
\end{equation}
where \(\lambda=\mathrm{Kn}\,L\) is the hard-sphere mean free path consistent with \(\mu\).
After these four conditions are imposed, the whole field is rescaled so that the driven pressure antinode of the harmonic far field is recovered,
\begin{equation}
  \label{eq:ns_scale}
  \widehat{p}(0)=2A_p.
\end{equation}
The resulting \(\widehat{T}_{\mathrm{NS}}(x)\) is the curve compared with LUGKS.

The same solution also tells how much of the wave bounces back.
Label \(k_{a+}\) the acoustic root with \(\mathrm{Re}\,k>0\) (right-going, towards the wall) and \(k_{a-}=-k_{a+}\).
On each of these two roots the momentum equation in Eq.~\eqref{eq:ns_lin} gives the travelling-wave pressure from the velocity amplitude \(C_{a\pm}\),
\begin{equation}
  \label{eq:p_pm}
  \widehat{p}_\pm(x)
  =\Bigl(\frac{\omega\rho_0}{k_{a\pm}}+\mathrm{i}\mu_e k_{a\pm}\Bigr)
  C_{a\pm}\,e^{ik_{a\pm}x}.
\end{equation}
The coefficient \(C_{a-}\) of the left-going (returning) wave is not assigned by hand: it is the unknown in Eq.~\eqref{eq:ns_modes} that, together with \(C_{a+}\) and the wall thermal mode, satisfies Eqs.~\eqref{eq:ns_inlet}--\eqref{eq:ns_scale}.
Evaluating Eq.~\eqref{eq:p_pm} at \(x=L\) then yields the incident and reflected acoustic pressures at the wall, and
\begin{equation*}
  R_w
  =\frac{\widehat{p}_-(L)}{\widehat{p}_+(L)}.
\end{equation*}
If the wall were inviscid and impermeable it would be a pressure antinode, everything that arrives would leave, and \(R_w=1\).
The Maxwell slip/jump conditions, Eq.~\eqref{eq:ns_maxwell}, absorb part of the incident energy in the Stokes layers, so \(|R_w|<1\): the two acoustic waves are unequal and no longer cancel at \(x/L=1/4\) and \(3/4\).
The RMS cusps then sit at
\begin{equation}
  \label{eq:T_node}
  \frac{T_{\mathrm{rms,node}}}{A_T}
  =\frac{\bigl||\widehat{p}_+|-|\widehat{p}_-|\bigr|}{\sqrt{2}\,A_p}.
\end{equation}
At \(\mathrm{Kn}=10^{-3}\), \(5\times10^{-3}\), \(10^{-2}\) and \(2\times10^{-2}\) one has
\(\alpha L\approx 0.028\), \(0.14\), \(0.28\) and \(0.57\), and
\(\delta_{\mathrm{th}}/L\approx 0.017\), \(0.038\), \(0.054\) and \(0.077\), respectively.

\subsubsection{Results}
\label{sec:sound_wall_results}

\begin{figure}[!htbp]
  \centering
  \begin{subfigure}[t]{0.4\textwidth}
    \includegraphics[width=\textwidth]{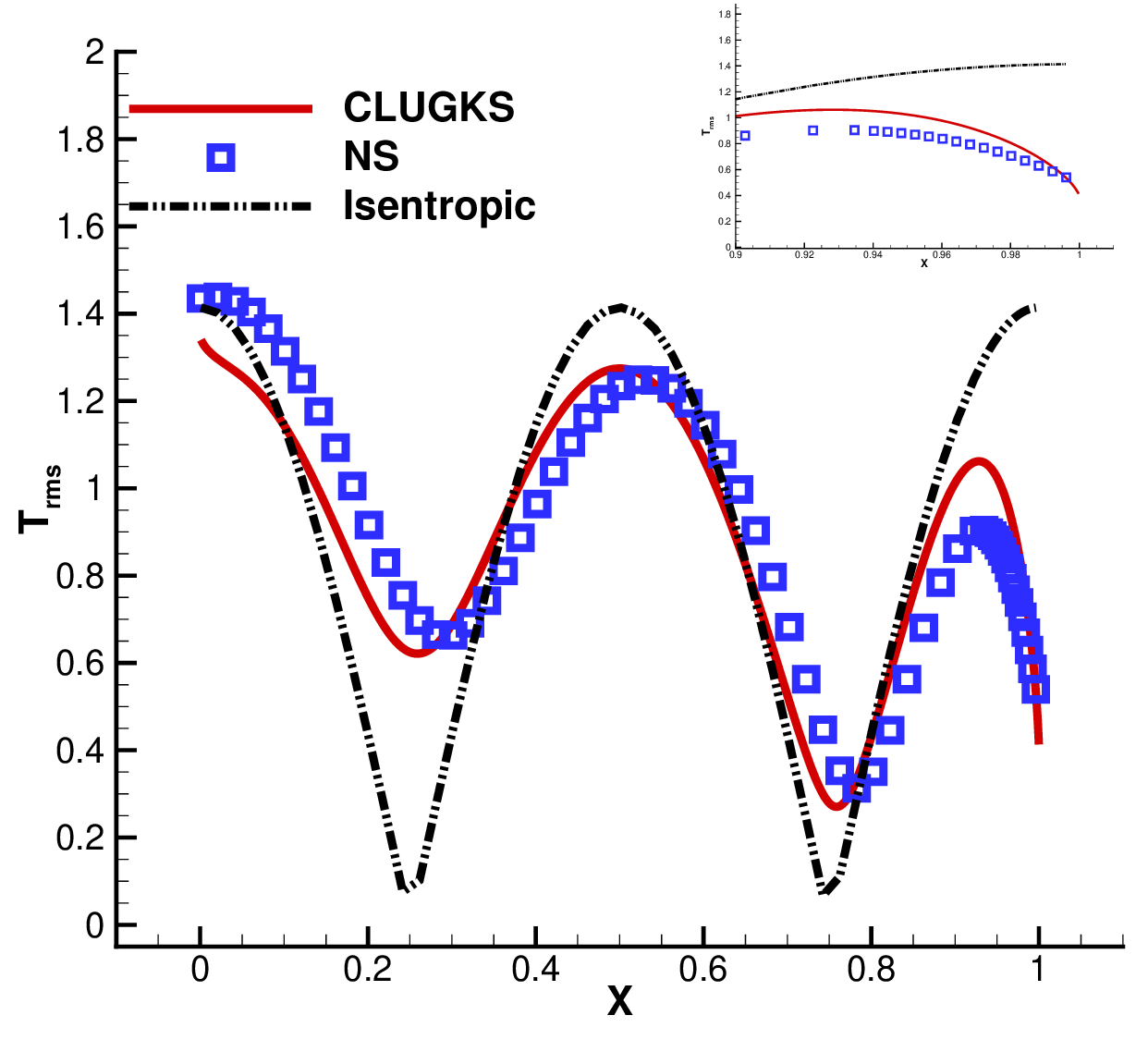}
    \caption{$\mathrm{Kn}=2\times10^{-2}$}
  \end{subfigure}
  \quad
  \begin{subfigure}[t]{0.4\textwidth}
    \includegraphics[width=\textwidth]{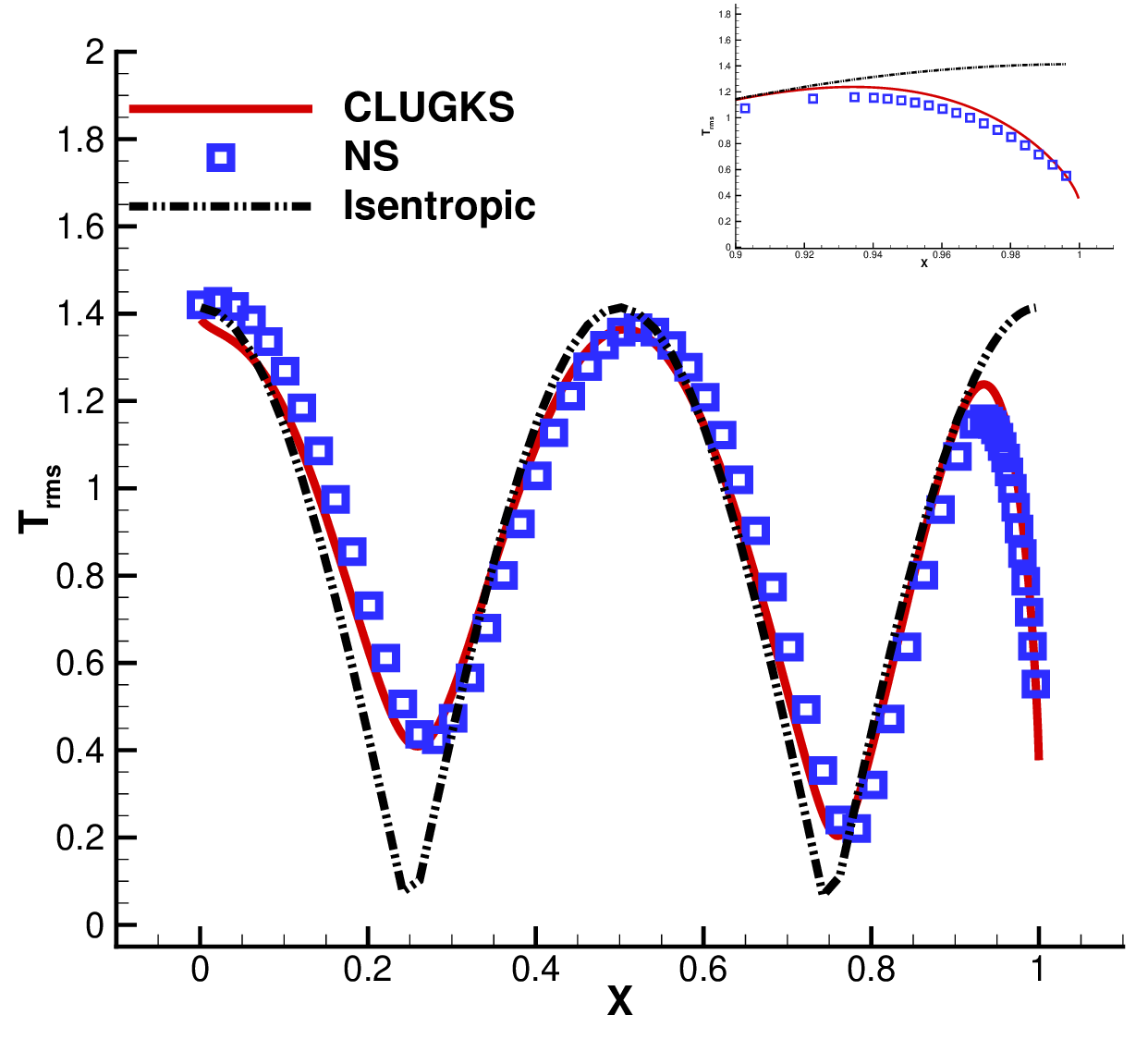}
    \caption{$\mathrm{Kn}=10^{-2}$}
  \end{subfigure}\\[0.6em]
  \begin{subfigure}[t]{0.4\textwidth}
    \includegraphics[width=\textwidth]{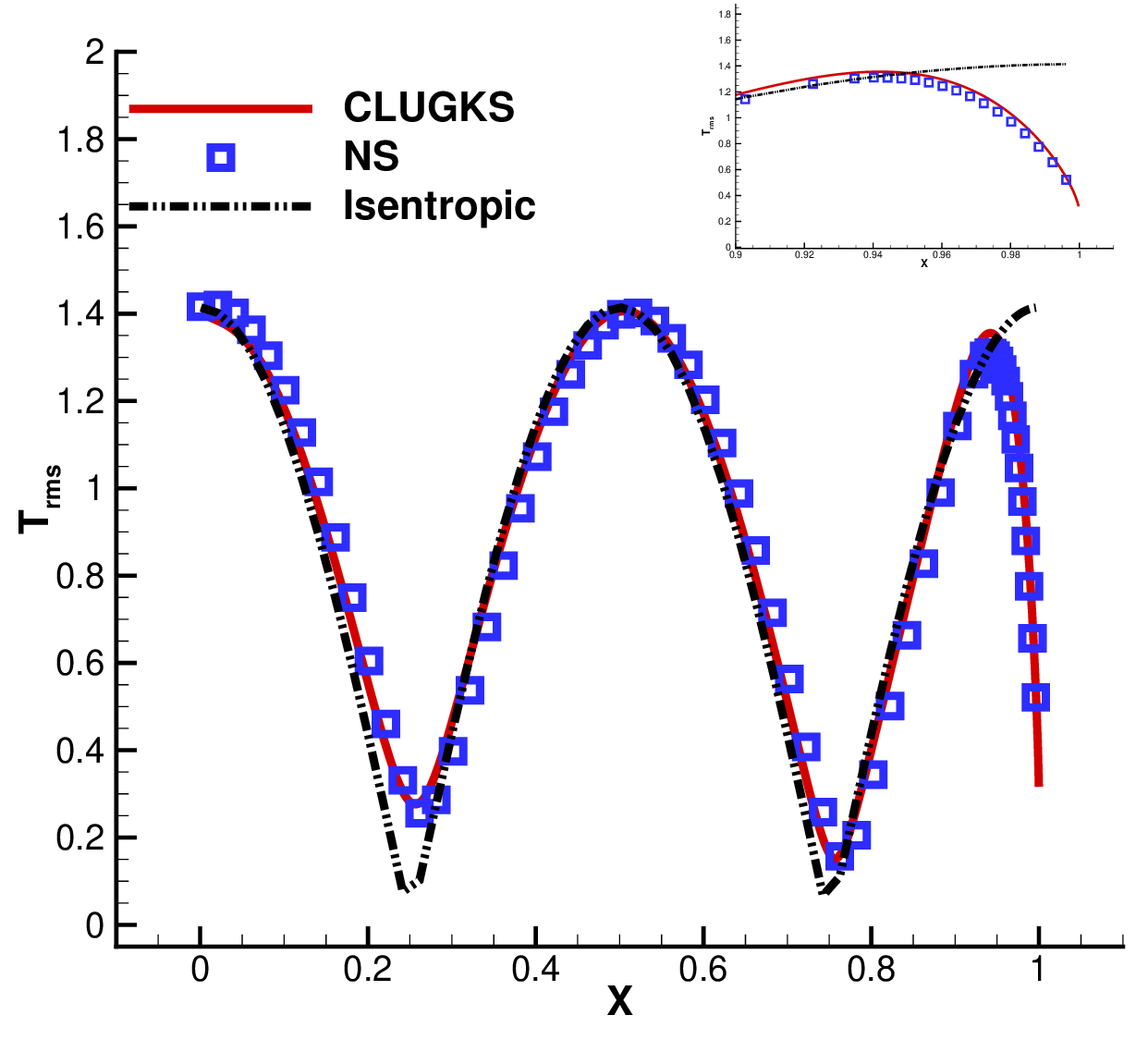}
    \caption{$\mathrm{Kn}=5\times10^{-3}$}
  \end{subfigure}
  \quad
  \begin{subfigure}[t]{0.4\textwidth}
    \includegraphics[width=\textwidth]{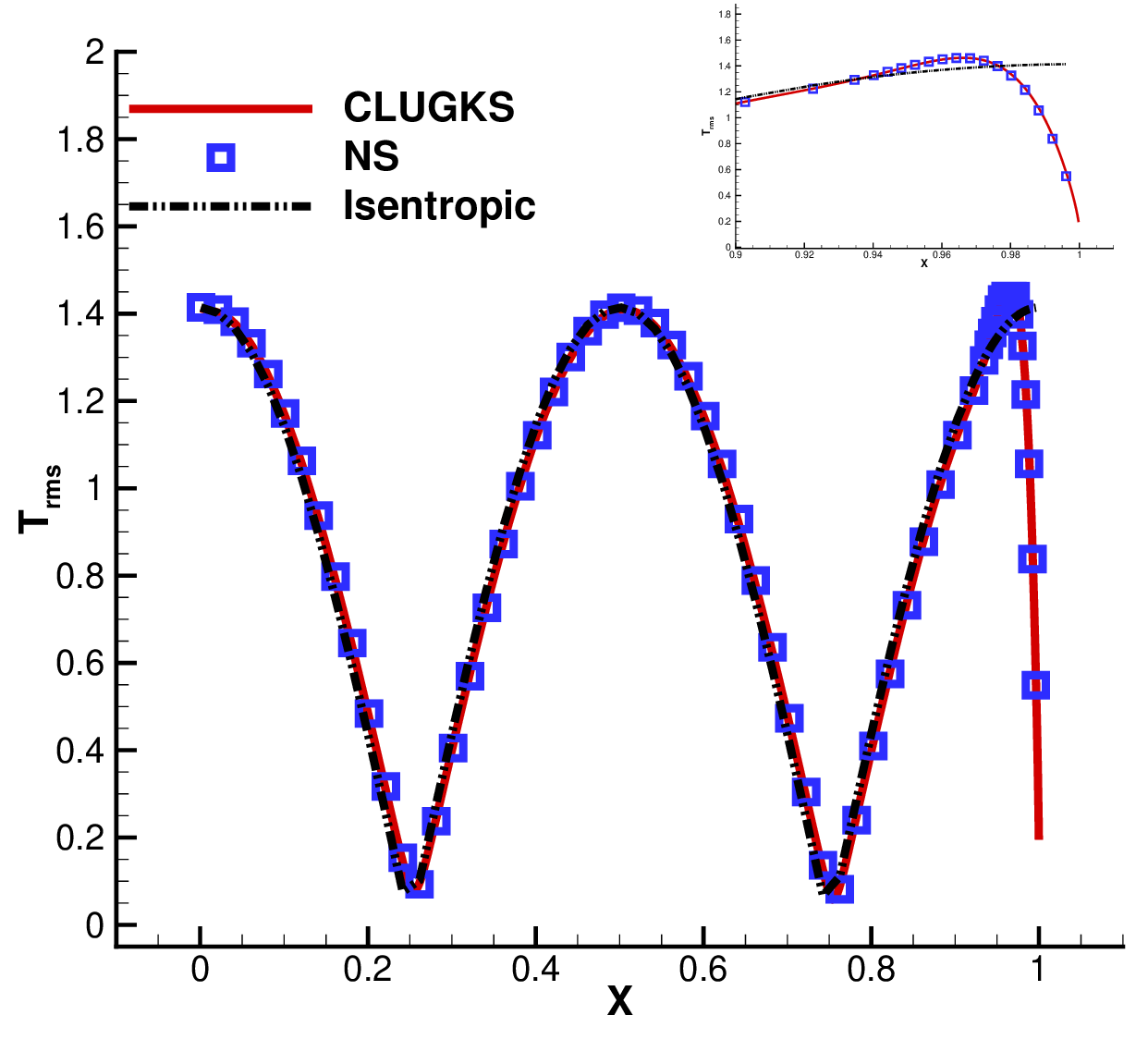}
    \caption{$\mathrm{Kn}=10^{-3}$}
  \end{subfigure}
  \caption{Standing acoustic wave at an isothermal diffuse-reflection wall:
  RMS temperature fluctuation \(T_{\mathrm{rms}}/A_T\) from the complex LUGKS (solid),
  the inviscid isentropic standing wave (dash-dotted) and the linearized NS solution, Eqs.~\eqref{eq:ns_modes}--\eqref{eq:ns_scale} (symbols).
  Right-hand frames zoom on the wall layer.}
  \label{fig:soundWall}
\end{figure}

Figure~\ref{fig:soundWall} compares \(T_{\mathrm{rms}}/A_T\) from the complex LUGKS with the isentropic standing wave and with Eqs.~\eqref{eq:ns_modes}--\eqref{eq:ns_scale}.
The visible mismatch of the inviscid profile is at the cusps \(x/L=1/4\) and \(3/4\): the kinetic solution does not reach zero there, and the lift grows with \(\mathrm{Kn}\) as Eq.~\eqref{eq:T_node}.
The four-mode NS field with \(|R_w|<1\) recovers that filling; grafting a Stokes layer onto an equal-amplitude standing wave (\(R_w=1\)) would still pin the nodes to zero.
As \(\mathrm{Kn}\) is reduced, the inlet amplitude approaches the isentropic peak \(\sqrt{2}\):
the relative deficit in \(T_{\mathrm{rms}}\) at the first cell falls from about \(5\%\) at \(\mathrm{Kn}=2\times10^{-2}\) and \(2\%\) at \(\mathrm{Kn}=10^{-2}\) to well below \(1\%\) at \(\mathrm{Kn}=10^{-3}\),
while \(P_{\mathrm{rms}}/A_p\) agrees with \(\sqrt{2}\) to three figures already at \(\mathrm{Kn}=5\times10^{-3}\).
Near \(x=L\) the temperature drops over a width consistent with Eq.~\eqref{eq:delta_th}.
The wall-adjacent values \(T_{\mathrm{rms}}/A_T\approx 0.41\), \(0.38\), \(0.32\) and \(0.20\) at \(\mathrm{Kn}=2\times10^{-2}\), \(10^{-2}\), \(5\times10^{-3}\) and \(10^{-3}\) follow the kinetic temperature jump of diffuse reflection rather than \(\widehat{T}=0\), and the slip in \(\widehat{U}\) remains of order a few thousandths of \(A_u=a c_0\).

\subsection{Near-critical fast acoustic wave--shock interaction}
\label{sec:acoustic_incidence}

This second case concerns a Mach\,2 normal shock interacting with a near-critical upstream fast acoustic wave \(La1\).
Unlike the inviscid Euler computation of He and Su~\cite{he2026revisiting}, the present simulation retains a kinetic shock structure, so that both the far-field modal amplitudes and the near-shock displacement can be examined on the same solution.
The complex LUGKS is first verified against their spatial linear interaction analysis (LIA) at \(k_2^*=10^{-3}\); the wavelength--mean-free-path ratio of the incident wave is then varied.

\subsubsection{Case configuration}
\label{sec:case2_config}

\paragraph{Continuum scaling}
Under a continuum (mean-free-path) nondimensionalization, the reference wave numbers and frequency \((k_1,k_2,\omega)=(0,1,0.5)\) of He and Su are mapped to
\begin{equation*}
  (k_1^*,k_2^*,\omega^*)=(0,\,0.001,\,0.0005),
\end{equation*}
which preserves the incidence angle
\begin{equation*}
  \theta_{Lev}=\arctan\!\Bigl(\frac{U_{1L}k_2^*}{\omega^*}\Bigr)=63.4^\circ
\end{equation*}
near the lower critical angle \(62.1^\circ\) at which the transmitted fast acoustic branch becomes evanescent.
The computational domain spans one transverse wavelength, \(L_y=2\pi/k_2^*\approx 6283.19\), with
\(x\in[-L_y,\,3L_y]\approx[-6283.19,\,18849.56]\), consistent with their DNS window after the same scaling.
In the streamwise direction the mesh is geometrically stretched about the shock at \(x=0\): within \(\pm 20\) mean free paths of the shock the spacing is refined to \(\Delta x_{\min}=0.5\), and outside this band it grows at a ratio \(1.1\) toward a far-field spacing \(\Delta x_{\max}=L_y/160\approx 39.3\), giving about \(790\) cells in \(x\).
The transverse direction is uniform with \(N_y=32\) cells per wavelength (\(\Delta y=L_y/N_y\)), and the \(y\)-boundaries are periodic.
The particle velocity space is discretized by a uniform Cartesian mesh of \(80\times 80\) points over \([-u_{\max},u_{\max}]^2\) with \(u_{\max}=4\), which covers several thermal velocities about both the upstream and downstream freestreams.

\begin{figure}[!htbp]
  \centering
  \includegraphics[width=0.45\textwidth]{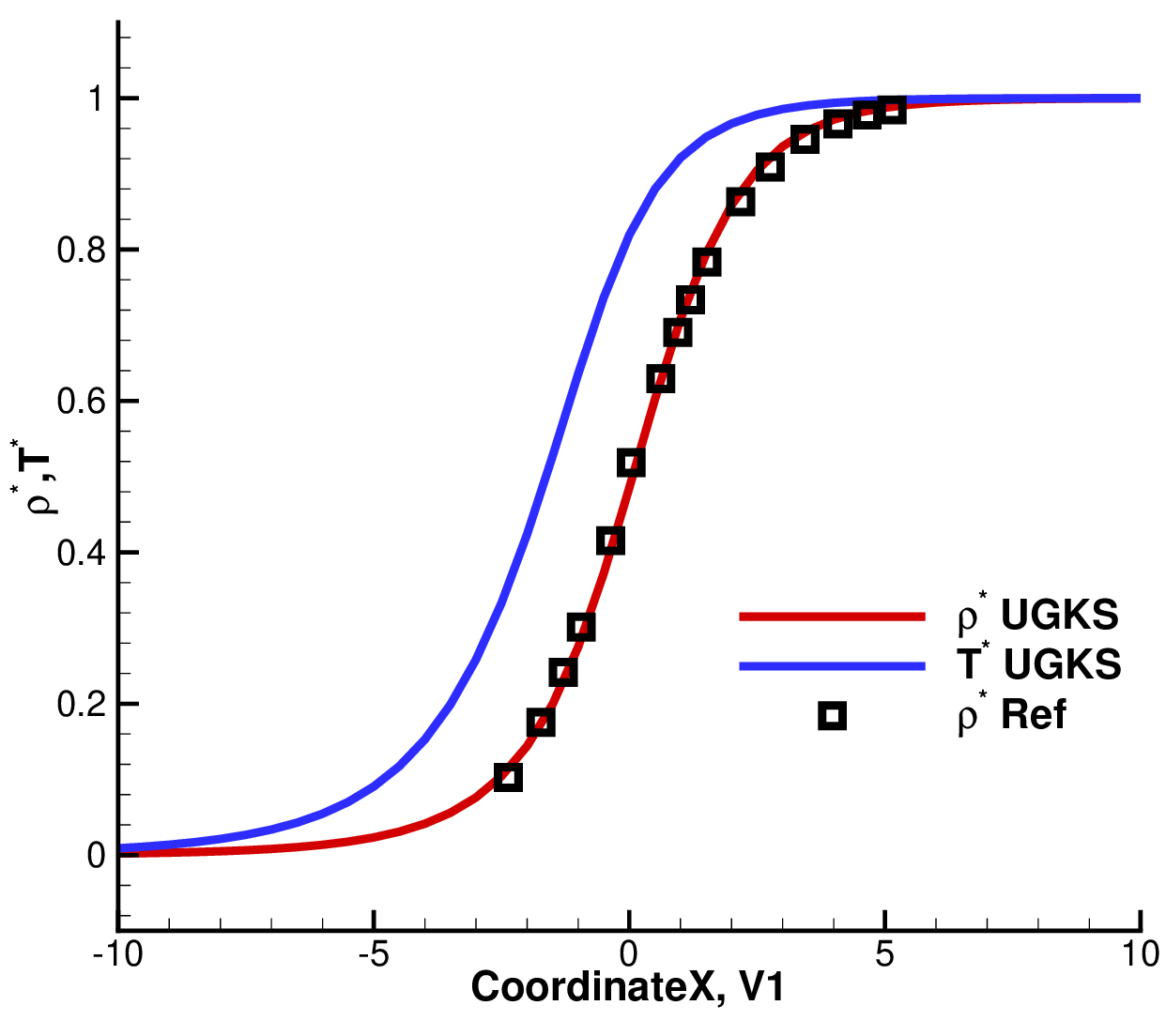}
  \caption{Mach\,2 base shock structure: normalized density and temperature across the kinetic shock layer, compared with the experimental density profile of Alsmeyer~\cite{alsmeyer1976}.}
  \label{fig:shockStructure}
\end{figure}

\paragraph{Gas and base shock}
The gas is treated as diatomic with \(\gamma=7/5\) and \(\mathrm{Pr}=0.72\), and follows the variable-hard-sphere collision model with viscosity exponent \(\omega=0.74\).
The gas constant is chosen as
\begin{equation*}
  R_g=\frac{1}{\gamma Ma_L^2}=\frac{5}{28},
\end{equation*}
so that the upstream Mach number is \(Ma_L=U_{1L}/\sqrt{\gamma R_g T_L}=2\) with \(U_{1L}=T_L=1\).
The reference viscosity at \(T_{\mathrm{ref}}=1\) is fixed by Bird's VHS mean-free-path relation with unit upstream mean free path \(\lambda_L=1\),
\begin{equation*}
  \mu_{\mathrm{ref}}
  =\frac{15}{2(5-2\omega)(7-2\omega)}\sqrt{2\pi R_g T_{\mathrm{ref}}},
\end{equation*}
whereby the shock thickness remains of order a few mean free paths~\cite{alsmeyer1976}.
The upstream freestream is
\begin{equation*}
  (\rho,U_1,U_2,p,T)_L=\Bigl(1,\,1,\,0,\,\tfrac{5}{28},\,1\Bigr),
\end{equation*}
and the Rankine--Hugoniot downstream state is
\begin{equation*}
  (\rho,U_1,U_2,p,T)_R=\Bigl(\tfrac{8}{3},\,\tfrac{3}{8},\,0,\,\tfrac{45}{56},\,\tfrac{27}{16}\Bigr).
\end{equation*}
A steady base flow is first obtained by UGKS from a discontinuous shock placed at \(x=0\), and is then held fixed for the linearized harmonic computation.
Figure~\ref{fig:shockStructure} shows the normalized density and temperature profiles
\(\rho^*=(\rho-\rho_L)/(\rho_R-\rho_L)\) and \(T^*=(T-T_L)/(T_R-T_L)\),
compared with the experimental density profile of Alsmeyer~\cite{alsmeyer1976}.

\paragraph{Incident wave and boundary conditions}
The complex LUGKS is driven by a single upstream fast acoustic wave \(La1\) of complex amplitude
\begin{equation*}
  a_{La1}=-10^{-4},
\end{equation*}
imposed at the upstream boundary and consistent with the initial disturbance in the upstream half of the domain.
The overall minus sign follows the phase convention of He and Su for this case.
The downstream boundary is non-reflecting and does not prescribe any outgoing modal content, so that the shock displacement is free.
The angular frequency is \(\omega^*=0.0005\).

\subsubsection{Data reduction}
\label{sec:case2_post}

From the complex amplitudes
\(\widehat{\rho}\), \(\widehat{U}_1\), \(\widehat{U}_2\), \(\widehat{p}\) and \(\widehat{T}\),
the following diagnostics are constructed.

\paragraph{Streamwise amplitude profiles}
At each \(x\), the transverse Fourier coefficient at wave number \(k_2^*\) is
\begin{equation*}
  \widehat{\varphi}_{k_2}(x)
  =\frac{1}{L_y}\int_0^{L_y}
  \widehat{\varphi}(x,y)\,e^{-i k_2^* y}\,\mathrm{d}y.
\end{equation*}
These profiles are compared with the LIA modal field for the same incident \(a_{La1}\).
Upstream, only \(La1\) is present; downstream, LIA comprises the transmitted (here evanescent) acoustic wave together with the entropy and vorticity waves generated at the shock.
Figure~\ref{fig:shockAcousticMode} plots \(\mathrm{Re}\,\widehat{\varphi}\), \(\mathrm{Im}\,\widehat{\varphi}\) and \(|\widehat{\varphi}|\) against \(x\) at the verification wave number \(k_2^*=10^{-3}\).

\paragraph{Shock displacement}
Within the kinetic shock layer the perturbation is dominated by a rigid displacement of the base profile,
\begin{equation*}
  \widehat{\varphi}(x)
  \approx
  -\hat{\eta}\,\partial_x \varphi_0(x),
\end{equation*}
where \(\varphi_0\) stands for any of \(\rho\), \(U_1\), \(p\) and \(T\) in the base flow.
Over the fitting window \(\mathcal{I}=\{x:|x|\le 15\}\), a least-squares estimate for each variable is
\begin{equation*}
  \hat{\eta}_{\varphi}
  =
  \frac{
  \displaystyle\int_{\mathcal{I}}
  \bigl(-\partial_x \varphi_0\bigr)^{\!*}\,
  \widehat{\varphi}\,\mathrm{d}x
  }{
  \displaystyle\int_{\mathcal{I}}
  \bigl|\partial_x \varphi_0\bigr|^{2}\,\mathrm{d}x
  },
\end{equation*}
and \(\hat{\eta}_{\mathrm{num}}\) is the average of \(\hat{\eta}_{\varphi}\) over \(\varphi\in\{\rho,U_1,p,T\}\).
The relative residual energy of the fit is
\begin{equation}
  \label{eq:eta_residual}
  \varepsilon_{\eta}
  =
  \frac{
  \displaystyle\sum_{\varphi}
  \int_{\mathcal{I}}
  \bigl|
  \widehat{\varphi}+\hat{\eta}_{\mathrm{num}}\,\partial_x \varphi_0
  \bigr|^{2}\,\mathrm{d}x
  }{
  \displaystyle\sum_{\varphi}
  \int_{\mathcal{I}}
  \bigl|\widehat{\varphi}\bigr|^{2}\,\mathrm{d}x
  }.
\end{equation}
Figure~\ref{fig:shockEta} shows \(\widehat{\rho}\) and \(\widehat{T}\) together with \(-\hat{\eta}_{\mathrm{num}}\partial_x \varphi_0\) (real, imaginary and absolute parts) at \(k_2^*=10^{-3}\):
inside the kinetic shock layer the computed amplitudes collapse onto the rigid-deformation mode associated with \(\hat{\eta}_{\mathrm{num}}\).

\paragraph{Downstream modal amplitudes}
On the downstream side the same Fourier amplitudes are projected onto the LIA right-going eigenmodes \(\{Ra1,Re,Rv1\}\) to obtain complex coefficients \(a_{Ra1}\), \(a_{Re}\) and \(a_{Rv1}\), referenced to the mean shock location.

\subsubsection{Verification against LIA}
\label{sec:case2_verify}

\begin{figure}[htbp]
  \centering
  \begin{subfigure}[t]{0.48\textwidth}
    \includegraphics[width=\textwidth]{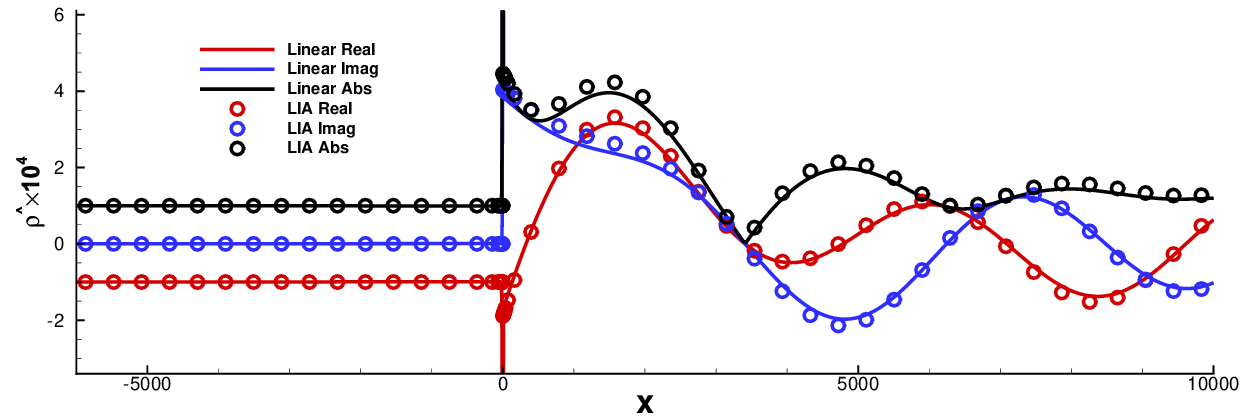}
    \caption*{$\widehat{\rho}$}
  \end{subfigure}
  \hfill
  \begin{subfigure}[t]{0.48\textwidth}
    \includegraphics[width=\textwidth]{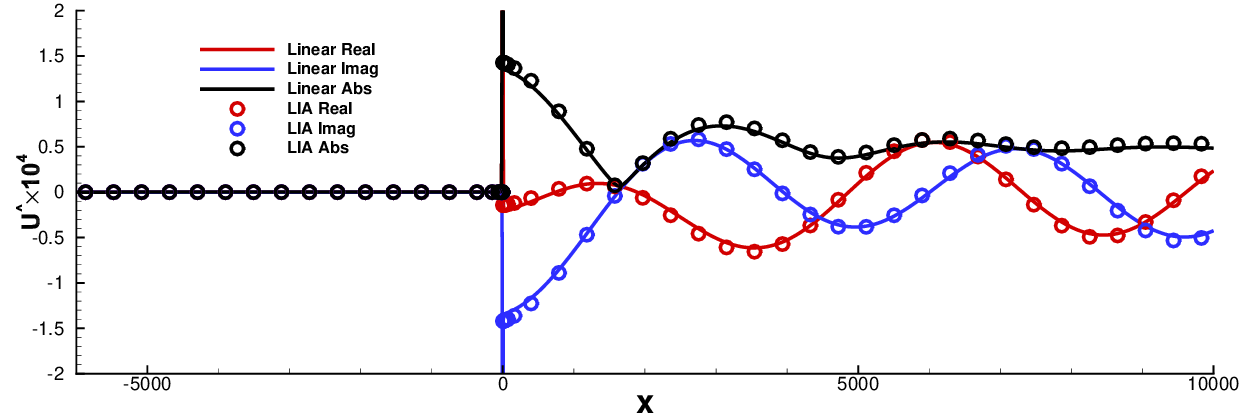}
    \caption*{$\widehat{U}_1$}
  \end{subfigure}\\[0.6em]
  \begin{subfigure}[t]{0.48\textwidth}
    \includegraphics[width=\textwidth]{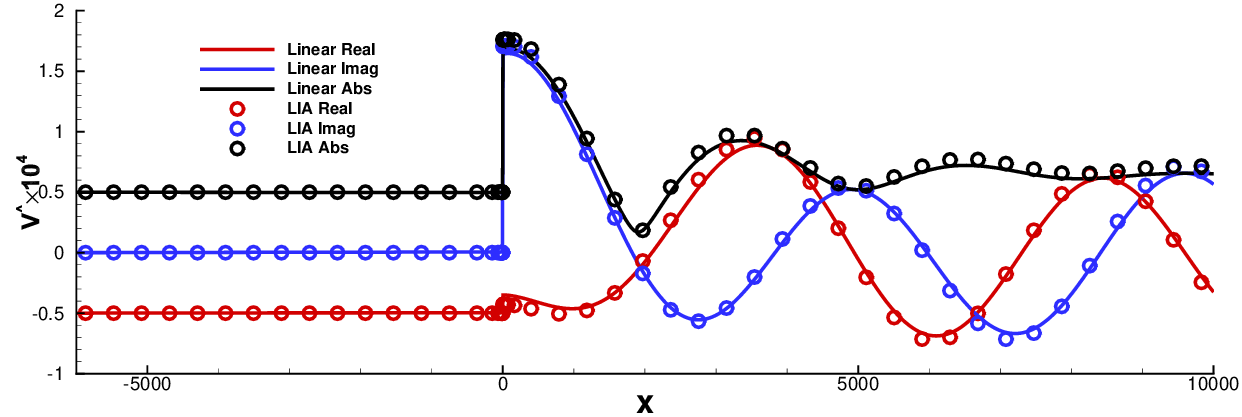}
    \caption*{$\widehat{U}_2$}
  \end{subfigure}
  \hfill
  \begin{subfigure}[t]{0.48\textwidth}
    \includegraphics[width=\textwidth]{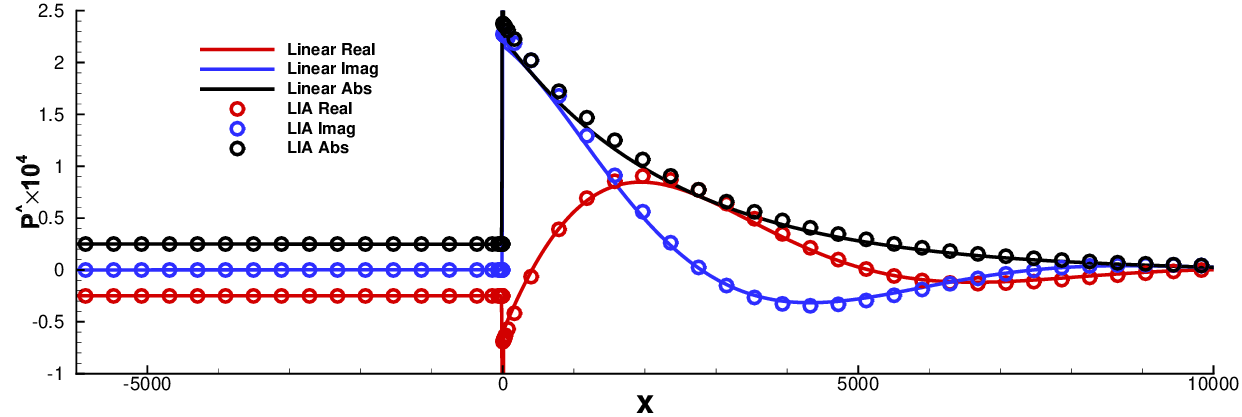}
    \caption*{$\widehat{p}$}
  \end{subfigure}
  \caption{Complex amplitude versus \(x\) (real, imaginary and absolute parts) compared with spatial LIA (markers) for \(\widehat{\rho}\), \(\widehat{U}_1\), \(\widehat{U}_2\) and \(\widehat{p}\) at \(k_2^*=10^{-3}\).}
  \label{fig:shockAcousticMode}
\end{figure}

The verification uses the continuum-scaled wave number \(k_2^*=10^{-3}\) of Sec.~\ref{sec:case2_config}, for which the inviscid spatial LIA is directly comparable.
The streamwise amplitudes in Fig.~\ref{fig:shockAcousticMode} agree closely with LIA in the uniform freestreams:
rapid decay of pressure behind the shock marks the evanescent transmitted acoustic branch, while the far-field density and velocity retain the entropy and vorticity waves generated at the shock.

\begin{figure}[htbp]
  \centering
  \begin{subfigure}[t]{0.48\textwidth}
    \includegraphics[width=\textwidth]{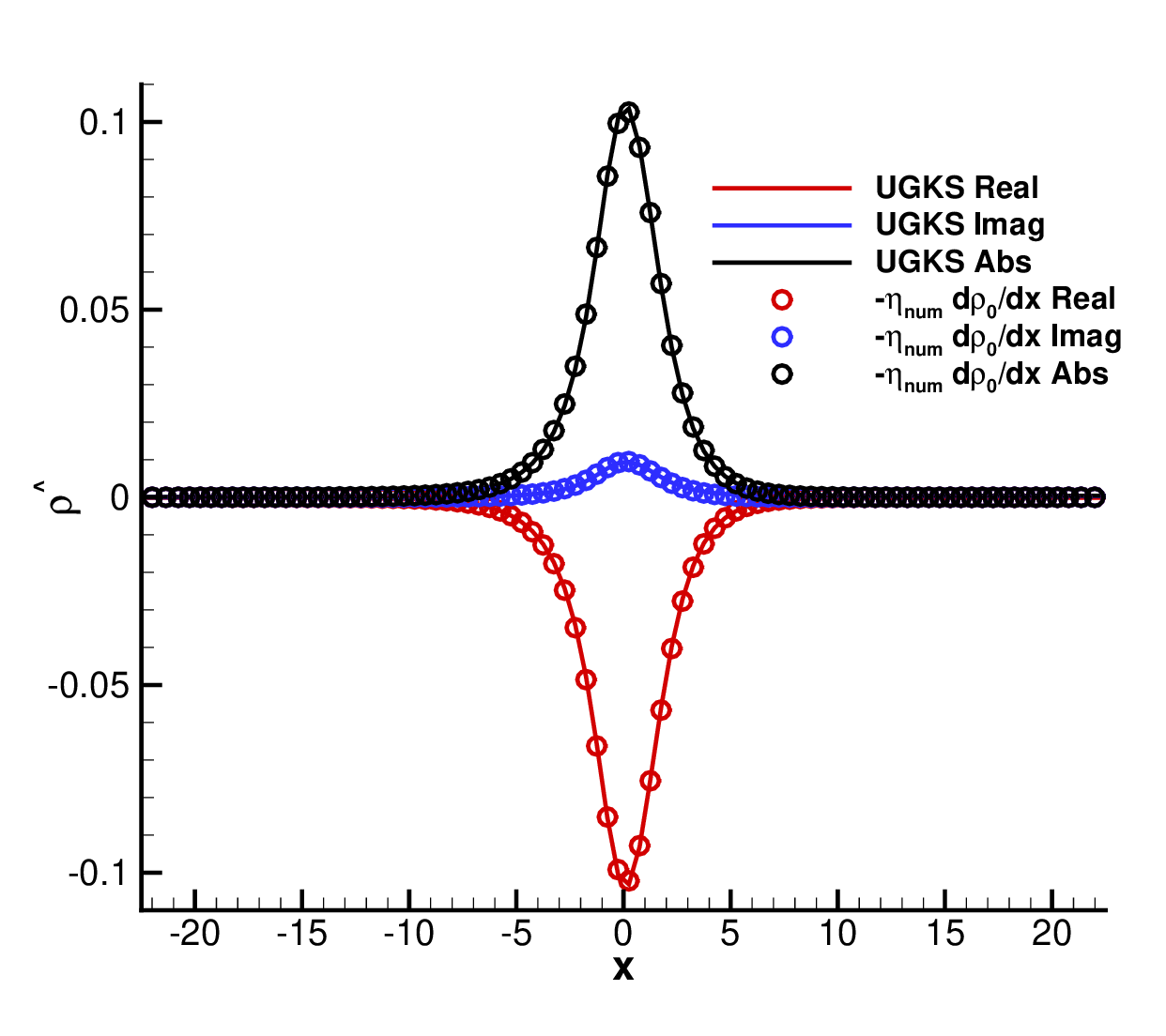}
    \caption*{$\widehat{\rho}$}
  \end{subfigure}
  \hfill
  \begin{subfigure}[t]{0.48\textwidth}
    \includegraphics[width=\textwidth]{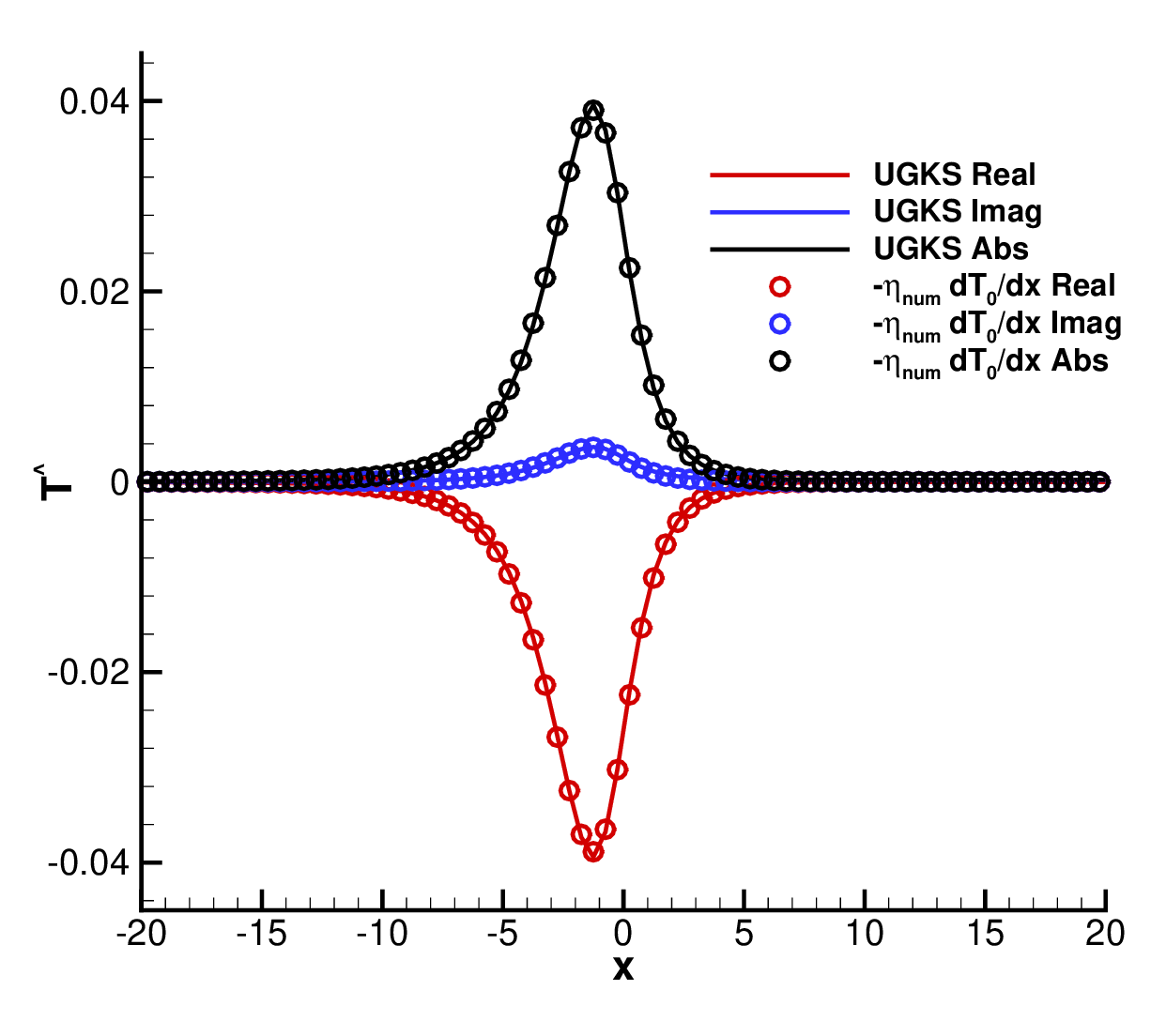}
    \caption*{$\widehat{T}$}
  \end{subfigure}
  \caption{Shock deformation diagnostic at \(k_2^*=10^{-3}\): UGKS amplitudes (solid) versus the numerical deformation mode \(-\hat{\eta}_{\mathrm{num}}\partial_x \varphi_0\) (dashed) for density and temperature (real, imaginary and absolute parts).
  Within the shock layer the computed field coincides with the deformation mode associated with \(\hat{\eta}_{\mathrm{num}}\).}
  \label{fig:shockEta}
\end{figure}

Downstream of the shock, the \(k_2^*\) field is projected onto the LIA eigenmodes to recover the complex amplitudes of the transmitted fast acoustic wave \(Ra1\), the entropy wave \(Re\) and the vorticity wave \(Rv1\), referred to the shock location through the phase factor \(e^{-ik_1(x-x_s)}\).
For the prescribed incident wave \(a_{La1}=-10^{-4}\), the UGKS modal amplitudes and shock displacement agree with the spatial LIA to within a few percent; this is seen directly in the streamwise profiles of Figs.~\ref{fig:shockAcousticMode} and~\ref{fig:shockEta}.
The relative residual, Eq.~\eqref{eq:eta_residual}, of the displacement fit is \(\varepsilon_{\eta}\approx 1.9\times10^{-4}\) (about \(0.02\%\)).

\subsubsection{Effect of the wavelength--mean-free-path ratio}
\label{sec:case2_k}

With the scheme verified at \(k_2^*=10^{-3}\), for which the acoustic wavelength \(\lambda\) greatly exceeds the molecular mean free path \(\ell\) and the inviscid LIA remains appropriate, the same incident \(La1\) configuration is repeated at larger continuum-scaled wave numbers \(k_2^*=0.01\), \(0.02\) and \(0.05\), with \(\omega^*\) scaled so that the incidence angle \(\theta_{Lev}\) is unchanged.
These choices reduce the ratio \(\lambda/\ell\) (equivalently, increase \(k_2^*\) relative to the inverse mean free path), so that the interaction leaves the continuum long-wave regime: deviations from the inviscid LIA become more visible near the shock, while the far-field modal structure remains recognisable.

\begin{figure}[htbp]
  \centering
  \begin{subfigure}[t]{0.48\textwidth}
    \includegraphics[width=\textwidth]{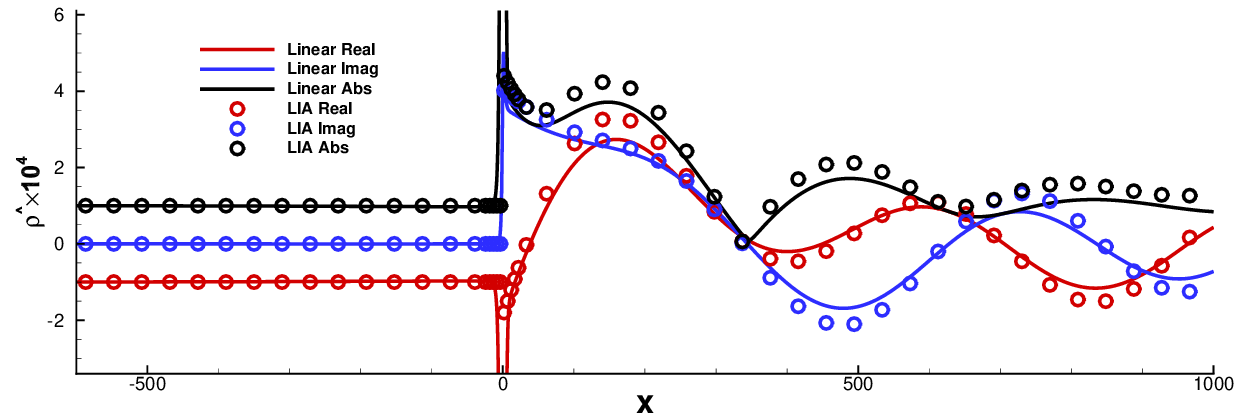}
    \caption*{$k_2^*=0.01$, $\widehat{\rho}$}
  \end{subfigure}
  \hfill
  \begin{subfigure}[t]{0.48\textwidth}
    \includegraphics[width=\textwidth]{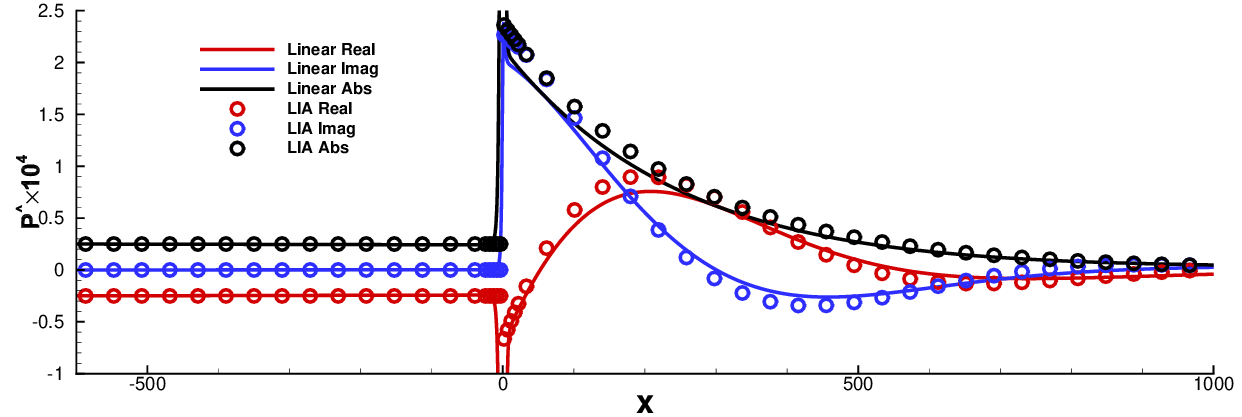}
    \caption*{$k_2^*=0.01$, $\widehat{p}$}
  \end{subfigure}\\[0.6em]
  \begin{subfigure}[t]{0.48\textwidth}
    \includegraphics[width=\textwidth]{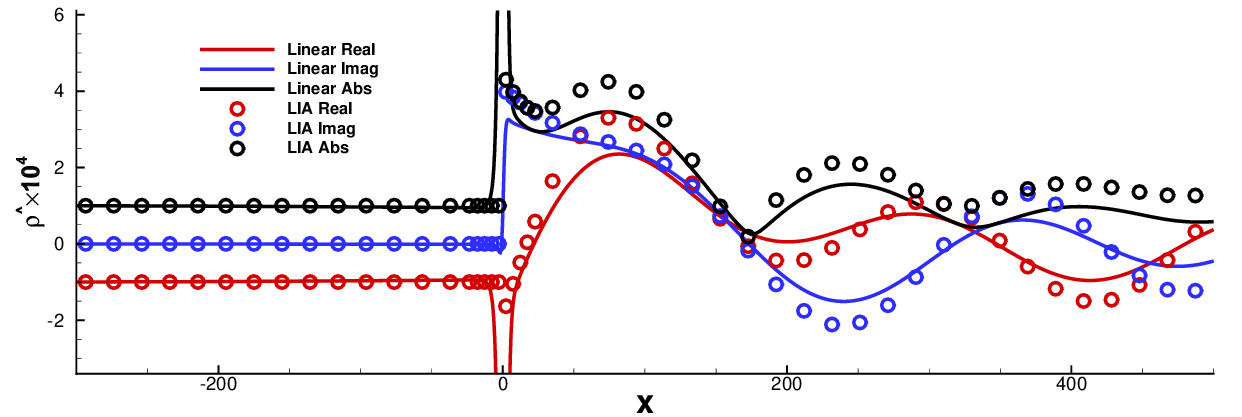}
    \caption*{$k_2^*=0.02$, $\widehat{\rho}$}
  \end{subfigure}
  \hfill
  \begin{subfigure}[t]{0.48\textwidth}
    \includegraphics[width=\textwidth]{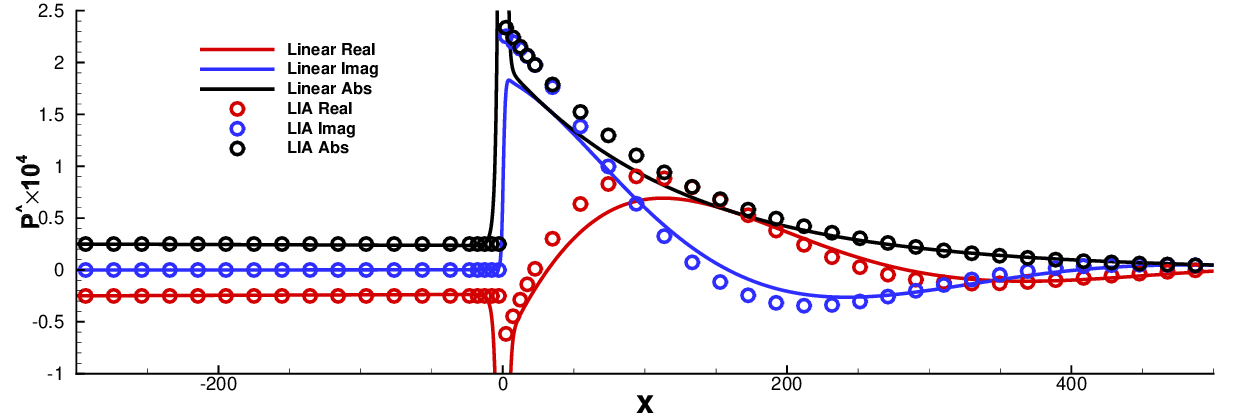}
    \caption*{$k_2^*=0.02$, $\widehat{p}$}
  \end{subfigure}\\[0.6em]
  \begin{subfigure}[t]{0.48\textwidth}
    \includegraphics[width=\textwidth]{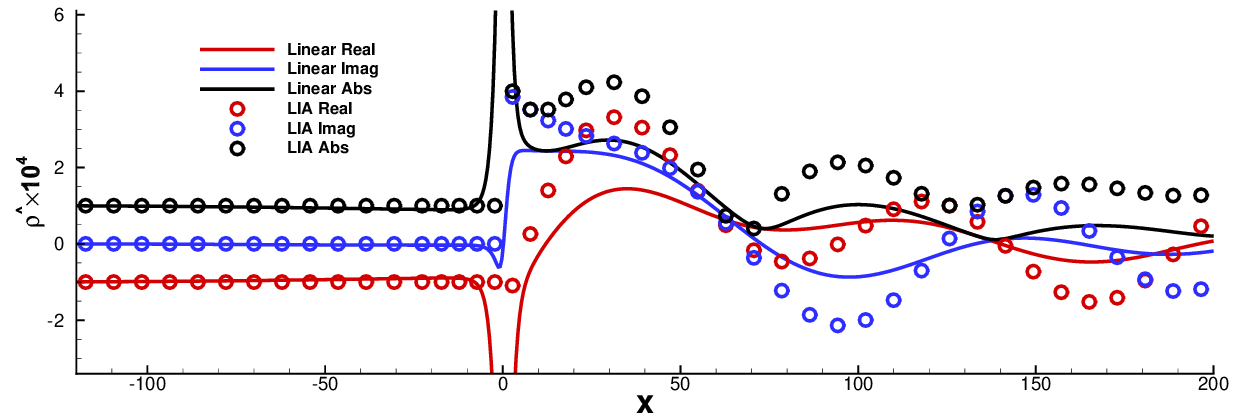}
    \caption*{$k_2^*=0.05$, $\widehat{\rho}$}
  \end{subfigure}
  \hfill
  \begin{subfigure}[t]{0.48\textwidth}
    \includegraphics[width=\textwidth]{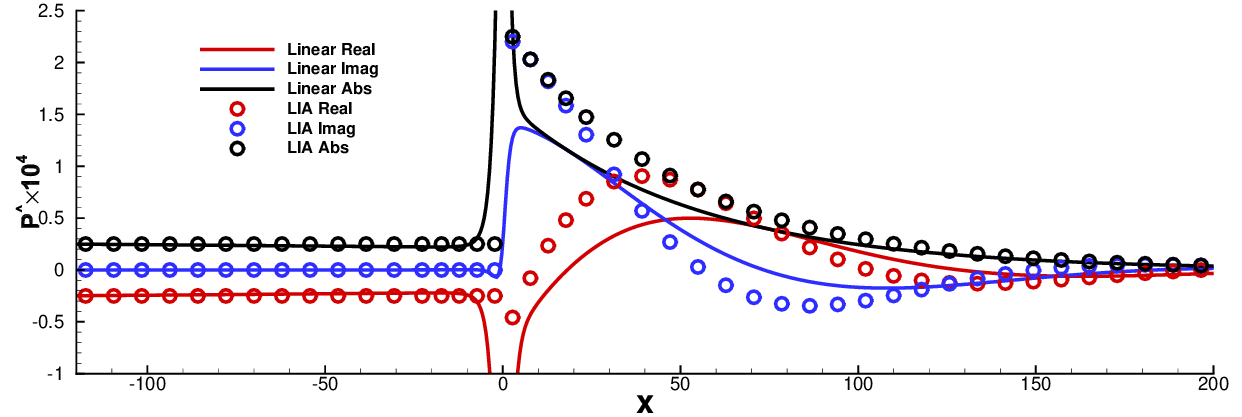}
    \caption*{$k_2^*=0.05$, $\widehat{p}$}
  \end{subfigure}
  \caption{Complex amplitude versus \(x\) for decreasing wavelength--mean-free-path ratio (\(k_2^*=0.01\), \(0.02\) and \(0.05\)): density (left) and pressure (right), compared with spatial LIA (markers).}
  \label{fig:shockAcousticMode_k}
\end{figure}

Figure~\ref{fig:shockAcousticMode_k} compares the streamwise density and pressure amplitudes with LIA at the three wave numbers.
Upstream of the shock the incident \(La1\) field remains flat and matches LIA at every \(k_2^*\), as expected for a prescribed freestream mode.
Across the kinetic shock layer both \(\widehat{\rho}\) and \(\widehat{p}\) show a sharp peak associated with the shock displacement; this near-shock spike is absent from the inviscid LIA markers and grows more conspicuous as \(k_2^*\) increases.
Downstream, pressure decays within a short distance of the shock, consistent with an evanescent transmitted acoustic branch at this near-critical incidence, whereas density retains longer-lived oscillations carried by the entropy (and vorticity) waves.
At \(k_2^*=0.01\) the UGKS profiles still follow the LIA markers closely outside the shock layer.
At \(k_2^*=0.02\) and especially \(0.05\), the post-shock density amplitude from UGKS falls below the inviscid prediction and the phase of the oscillations drifts, while the pressure signal damps even faster than LIA suggests.
These trends show that, once the acoustic wavelength approaches a few tens of mean free paths, collisional dissipation inside the finite-thickness shock and in the post-shock gas weakens the transmitted modal content relative to the discontinuous-shock LIA---an effect that only a kinetic description captures.

\subsection{Receptivity of hypersonic flow over a blunt wedge}
\label{sec:blunt_receptivity}

The third case examines freestream acoustic receptivity of a hypersonic blunt-body flow, following the geometry and freestream conditions of Li et~al.~\cite{li2026receptivity}.
A steady bow-shock base flow about a cylindrical nose with a tangent wedge is first obtained by UGKS; small harmonic freestream disturbances are then imposed and the complex LUGKS is used to compute the near-wall response.
The DNS of Ref.~\cite{li2026receptivity} provides the reference solution for the wall response.
Unlike the planar-shock configuration of Sec.~\ref{sec:acoustic_incidence}, both the curved shock and the stagnation-region entropy/vorticity layer enter the receptivity path.

\paragraph{Geometry and freestream}
The body is a blunt wedge formed by a circular cylinder of nose radius \(R_n=0.1\,\mathrm{mm}\) joined to a tangent half-angle wedge of \(\theta=9^\circ\).
The computational domain includes the bow shock.
Cartesian coordinates \((x,y)\) and body-fitted curvilinear coordinates \((\xi,\eta)\) are both employed, with the origin at the stagnation point.
The freestream is hypersonic at zero angle of attack,
\begin{equation*}
  Ma_\infty=5,
  \qquad
  T_\infty=216.7\,\mathrm{K},
  \qquad
  Re_\infty'=9.2\times10^{6}\,\mathrm{m}^{-1},
\end{equation*}
where \(Re_\infty'\) is the unit Reynolds number based on freestream quantities.
Taking \(R_n\) as the reference length gives the nose Reynolds number
\begin{equation*}
  Re_{R_n}
  =Re_\infty' R_n
  =920.
\end{equation*}
All flow variables are nondimensionalized by the corresponding freestream values, except that pressure is scaled by \(\rho_\infty u_\infty^2\).

\paragraph{Wall condition}
The wall is isothermal with
\begin{equation*}
  T_w=300\,\mathrm{K},
\end{equation*}
so that the wall-to-freestream temperature ratio is \(T_w/T_\infty\approx 1.384\).
A fully accommodating isothermal diffuse-reflection condition is used on the solid surface.
The far-field boundary lies outside the bow shock and admits the freestream together with the prescribed harmonic disturbance.

\paragraph{Incident freestream wave}
The freestream disturbance is a time-harmonic fast acoustic wave of frequency
\begin{equation*}
  f=300\,\mathrm{kHz},
  \qquad
  \omega=2\pi f,
\end{equation*}
imposed at the inflow with a small complex amplitude.
In the freestream the corresponding streamwise wavelength is
\begin{equation*}
  \lambda_\infty
  =\frac{u_\infty+a_\infty}{f},
\end{equation*}
where \(a_\infty\) is the freestream sound speed.

\begin{figure}[!htbp]
  \centering
  \includegraphics[width=0.55\textwidth]{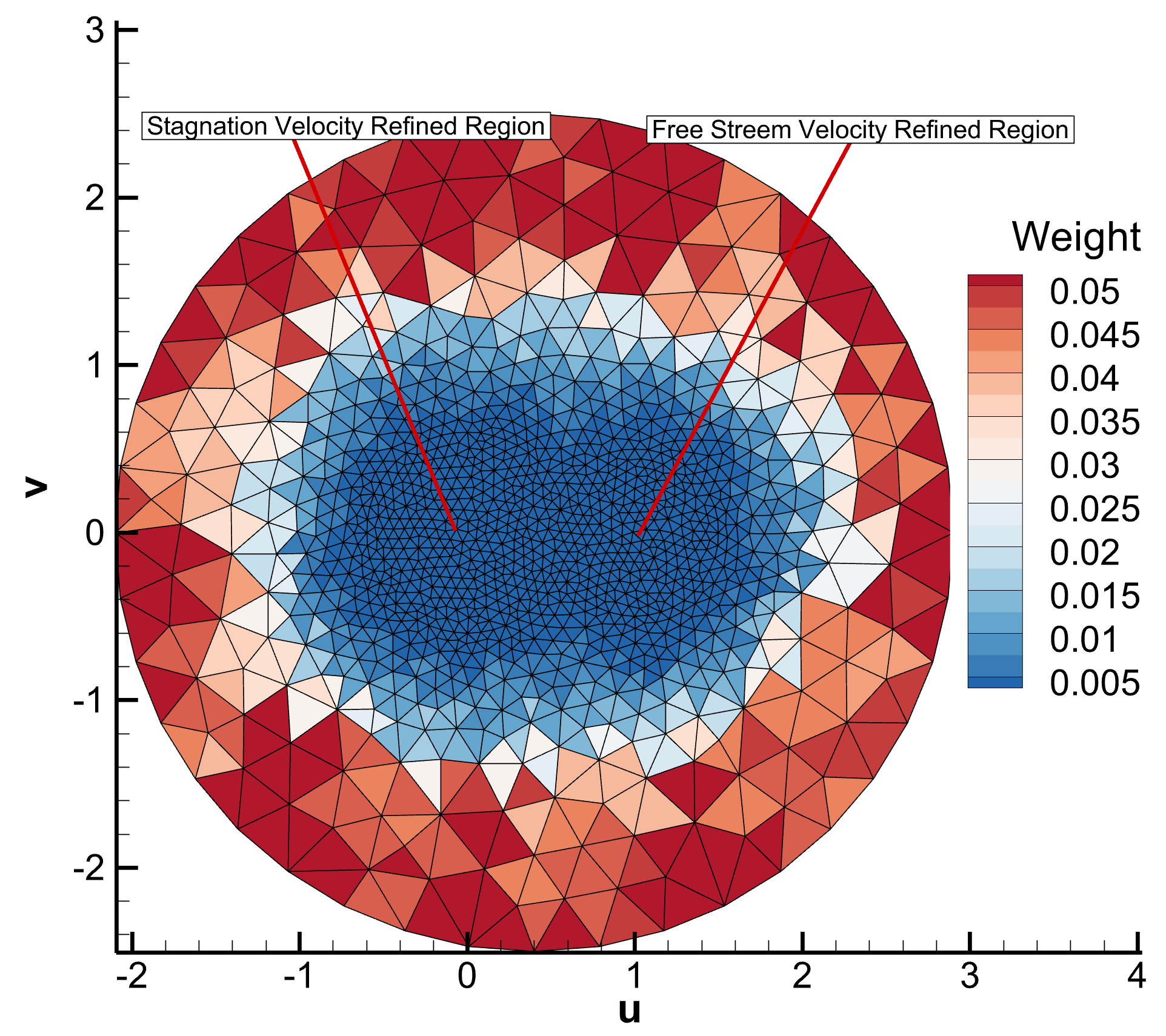}
  \caption{Unstructured discrete-velocity mesh for the blunt-wedge case (\(2152\) triangular elements).
  Colour indicates the quadrature weight.
  The DVS is refined near the zero-velocity and freestream points.}
  \label{fig:bluntnessDVS}
\end{figure}

\paragraph{Physical and velocity-space meshes}
The physical mesh is body-fitted.
In the wall-normal direction \(401\) points are used between the wall and the outer boundary (including the bow shock).
In the streamwise direction the local spacing is set to one twenty-fifth of the freestream acoustic wavelength,
\begin{equation*}
  \Delta s
  =\frac{\lambda_\infty}{25}.
\end{equation*}
The discrete velocity space (DVS) is discretized in a circular region centered at \(0.4\times(U_\infty,0)\) with a total radius of \(6\sqrt{R_g T_s}\), where \(T_s\) is the stagnation temperature of the freestream flow (Fig.~\ref{fig:bluntnessDVS}).
The unstructured triangular mesh comprises \(2152\) elements and is refined at the zero-velocity point within a radius of \(3\sqrt{R_g T_w}\) and at the freestream point within a radius of \(3\sqrt{R_g T_\infty}\).

\begin{figure}[!htbp]
  \centering
  \begin{subfigure}[t]{0.48\textwidth}
    \includegraphics[width=\textwidth]{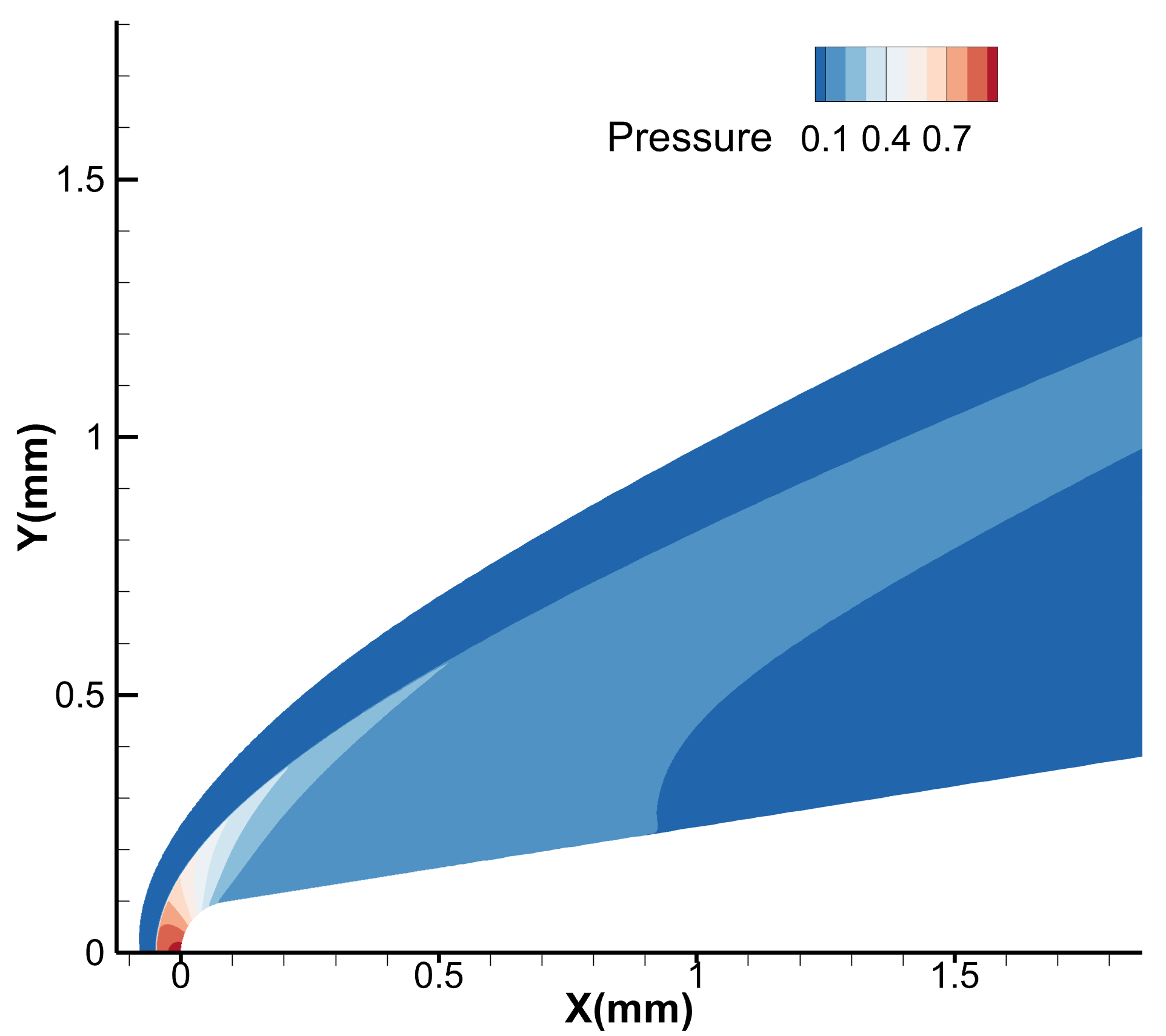}
    \caption*{Pressure}
  \end{subfigure}
  \hfill
  \begin{subfigure}[t]{0.48\textwidth}
    \includegraphics[width=\textwidth]{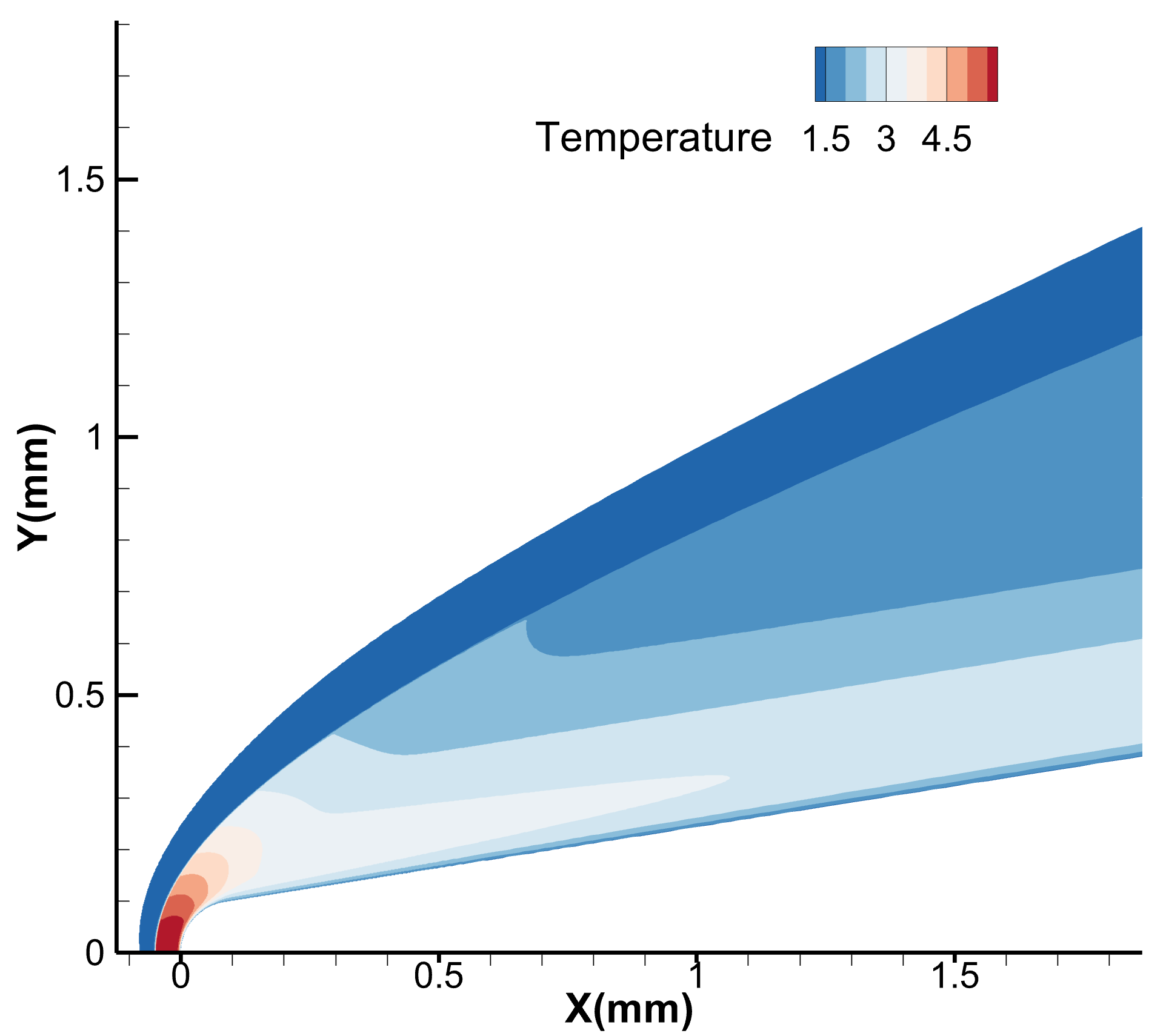}
    \caption*{Temperature}
  \end{subfigure}
  \caption{Steady base flow about the blunt wedge: nondimensional pressure (left) and temperature (right).
  The bow shock and the stagnation hot spot are clearly visible; the wall is isothermal at \(T_w=300\,\mathrm{K}\).}
  \label{fig:bluntnessBaseContour}
\end{figure}

\begin{figure}[!htbp]
  \centering
  \begin{subfigure}[t]{0.48\textwidth}
    \includegraphics[width=\textwidth]{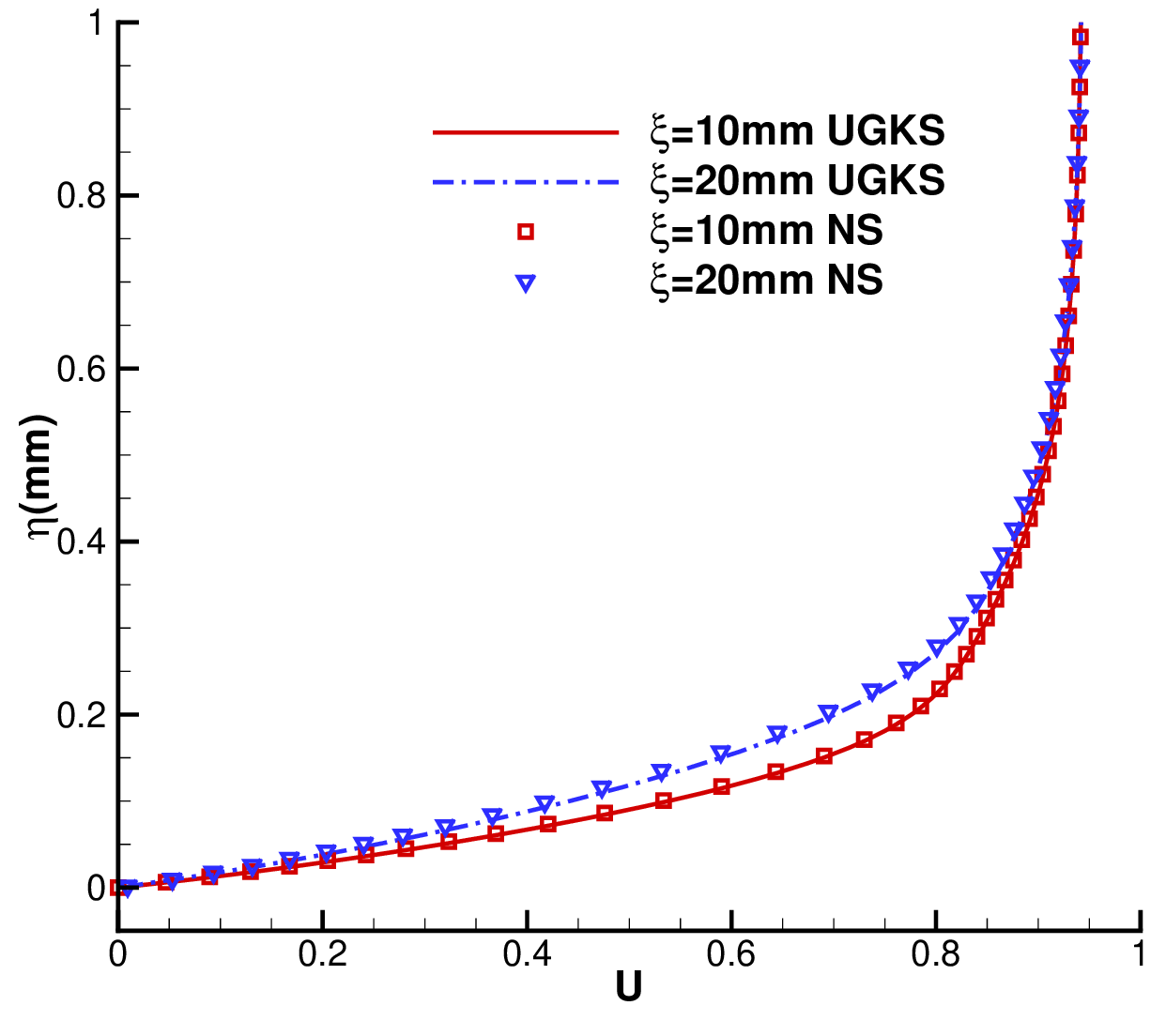}
    \caption*{Velocity}
  \end{subfigure}
  \hfill
  \begin{subfigure}[t]{0.48\textwidth}
    \includegraphics[width=\textwidth]{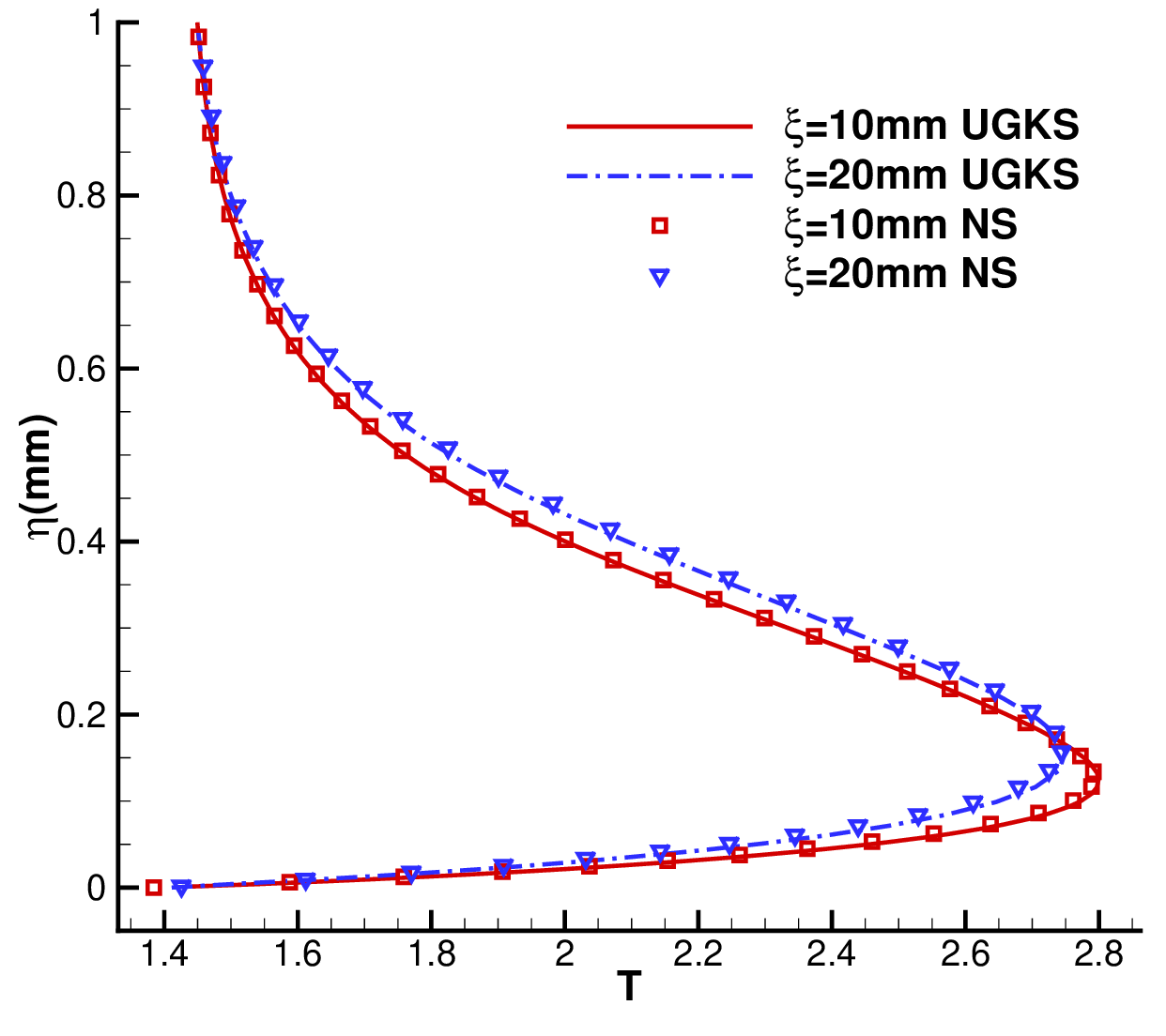}
    \caption*{Temperature}
  \end{subfigure}
  \caption{Wall-normal base-flow profiles of streamwise velocity (left) and temperature (right) versus \(\eta\).}
  \label{fig:bluntnessBaseProfile}
\end{figure}

\paragraph{Base flow and linearized solve}
A steady UGKS base flow is first computed on this mesh and then frozen for the complex LUGKS.
Figure~\ref{fig:bluntnessBaseContour} shows the nondimensional pressure and temperature fields: a detached bow shock stands ahead of the nose, with peaks of \(p\) and \(T\) at the stagnation point and a steep temperature drop toward the isothermal wall.
Wall-normal profiles of the streamwise velocity and temperature along a cut through the stagnation region are given in Fig.~\ref{fig:bluntnessBaseProfile}, where \(\eta\) is the wall-normal distance in the body-fitted frame.
The wall response (pressure and heat-flux transfer functions) and the disturbance field between the shock and the body are extracted from the converged real and imaginary amplitudes of the complex LUGKS.
The complex LUGKS computation was run on six dual-socket nodes interconnected by a \(100\,\mathrm{Gb}\) InfiniBand fabric; each node has two AMD EPYC~7542 processors (\(2\times 32\) cores at \(2.9\,\mathrm{GHz}\)) and \(256\,\mathrm{GB}\) of memory, for a total of \(384\) cores.
The job used about \(131\,\mathrm{GB}\) of memory.
One hundred pseudo-time steps required \(2211\,\mathrm{s}\) of wall-clock time (\(\approx 22.1\,\mathrm{s}\) per step); the reported solution was obtained after \(4000\) steps (\(\approx 24.6\,\mathrm{h}\)).

\begin{figure}[!htbp]
  \centering
  \includegraphics[width=0.7\textwidth]{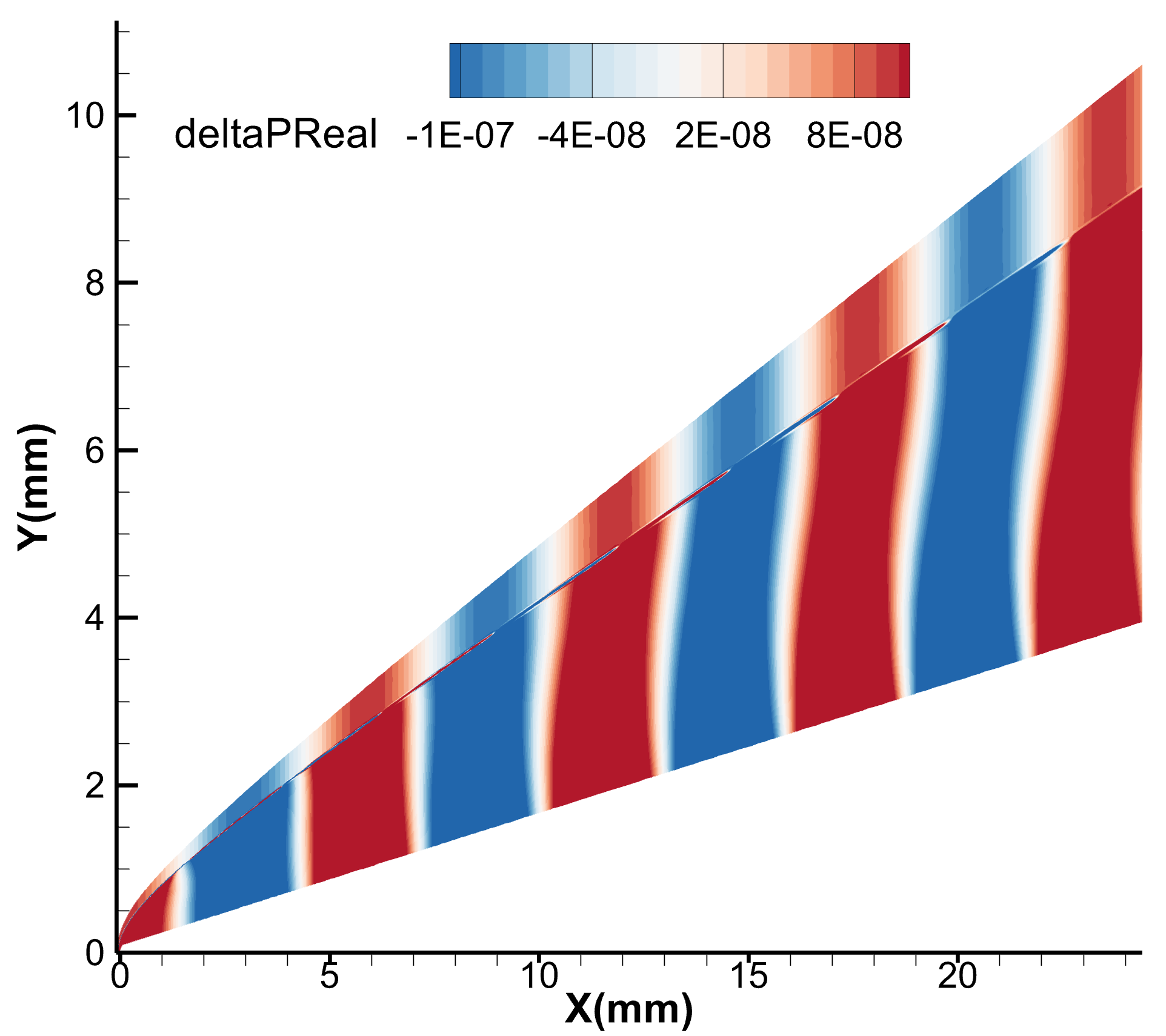}
  \caption{Real part of the pressure perturbation \(\mathrm{Re}\{\widehat{p}\}\) for freestream fast-acoustic incidence at \(f=300\,\mathrm{kHz}\).
  Periodic bands outside the bow shock are the incident wave; the transmitted field fills the shock layer down to the wall.}
  \label{fig:bluntnessDeltaPContour}
\end{figure}

\begin{figure}[!htbp]
  \centering
  \includegraphics[width=0.55\textwidth]{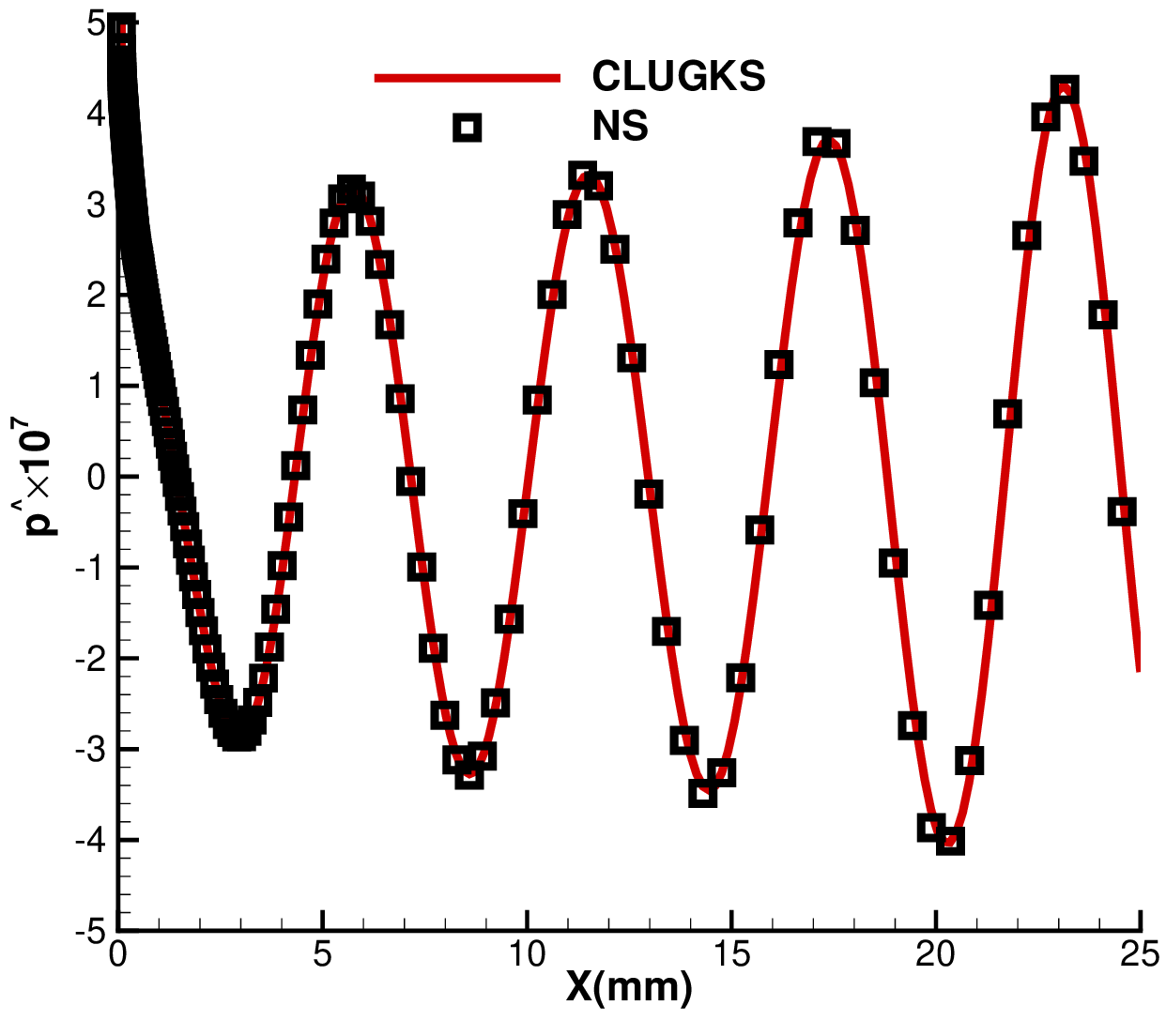}
  \caption{Wall-pressure perturbation along the blunt-wedge surface for freestream fast-acoustic forcing at \(f=300\,\mathrm{kHz}\), compared with the DNS of Li et~al.~\cite{li2026receptivity}.}
  \label{fig:bluntnessDeltaPWall}
\end{figure}

\paragraph{Disturbance field}
Figure~\ref{fig:bluntnessDeltaPContour} shows the real part of the pressure perturbation for the incident freestream fast acoustic wave at \(f=300\,\mathrm{kHz}\).
Upstream of the bow shock the incident wave appears as nearly periodic bands; across the shock the wave is transmitted into the shock layer with a change of wavelength and amplitude, and the disturbance reaches the body surface.
The corresponding wall-pressure response along the surface is plotted in Fig.~\ref{fig:bluntnessDeltaPWall} and compared with the DNS of Li et~al.~\cite{li2026receptivity}.

\section{Hypersonic argon flow over a blunt wedge at 40\,km}
\label{sec:argon40km}

Hypersonic argon flow over a blunt wedge is next considered at an altitude equivalent of \(40\,\mathrm{km}\).
The geometry is the same as in Sec.~\ref{sec:blunt_receptivity}: a circular cylinder of nose radius \(R_n=0.1\,\mathrm{mm}\) joined to a tangent half-angle wedge of \(\theta=9^\circ\), at zero angle of attack.
Argon is chosen as the working gas and treated as a monatomic perfect gas with
\begin{equation*}
  \gamma=\frac{5}{3},
  \qquad
  R_g=208.13\,\mathrm{J\,kg^{-1}\,K^{-1}},
  \qquad
  \mathrm{Pr}=\frac{2}{3},
\end{equation*}
so that rotational and vibrational internal degrees of freedom do not enter the receptivity response and the comparison between UGKS and Navier--Stokes isolates translational non-equilibrium.
The freestream Mach number and the dimensional freestream temperature, sound speed and velocity are
\begin{equation*}
  Ma_\infty=5,
  \qquad
  T^*=250\,\mathrm{K},
  \qquad
  a^*\approx 294.48\,\mathrm{m\,s^{-1}},
  \qquad
  U^*\approx 1472.4\,\mathrm{m\,s^{-1}},
\end{equation*}
where \(a^*=\sqrt{\gamma R_g T^*}\) and \(U^*=Ma_\infty a^*\).
The freestream density corresponds to argon at the \(40\,\mathrm{km}\) level of the U.S. Standard Atmosphere~\cite{us1976atmosphere}, giving
\begin{equation*}
  Re_{R_n}=39.
\end{equation*}
Compared with the validation case of Sec.~\ref{sec:blunt_receptivity} (\(Re_{R_n}=920\)), the present flow is more rarefied, so that slip and temperature jump near the wall are no longer negligible.
The wall is isothermal with a fully accommodating diffuse-reflection condition.
A freestream fast acoustic wave with nondimensional angular frequency
\begin{equation*}
  \omega=0.11875
\end{equation*}
(based on \(U^*\) and \(R_n\)) is imposed after the base flow is frozen, corresponding to
\begin{equation*}
  f=\frac{\omega U^*}{2\pi R_n}\approx 278.3\,\mathrm{kHz}.
\end{equation*}
The physical mesh is body-fitted, with \(401\) wall-normal points between the wall and the outer boundary and \(743\) streamwise points on the upper half (\(y>0\)); a coarse lower half (\(41\) points) is retained so that a symmetry boundary is not required.
The discrete velocity space is a circular unstructured mesh of \(2210\) triangles, centred at \(0.4\times(U_\infty,0)\) with radius \(5\sqrt{2R_g T_s}\), where \(T_s\) is the freestream stagnation temperature.

\subsection{Base flow}
\label{sec:argon40km_base}

\begin{figure}[!htbp]
  \centering
  \begin{subfigure}[t]{0.48\textwidth}
    \includegraphics[width=\textwidth]{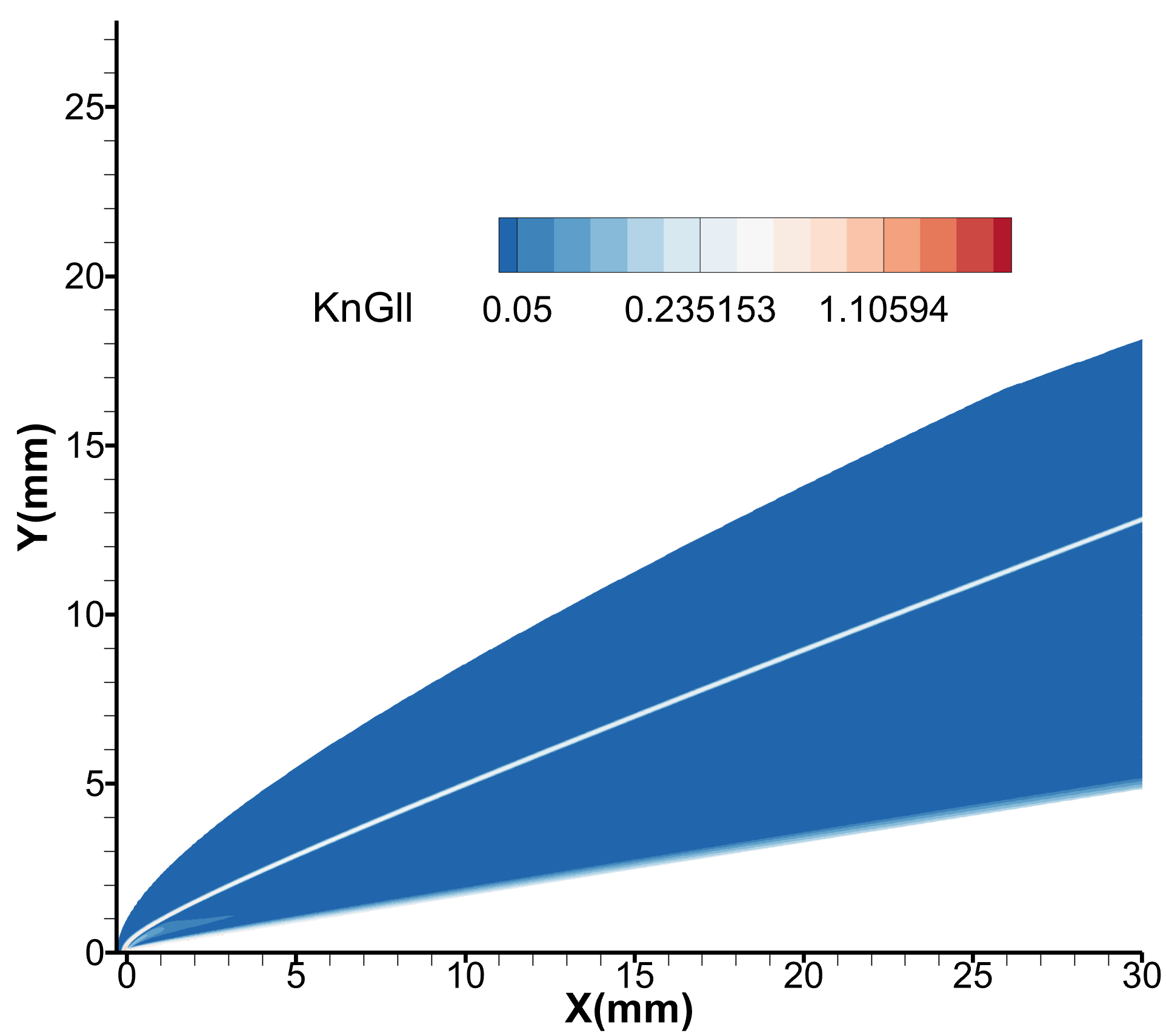}
    \caption{Full field}
    \label{fig:argon40km_KnGll_a}
  \end{subfigure}
  \hfill
  \begin{subfigure}[t]{0.48\textwidth}
    \includegraphics[width=\textwidth]{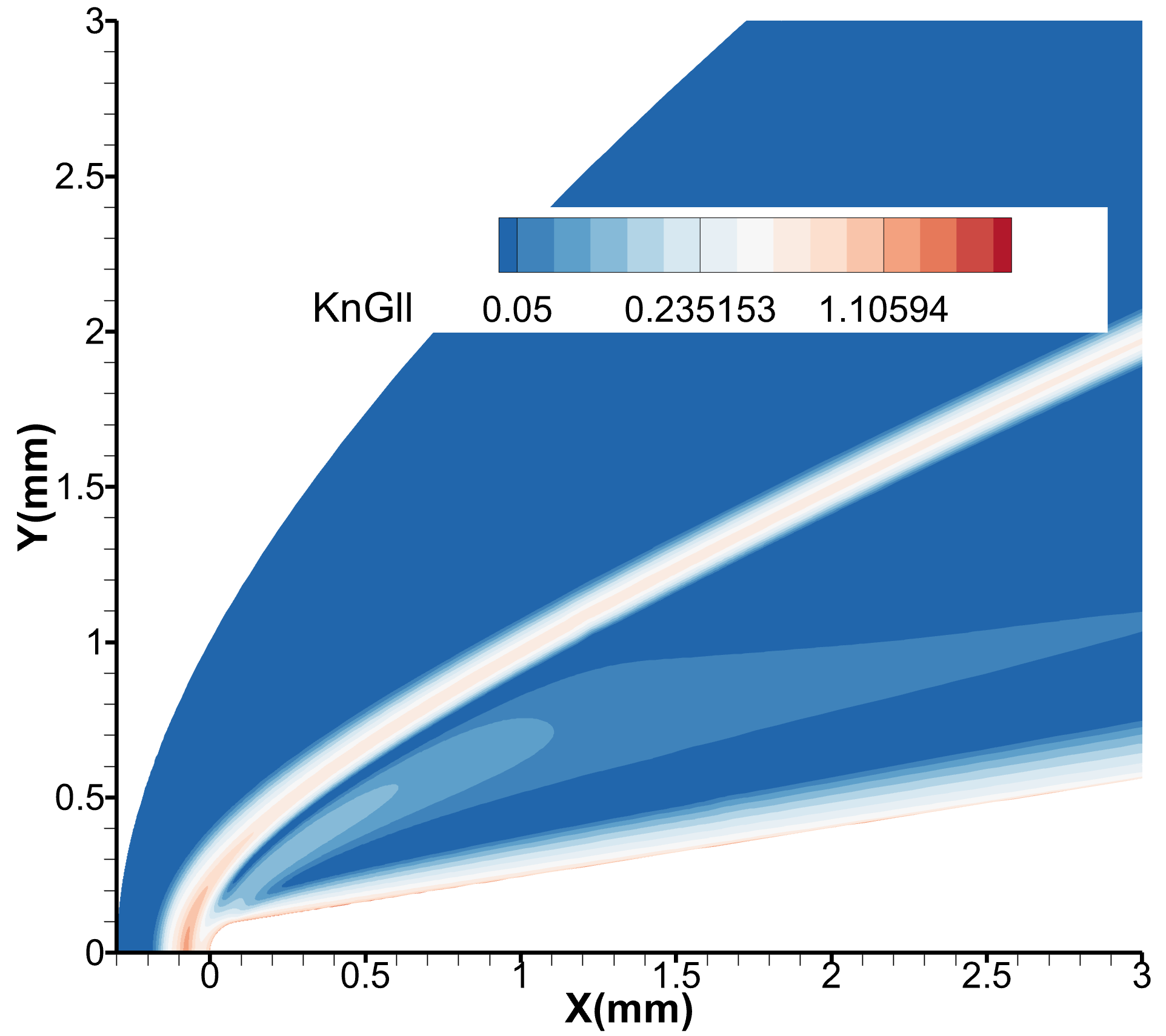}
    \caption{Local near nose}
    \label{fig:argon40km_KnGll_b}
  \end{subfigure}
  \caption{Gradient-length local Knudsen number \(\mathrm{Kn}_{\mathrm{Gll}}=l\,|\nabla\rho|/\rho\) for the argon base flow at \(Re_{R_n}=39\):
  (a)~full field; (b)~local view near the nose.
  Elevated \(\mathrm{Kn}_{\mathrm{Gll}}\) is limited to the bow shock and the near-wall layer.}
  \label{fig:argon40km_KnGll}
\end{figure}

\begin{figure}[!htbp]
  \centering
  \begin{subfigure}[t]{0.32\textwidth}
    \includegraphics[width=\textwidth]{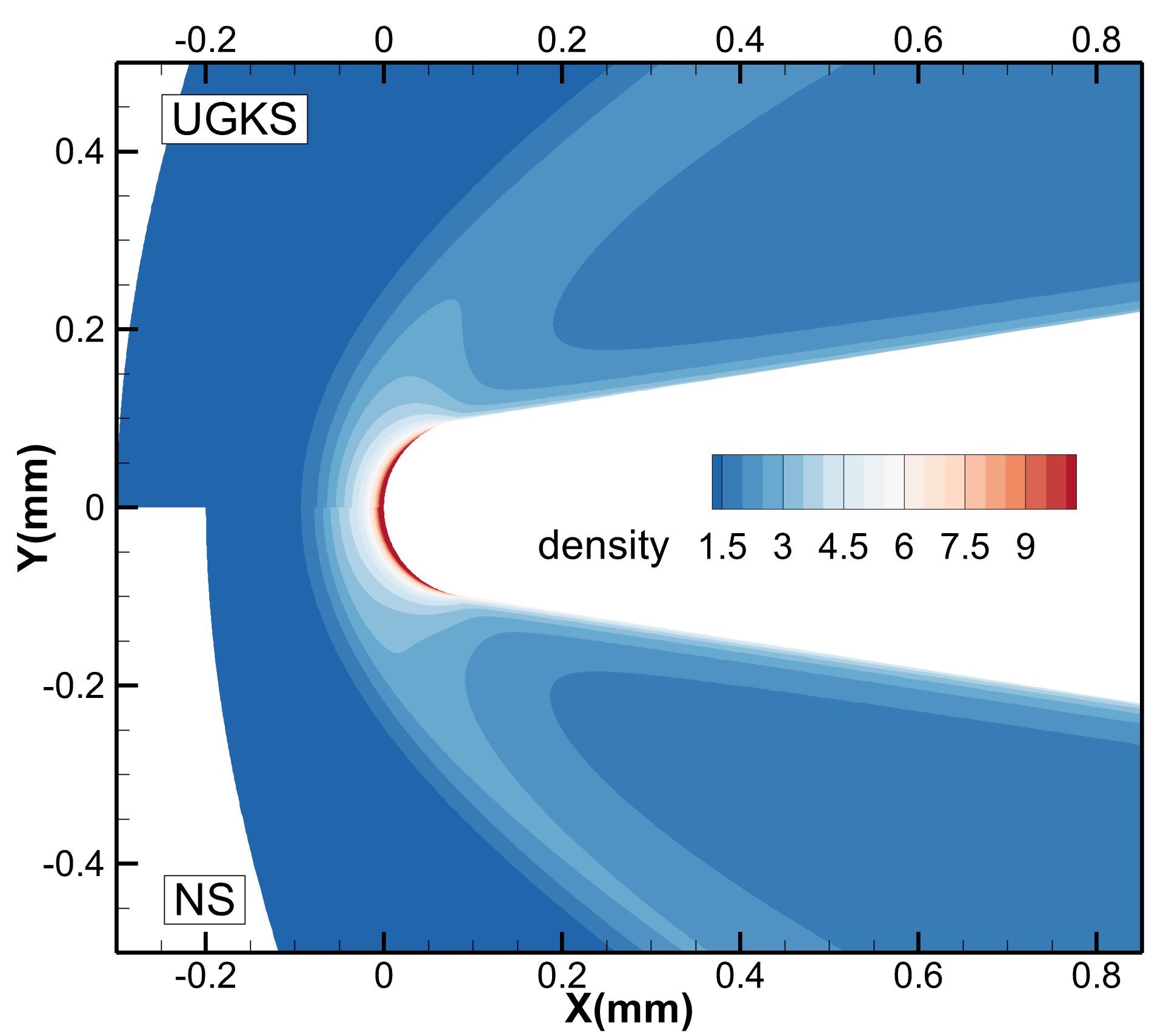}
    \caption{Density}
    \label{fig:argon40km_vsNS_a}
  \end{subfigure}
  \hfill
  \begin{subfigure}[t]{0.32\textwidth}
    \includegraphics[width=\textwidth]{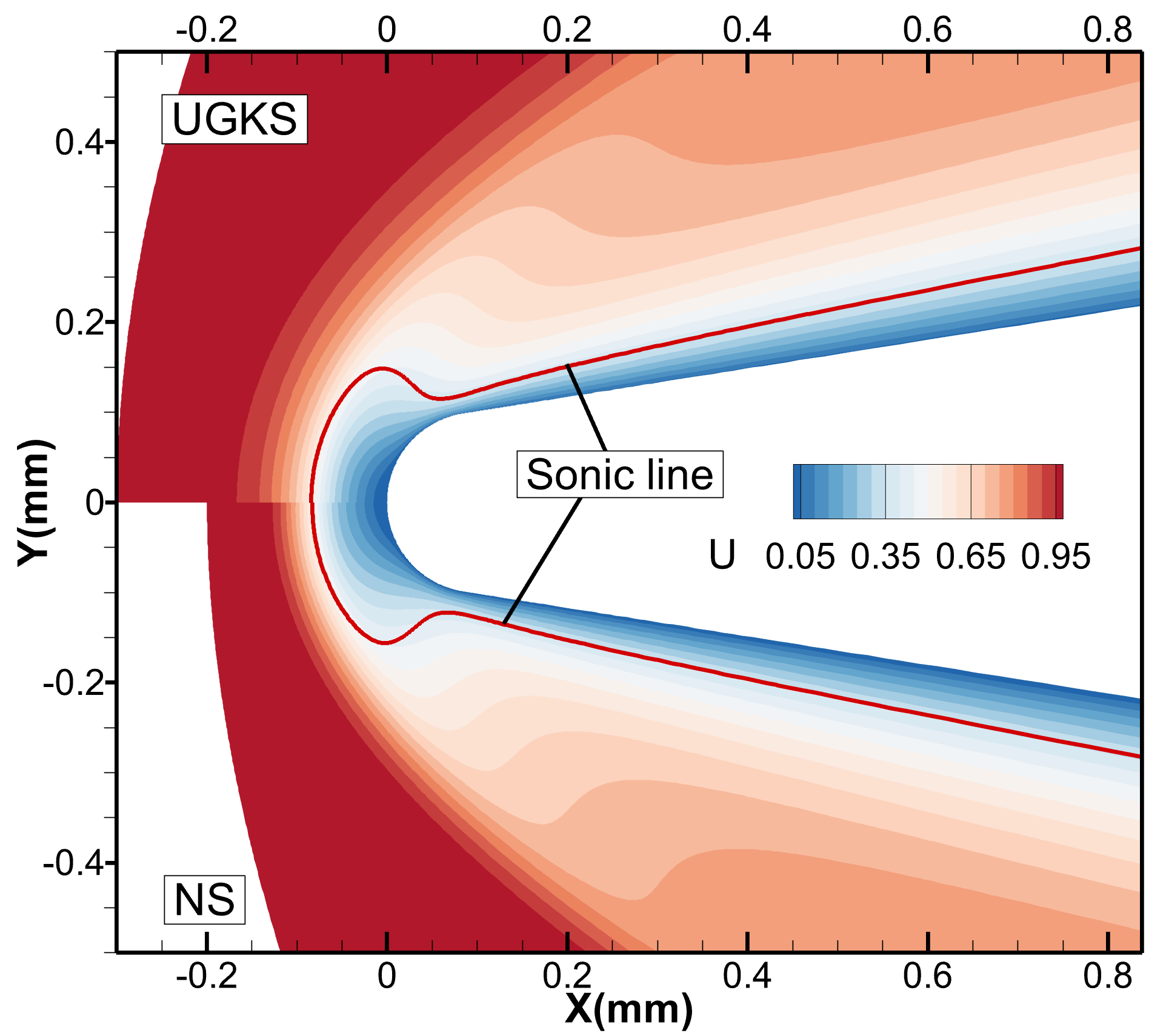}
    \caption{$U$}
    \label{fig:argon40km_vsNS_b}
  \end{subfigure}
  \hfill
  \begin{subfigure}[t]{0.32\textwidth}
    \includegraphics[width=\textwidth]{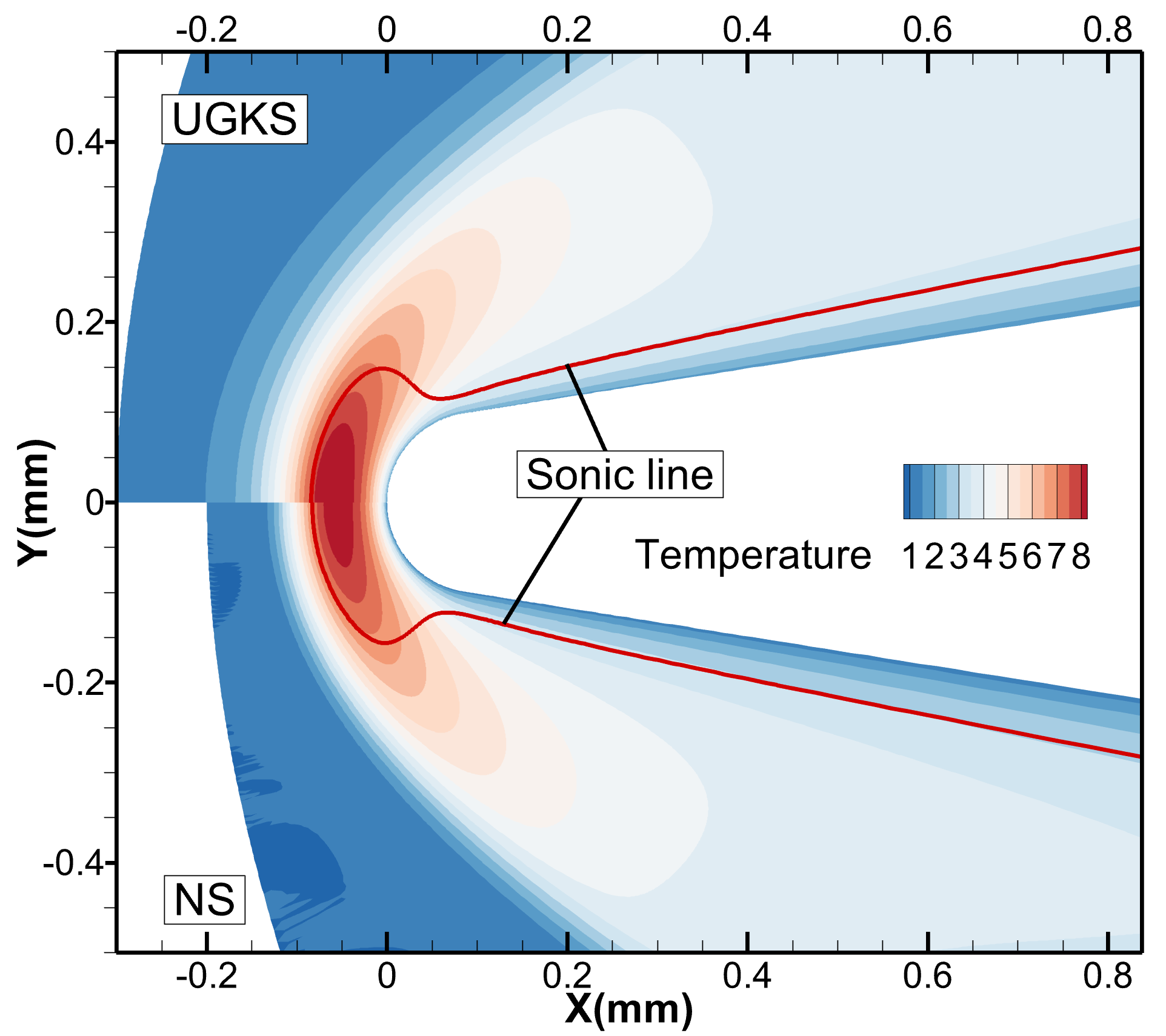}
    \caption{Temperature}
    \label{fig:argon40km_vsNS_c}
  \end{subfigure}
  \caption{Steady base flow for argon at 40\,km with \(Re_{R_n}=39\): (a)~density, (b)~streamwise velocity and (c)~temperature.
  Upper half: UGKS; lower half: NS.}
  \label{fig:argon40km_vsNS}
\end{figure}

\begin{figure}[!htbp]
  \centering
  \begin{subfigure}[t]{0.32\textwidth}
    \includegraphics[width=\textwidth]{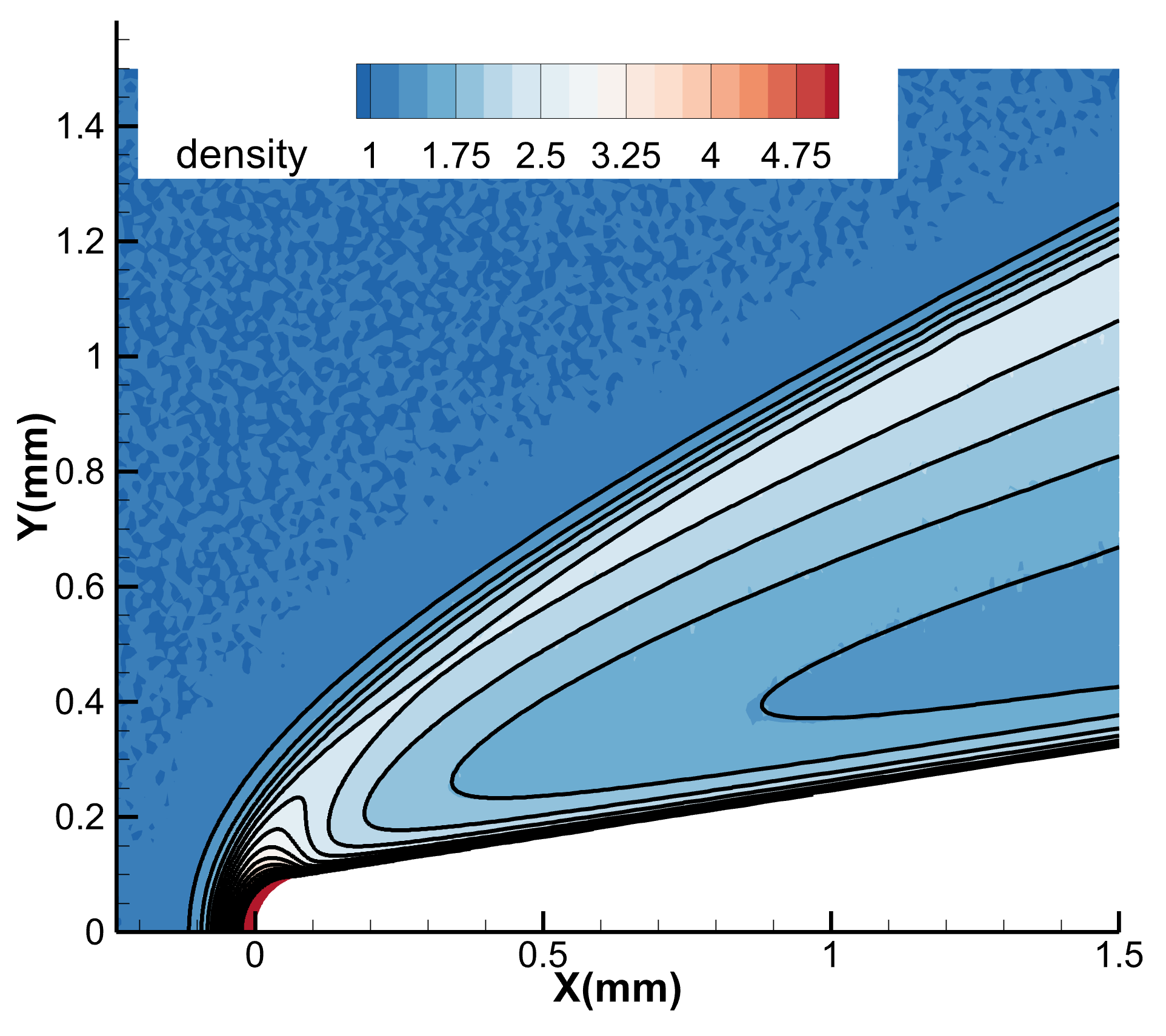}
    \caption{Density}
    \label{fig:argon40km_vsDSMC_a}
  \end{subfigure}
  \hfill
  \begin{subfigure}[t]{0.32\textwidth}
    \includegraphics[width=\textwidth]{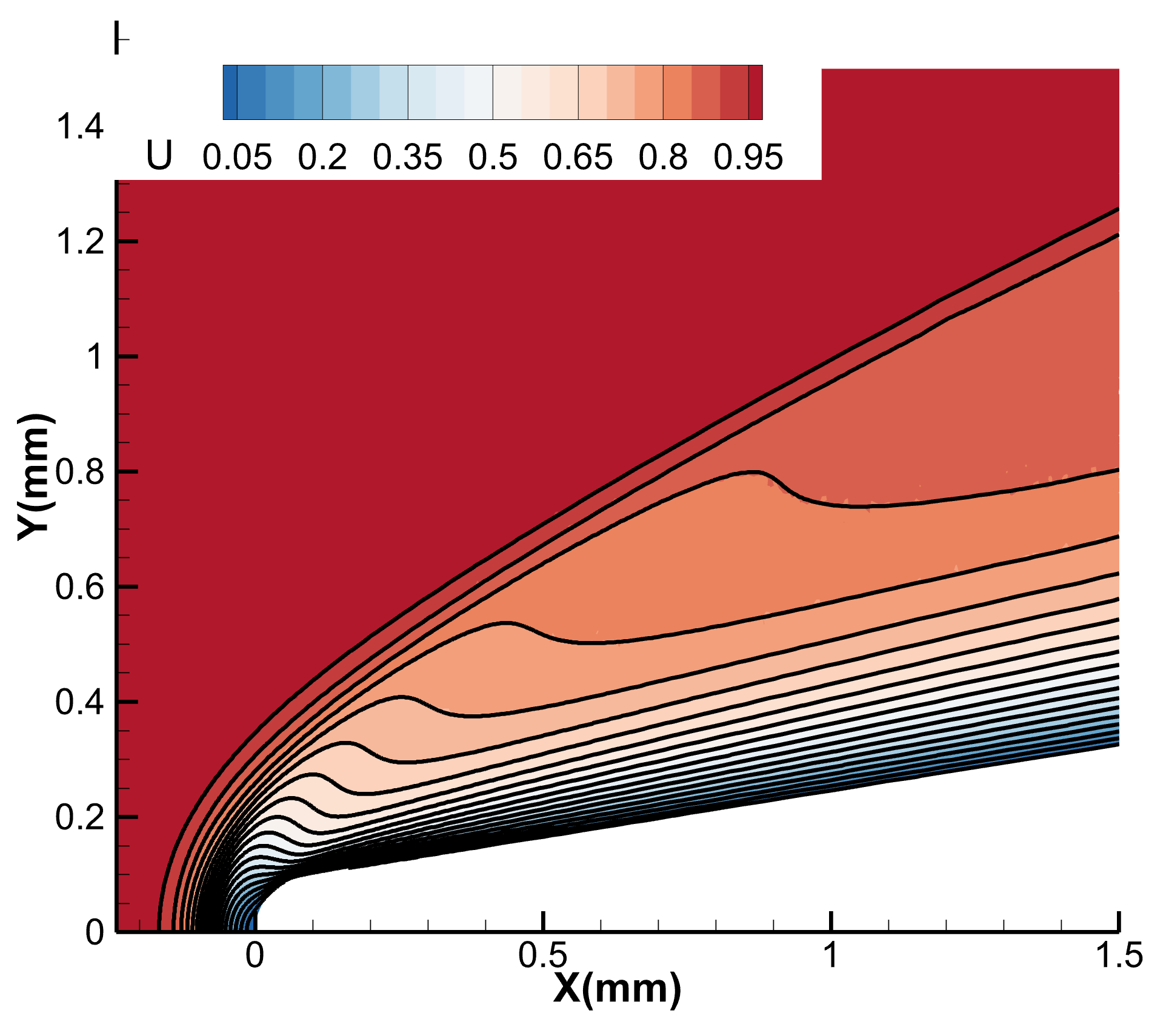}
    \caption{$U$}
    \label{fig:argon40km_vsDSMC_b}
  \end{subfigure}
  \hfill
  \begin{subfigure}[t]{0.32\textwidth}
    \includegraphics[width=\textwidth]{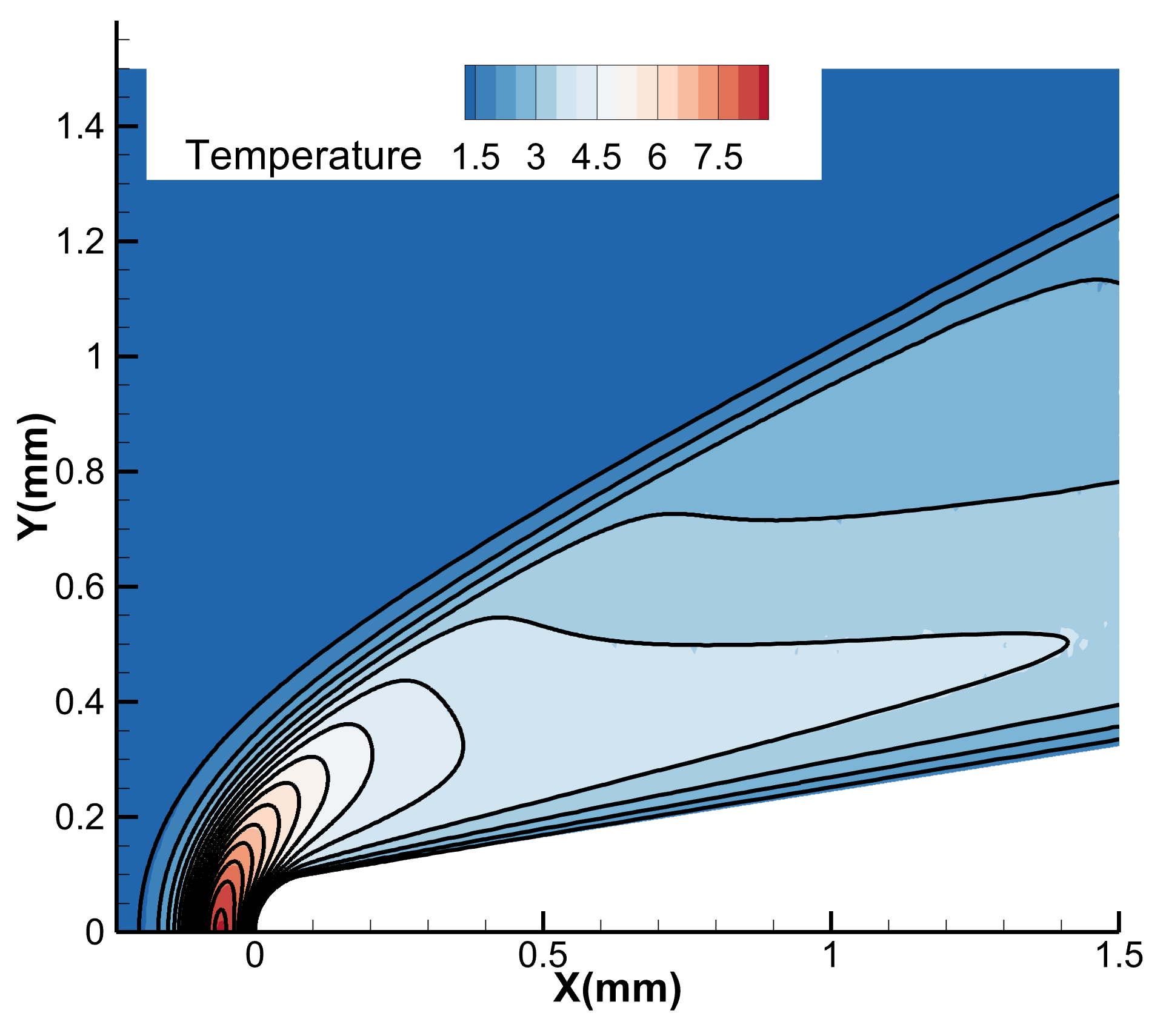}
    \caption{Temperature}
    \label{fig:argon40km_vsDSMC_c}
  \end{subfigure}
  \caption{Same base flow compared with DSMC (DS2V)~\cite{bird1994molecular}: (a)~density, (b)~streamwise velocity and (c)~temperature.
  Colour contours: DSMC; black isolines: UGKS.}
  \label{fig:argon40km_vsDSMC}
\end{figure}

\begin{figure}[!htbp]
  \centering
  \begin{subfigure}[t]{0.24\textwidth}
    \includegraphics[width=\textwidth]{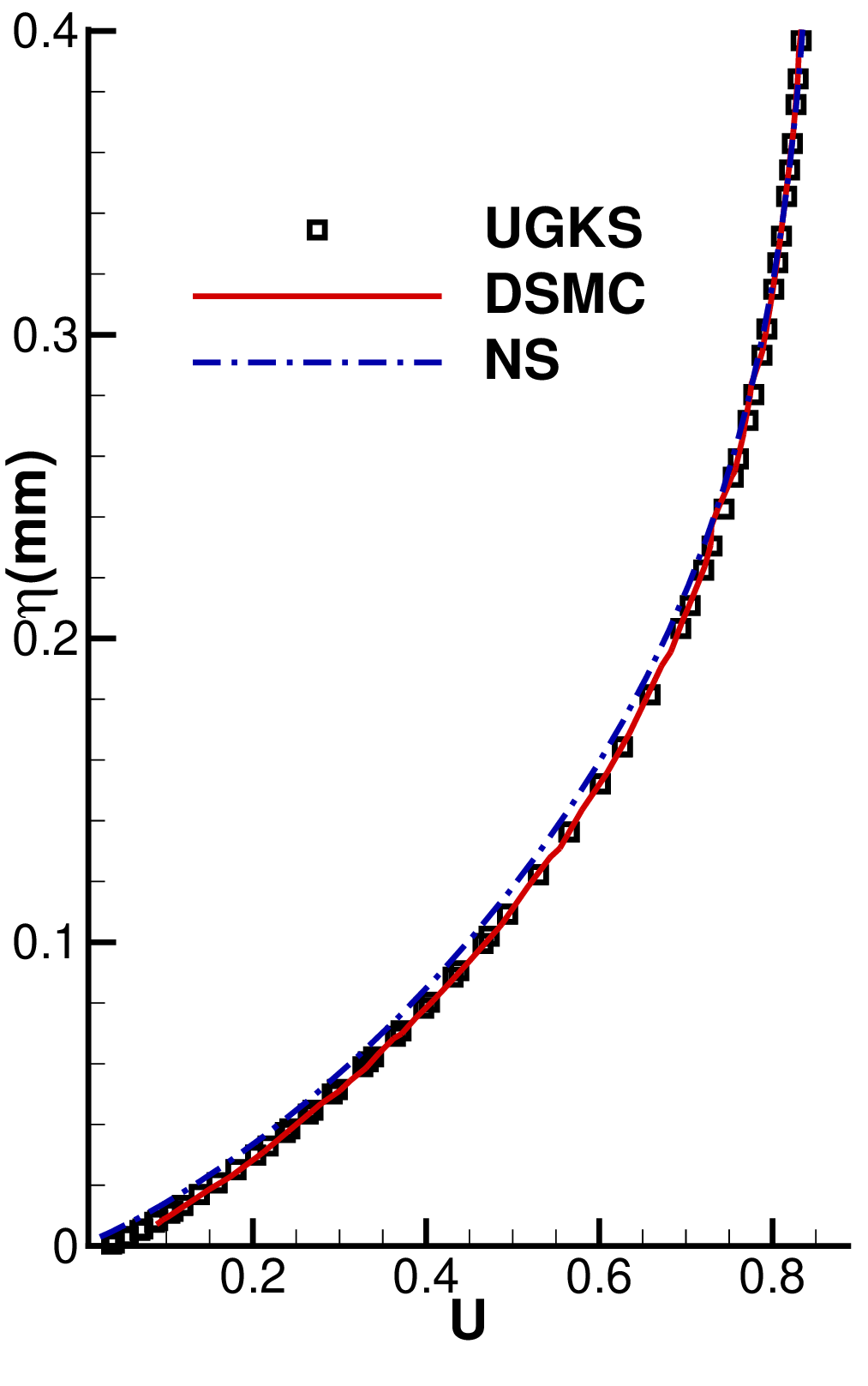}
    \caption{$U$, $\xi=1\,\mathrm{mm}$}
    \label{fig:argon40km_profile_a}
  \end{subfigure}
  \hfill
  \begin{subfigure}[t]{0.24\textwidth}
    \includegraphics[width=\textwidth]{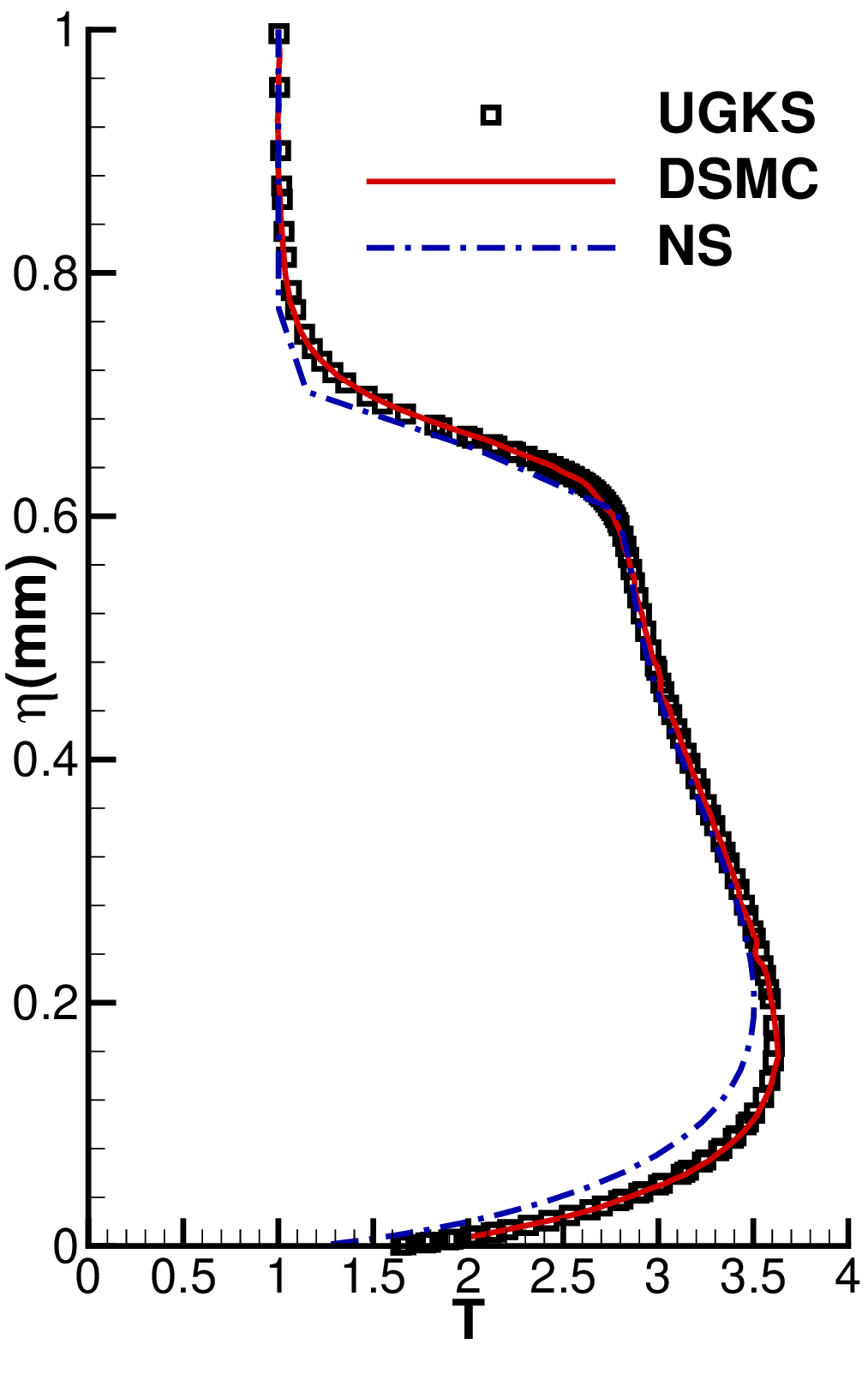}
    \caption{$T$, $\xi=1\,\mathrm{mm}$}
    \label{fig:argon40km_profile_b}
  \end{subfigure}
  \hfill
  \begin{subfigure}[t]{0.24\textwidth}
    \includegraphics[width=\textwidth]{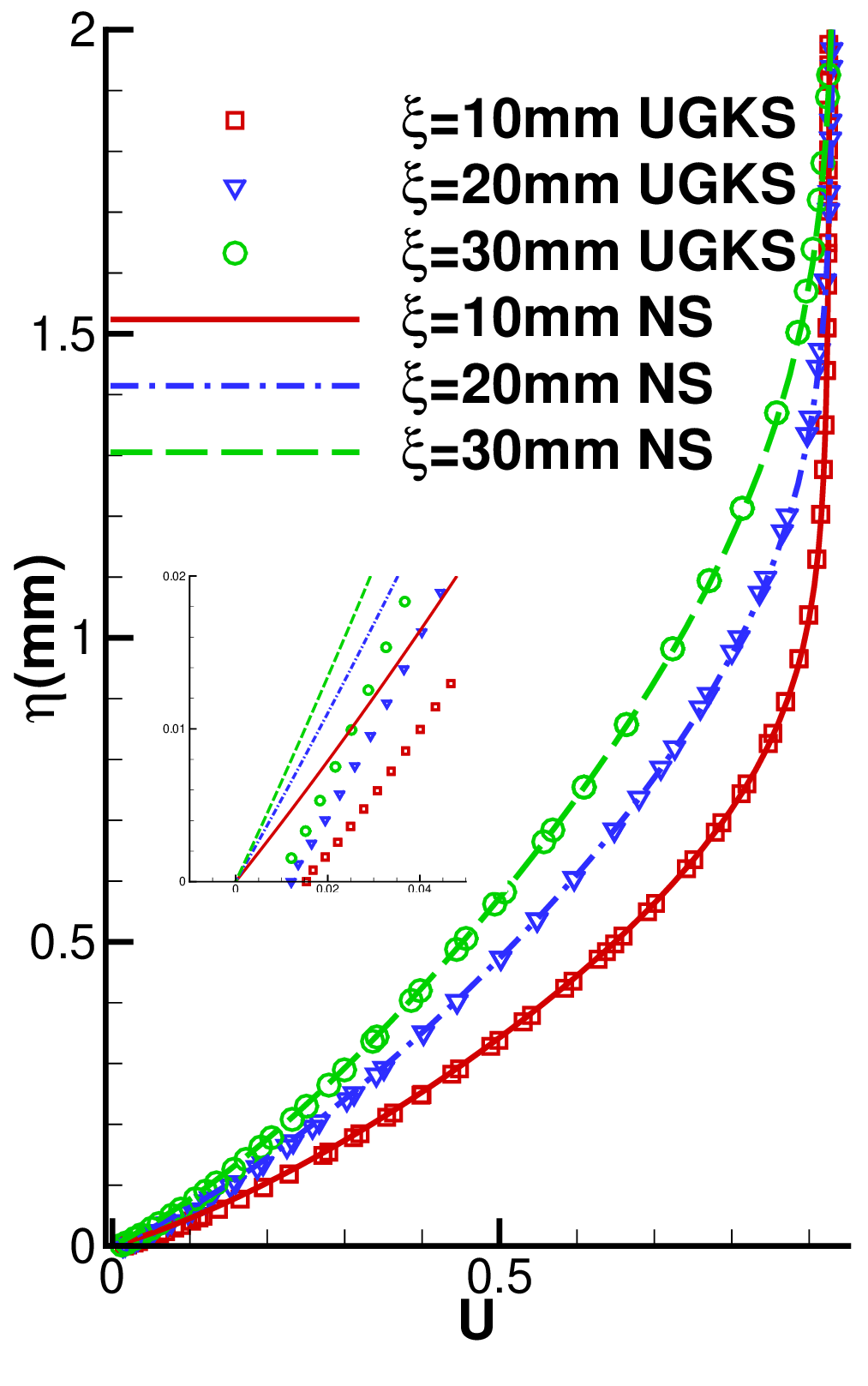}
    \caption{$U$, multi-station}
    \label{fig:argon40km_profile_c}
  \end{subfigure}
  \hfill
  \begin{subfigure}[t]{0.24\textwidth}
    \includegraphics[width=\textwidth]{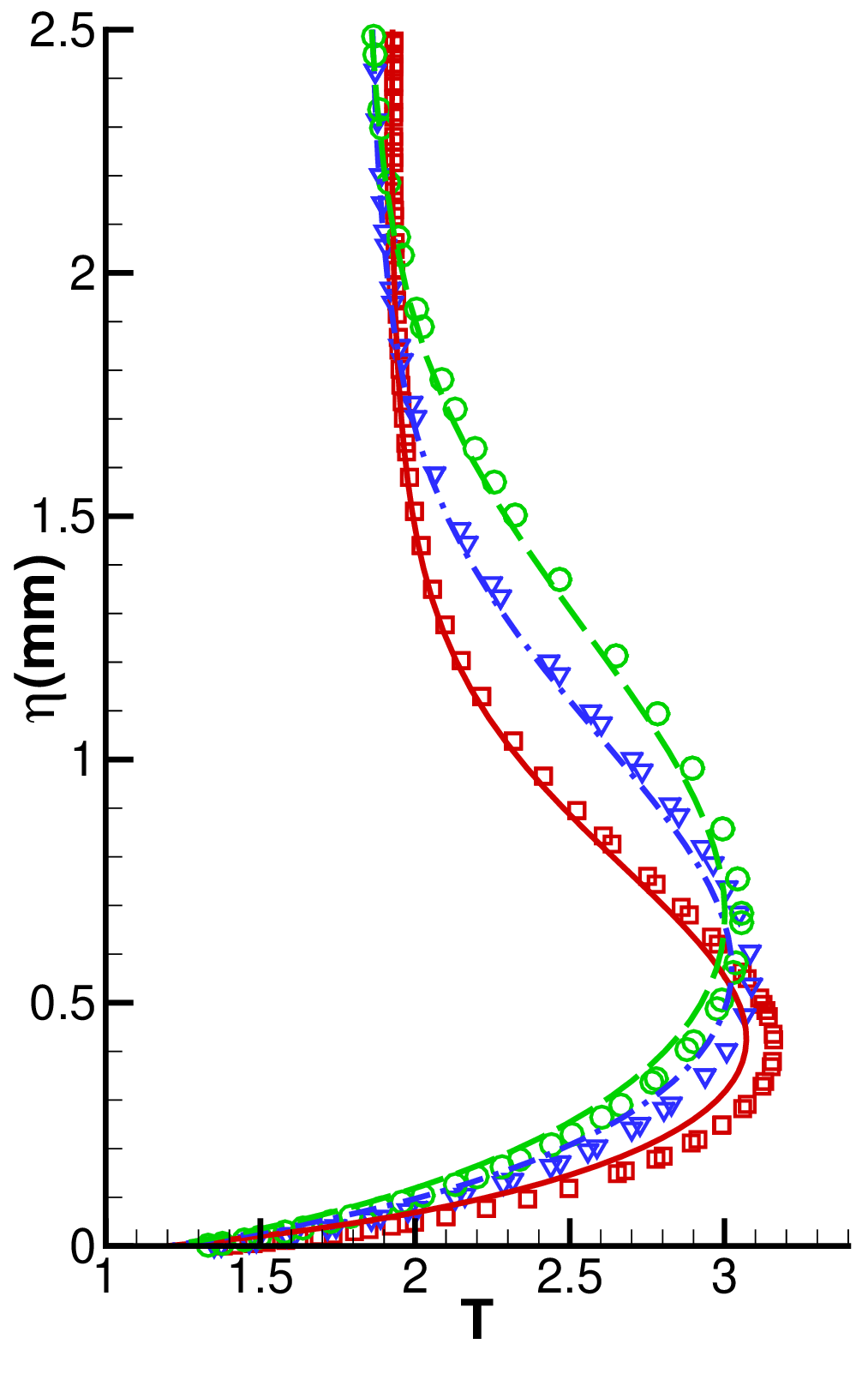}
    \caption{$T$, multi-station}
    \label{fig:argon40km_profile_d}
  \end{subfigure}
  \caption{Wall-normal base-flow profiles at \(Re_{R_n}=39\):
  (a)--(b)~streamwise velocity and temperature at \(\xi=1\,\mathrm{mm}\);
  (c)--(d)~profiles at downstream stations.}
  \label{fig:argon40km_profile}
\end{figure}

A steady UGKS base flow is computed first and then frozen for the complex LUGKS.
The UGKS wall uses fully accommodating isothermal diffuse reflection, whereas the companion NS solver uses a no-slip isothermal wall.
Local rarefaction is measured by the gradient-length local Knudsen number \(\mathrm{Kn}_{\mathrm{Gll}}=l\,|\nabla\rho|/\rho\), where \(l\) is the local mean free path.
Figure~\ref{fig:argon40km_KnGll} shows \(\mathrm{Kn}_{\mathrm{Gll}}\) over the full domain and near the nose: elevated values are confined to the bow shock and a thin near-wall layer, while the freestream and most of the post-shock flow remain near continuum.
This localized rarefaction is consistent with the close UGKS--NS agreement reported below, with kinetic effects mainly near the shock and the wall.
Figure~\ref{fig:argon40km_vsNS} compares density, streamwise velocity and temperature (UGKS upper half, NS lower half).
Away from the wall the shock stand-off and the outer shock layer agree closely.
Near the wall, the UGKS shows the expected slip and temperature jump.
The same base flow is compared with DSMC solutions obtained with Bird's DS2V code~\cite{bird1994molecular} in Fig.~\ref{fig:argon40km_vsDSMC}: the shock location and near-wall kinetic layer match well.
Wall-normal profiles are collected in Fig.~\ref{fig:argon40km_profile}.
At \(\xi=1\,\mathrm{mm}\), slip and temperature jump are clearest; farther downstream the velocity profiles of UGKS and NS approach each other, while the temperature profiles remain more separated.

\subsection{Disturbance field}
\label{sec:argon40km_disturb}

\begin{figure}[!htbp]
  \centering
  \begin{subfigure}[t]{0.48\textwidth}
    \includegraphics[width=\textwidth]{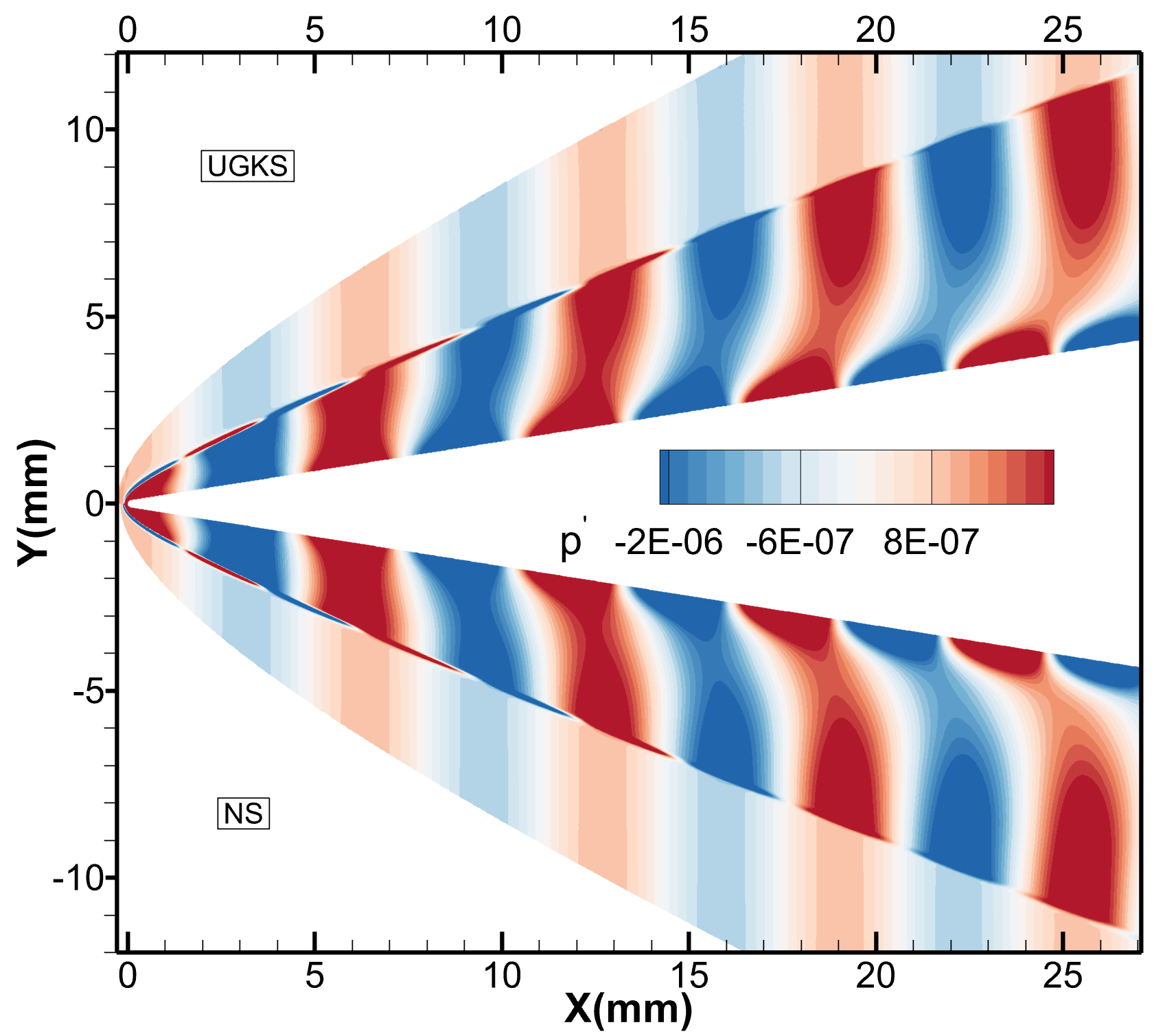}
    \caption{Full field}
    \label{fig:argon40km_deltaP_a}
  \end{subfigure}
  \hfill
  \begin{subfigure}[t]{0.48\textwidth}
    \includegraphics[width=\textwidth]{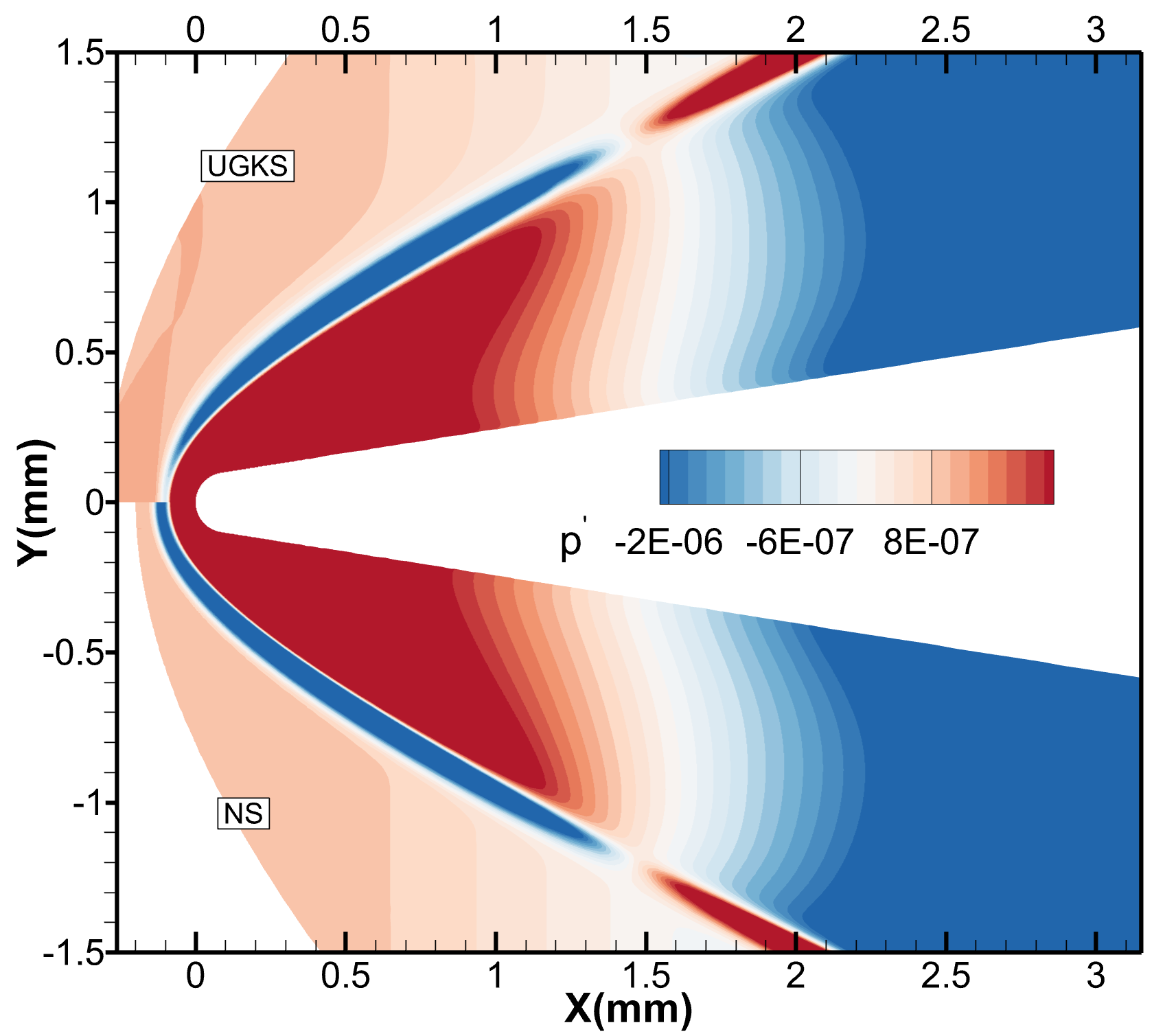}
    \caption{Local near nose}
    \label{fig:argon40km_deltaP_b}
  \end{subfigure}
  \caption{Real part of the pressure perturbation \(\mathrm{Re}\{p'\}\) for freestream fast-acoustic incidence on the argon blunt wedge at \(Re_{R_n}=39\):
  (a)~full field; (b)~local view near the nose.}
  \label{fig:argon40km_deltaP}
\end{figure}

With the base flow frozen, a freestream fast acoustic wave is imposed and the complex LUGKS is compared with a companion continuum DNS under the same forcing; the DNS disturbance is taken from a Fourier decomposition at the forcing frequency.
Figure~\ref{fig:argon40km_deltaP} shows \(\mathrm{Re}\{p'\}\) over the full domain and near the nose.
The overall shock-layer pattern of the two methods is similar: the incident wave is refracted across the bow shock and reaches the wall with a comparable wavelength and phase.
Differences remain mainly near the kinetic shock structure, where the UGKS shock is thicker than the continuum front.

\begin{figure}[!htbp]
  \centering
  \begin{subfigure}[t]{0.32\textwidth}
    \includegraphics[width=\textwidth]{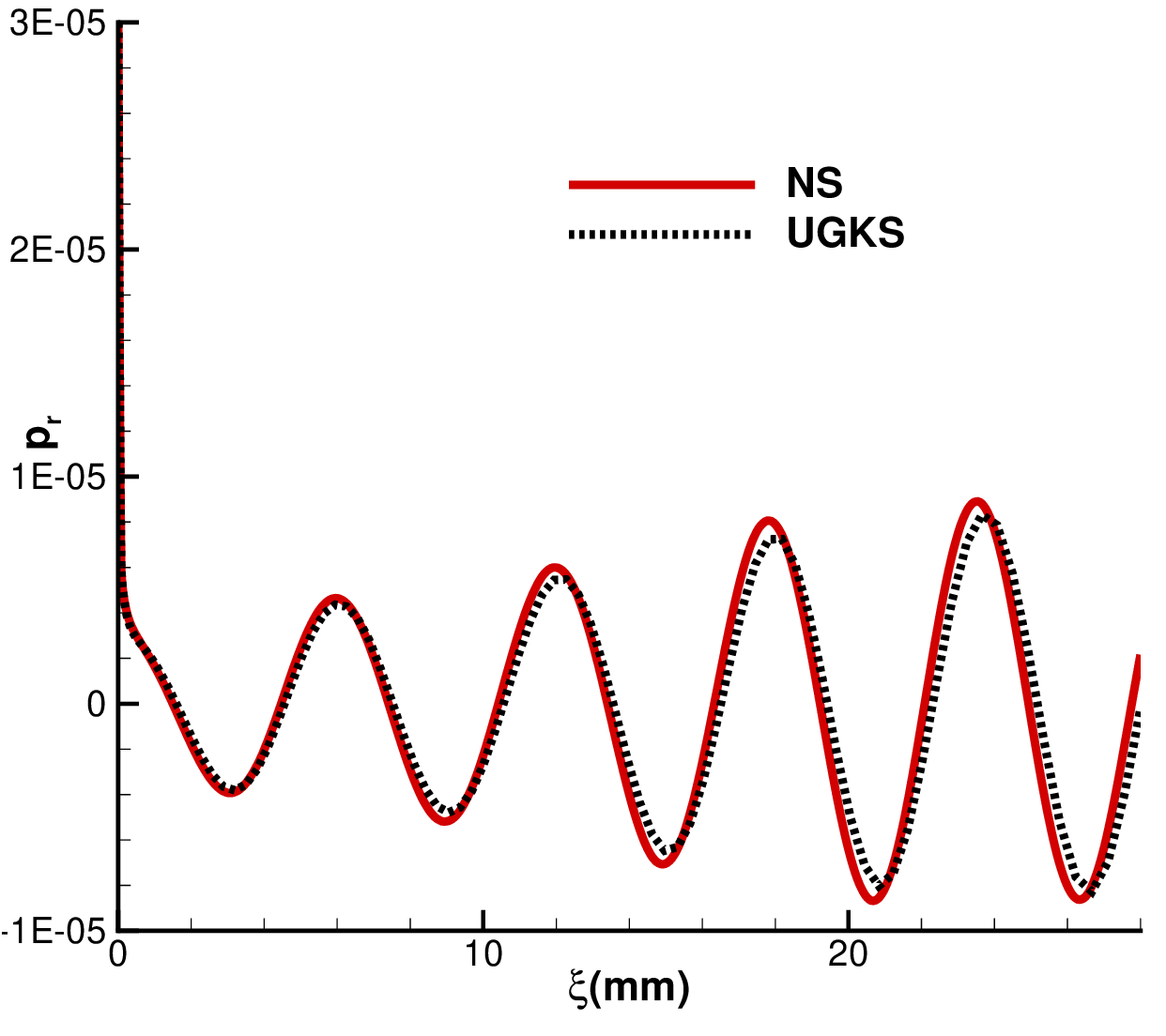}
    \caption{$\mathrm{Re}\{p'_w\}$}
    \label{fig:argon40km_wallDisp_a}
  \end{subfigure}
  \hfill
  \begin{subfigure}[t]{0.32\textwidth}
    \includegraphics[width=\textwidth]{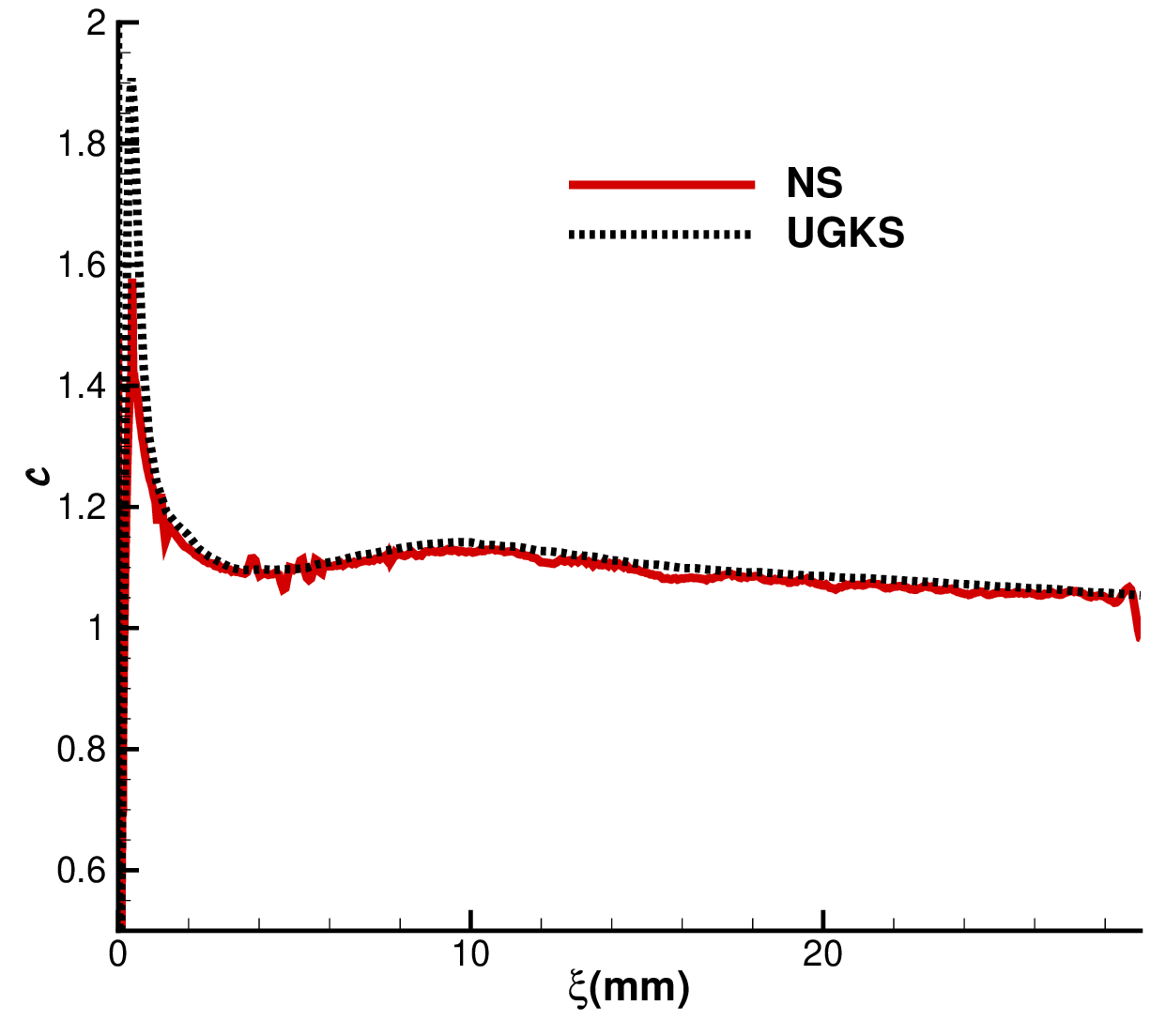}
    \caption{Phase speed}
    \label{fig:argon40km_wallDisp_b}
  \end{subfigure}
  \hfill
  \begin{subfigure}[t]{0.32\textwidth}
    \includegraphics[width=\textwidth]{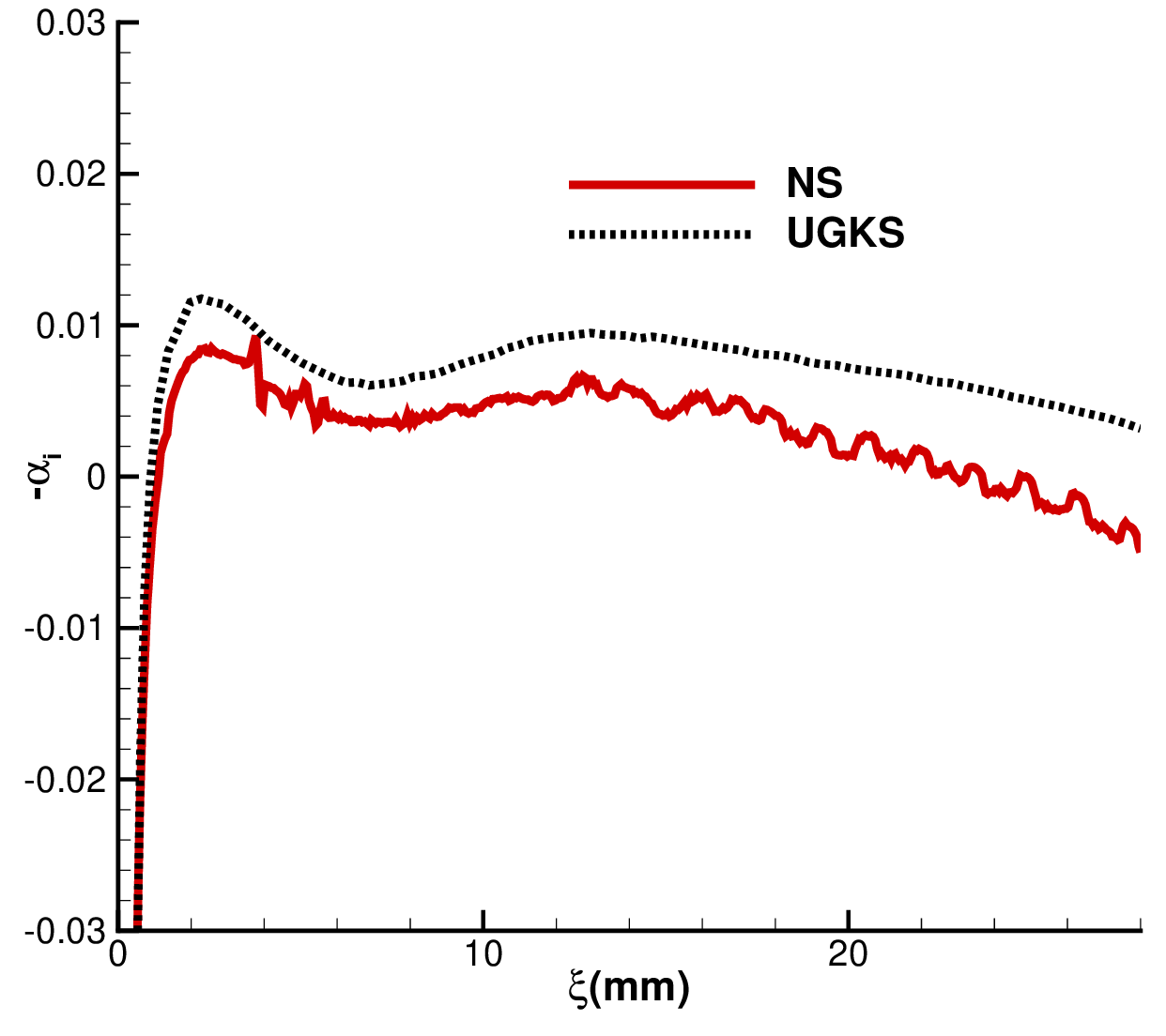}
    \caption{Growth rate}
    \label{fig:argon40km_wallDisp_c}
  \end{subfigure}
  \caption{Wall response for freestream fast-acoustic incidence at \(Re_{R_n}=39\):
  (a)~real part of the wall-pressure perturbation \(\mathrm{Re}\{p'_w\}\);
  (b)~wall-mode phase speed \(c_r\);
  (c)~spatial growth rate \(-\alpha_i\).
  Period and phase agree closely; after the nose peak, UGKS keeps a weakly positive growth-rate plateau while NS turns slightly negative farther downstream.}
  \label{fig:argon40km_wallDisp}
\end{figure}

Figure~\ref{fig:argon40km_wallDisp} compares the wall response of NS and UGKS.
Writing the complex wall-pressure amplitude as \(\widehat{p}_w(\xi)=|\widehat{p}_w|\,e^{i\theta(\xi)}\) along the surface coordinate \(\xi\), and nondimensionalizing \(\xi\) and \(\omega\) by \(R_n\) and \(U_\infty\), the local streamwise wave number, spatial growth rate and phase speed are
\begin{equation*}
  \alpha_r=\frac{\mathrm{d}\theta}{\mathrm{d}\xi},
  \qquad
  -\alpha_i=\frac{\mathrm{d}}{\mathrm{d}\xi}\ln|\widehat{p}_w|,
  \qquad
  c_r=\frac{\omega}{\alpha_r},
\end{equation*}
so that \(-\alpha_i>0\) denotes spatial amplification and \(c_r\) is the phase speed relative to \(U_\infty\).
In \(\mathrm{Re}\{p'_w\}\) (Fig.~\ref{fig:argon40km_wallDisp_a}), the two methods share nearly the same oscillation period and phase.
Near the blunt nose (\(\xi<5\,\mathrm{mm}\)), the first NS peak is slightly higher than that of UGKS, indicating a small difference in the initial energy deposited by shock--disturbance interaction and near-stagnation receptivity.
The phase speed \(c_r\) (Fig.~\ref{fig:argon40km_wallDisp_b}) shows a short apparent rise close to the stagnation point (\(\xi\to 0\)), with a peak near \(1.8\), caused by the geometric projection of a planar incident acoustic wave onto the curved surface.
For \(\xi>5\,\mathrm{mm}\), \(c_r\) levels off near \(1.1\) and the two solvers agree closely, consistent with a dominant fast acoustic boundary-layer mode.
The spatial growth rate \(-\alpha_i\) (Fig.~\ref{fig:argon40km_wallDisp_c}) first increases over the nose and reaches a peak of about \(0.012\) near \(\xi\approx 3\,\mathrm{mm}\).
Farther downstream (\(\xi>5\,\mathrm{mm}\)), the NS growth rate decreases gradually and crosses zero near \(\xi\approx 15\,\mathrm{mm}\), becoming weakly negative (spatial decay).
By contrast, after the peak the UGKS growth rate does not keep falling; over \(\xi=5\text{--}20\,\mathrm{mm}\) it stays on a positive plateau of about \(0.008\text{--}0.010\).
This gradual separation suggests that, under the weakly rarefied \(40\,\mathrm{km}\) condition, near-wall slip included in UGKS reduces viscous dissipation in the strong shear region relative to the continuum no-slip NS model, so that the disturbance mode remains weakly amplifying or nearly neutral over a longer streamwise distance.

\section{Hypersonic argon flow over a blunt wedge at 50\,km}
\label{sec:argon50km}

\begin{figure}[!htbp]
  \centering
  \begin{subfigure}[t]{0.48\textwidth}
    \includegraphics[width=\textwidth]{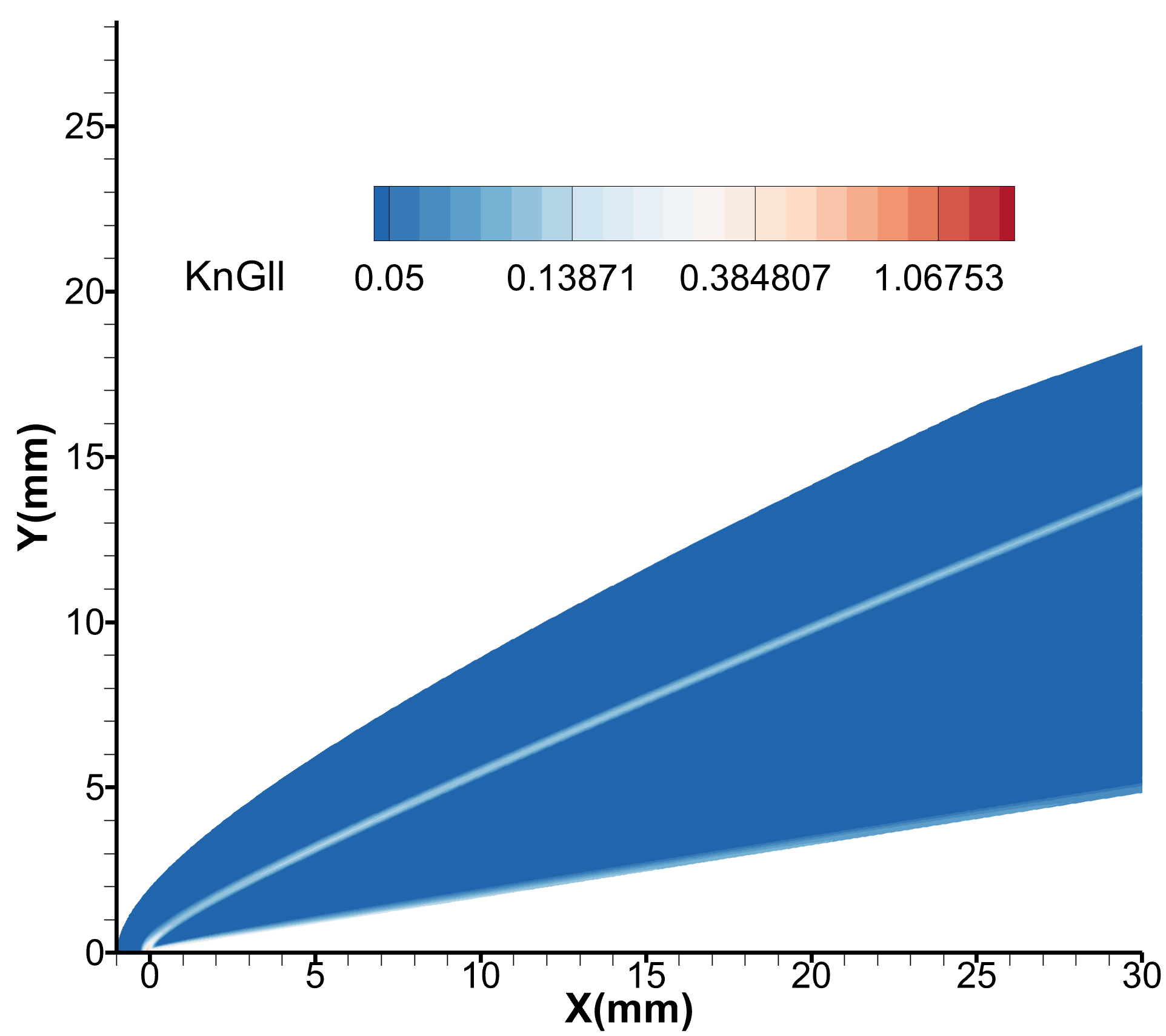}
    \caption{Full field}
    \label{fig:argon50km_KnGll_a}
  \end{subfigure}
  \hfill
  \begin{subfigure}[t]{0.48\textwidth}
    \includegraphics[width=\textwidth]{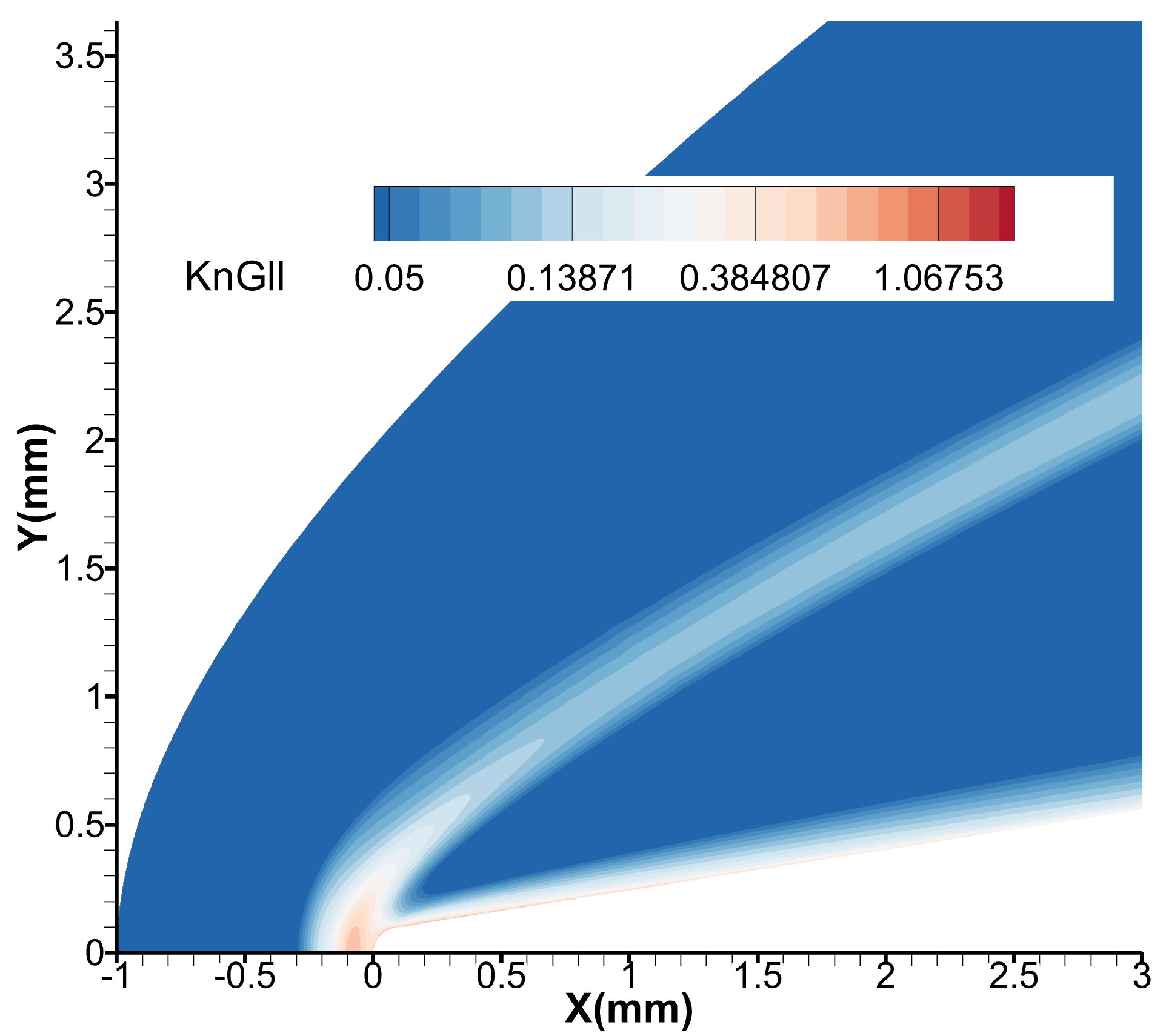}
    \caption{Local near nose}
    \label{fig:argon50km_KnGll_b}
  \end{subfigure}
  \caption{Gradient-length local Knudsen number \(\mathrm{Kn}_{\mathrm{Gll}}=l\,|\nabla\rho|/\rho\) for the argon base flow at 50\,km with \(Re_{R_n}=9.83\):
  (a)~full field; (b)~local view near the nose.}
  \label{fig:argon50km_KnGll}
\end{figure}

To further increase rarefaction, the same blunt-wedge geometry and argon gas model as in Sec.~\ref{sec:argon40km} are retained
(\(\gamma=5/3\), \(R_g=208.13\,\mathrm{J\,kg^{-1}\,K^{-1}}\), \(\mathrm{Pr}=2/3\), Bird VHS viscosity exponent \(0.81\)).
The freestream follows the U.S. Standard Atmosphere~\cite{us1976atmosphere} at \(50\,\mathrm{km}\), with
\begin{equation*}
  Ma_\infty=5,
  \qquad
  T^*=270.65\,\mathrm{K},
  \qquad
  a^*\approx 306.41\,\mathrm{m\,s^{-1}},
  \qquad
  U^*\approx 1532.0\,\mathrm{m\,s^{-1}},
\end{equation*}
and an argon freestream density obtained by keeping the same number density as air at that altitude,
\begin{equation*}
  \rho_\infty=\rho_{\mathrm{air}}\frac{M_{\mathrm{Ar}}}{M_{\mathrm{air}}},
\end{equation*}
where \(\rho_{\mathrm{air}}\) is the air mass density from the U.S. Standard Atmosphere, and \(M_{\mathrm{Ar}}\) and \(M_{\mathrm{air}}\) are the molar masses of argon and air.
The resulting nose Reynolds number is
\begin{equation*}
  Re_{R_n}=9.83.
\end{equation*}
In nondimensional form, \(R=1/(\gamma Ma_\infty^2)=0.024\) and \(\mu_{\mathrm{ref}}=1/Re_{R_n}\).
The wall remains isothermal with fully accommodating diffuse reflection; the wall temperature is fixed at \(T_w^*=300\,\mathrm{K}\), so
\begin{equation*}
  T_w/T_\infty\approx 1.108.
\end{equation*}
A freestream fast acoustic wave (\(k_2=k_3=0\)) of dimensional frequency \(f=300\,\mathrm{kHz}\) is imposed after the base flow is frozen, with nondimensional angular frequency and streamwise wave number
\begin{equation*}
  \omega=\frac{2\pi f R_n}{U^*}\approx 0.12304,
  \qquad
  k_1=\frac{\omega}{U_\infty+a_\infty}=\frac{\omega}{1.2}\approx 0.10253,
\end{equation*}
and a freestream amplitude scale of \(10^{-4}\).
The discrete velocity space is identical to that of Sec.~\ref{sec:argon40km}.
The physical mesh is similar as well, except that the outer boundary ahead of the nose is extended upstream to \(X=-1\,\mathrm{mm}\) so that the thicker detached shock and the temperature precursor remain inside the domain.
The corresponding \(\mathrm{Kn}_{\mathrm{Gll}}\) field is shown in Fig.~\ref{fig:argon50km_KnGll}: relative to the \(40\,\mathrm{km}\) case (Fig.~\ref{fig:argon40km_KnGll}), the elevated-\(\mathrm{Kn}_{\mathrm{Gll}}\) region in the bow shock and near the wall is broader and more intense, consistent with the lower freestream density.

\subsection{Base flow}
\begin{figure}[!htbp]
  \centering
  \begin{subfigure}[t]{0.32\textwidth}
    \includegraphics[width=\textwidth]{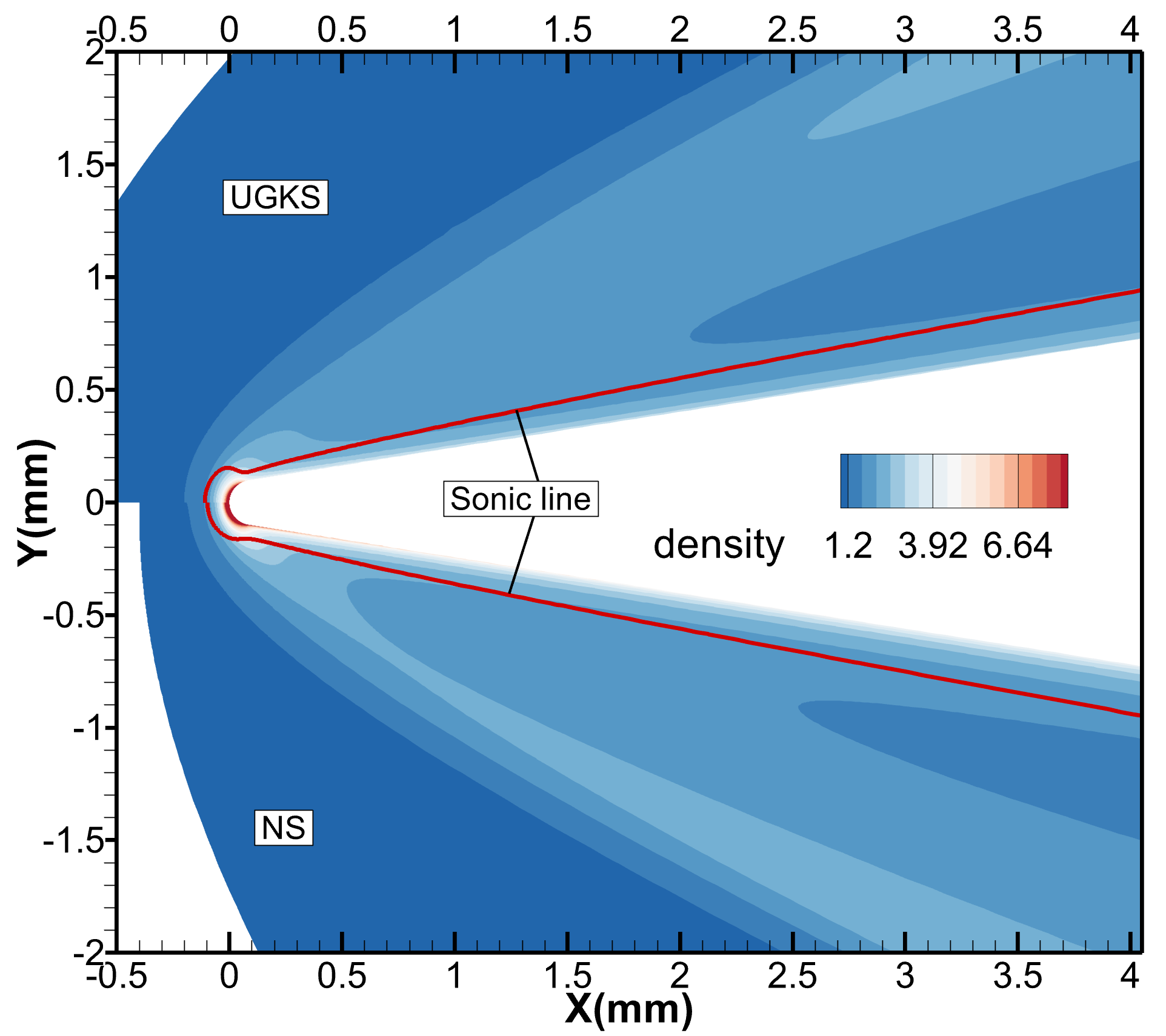}
    \caption{Density}
    \label{fig:argon50km_vsNS_a}
  \end{subfigure}
  \hfill
  \begin{subfigure}[t]{0.32\textwidth}
    \includegraphics[width=\textwidth]{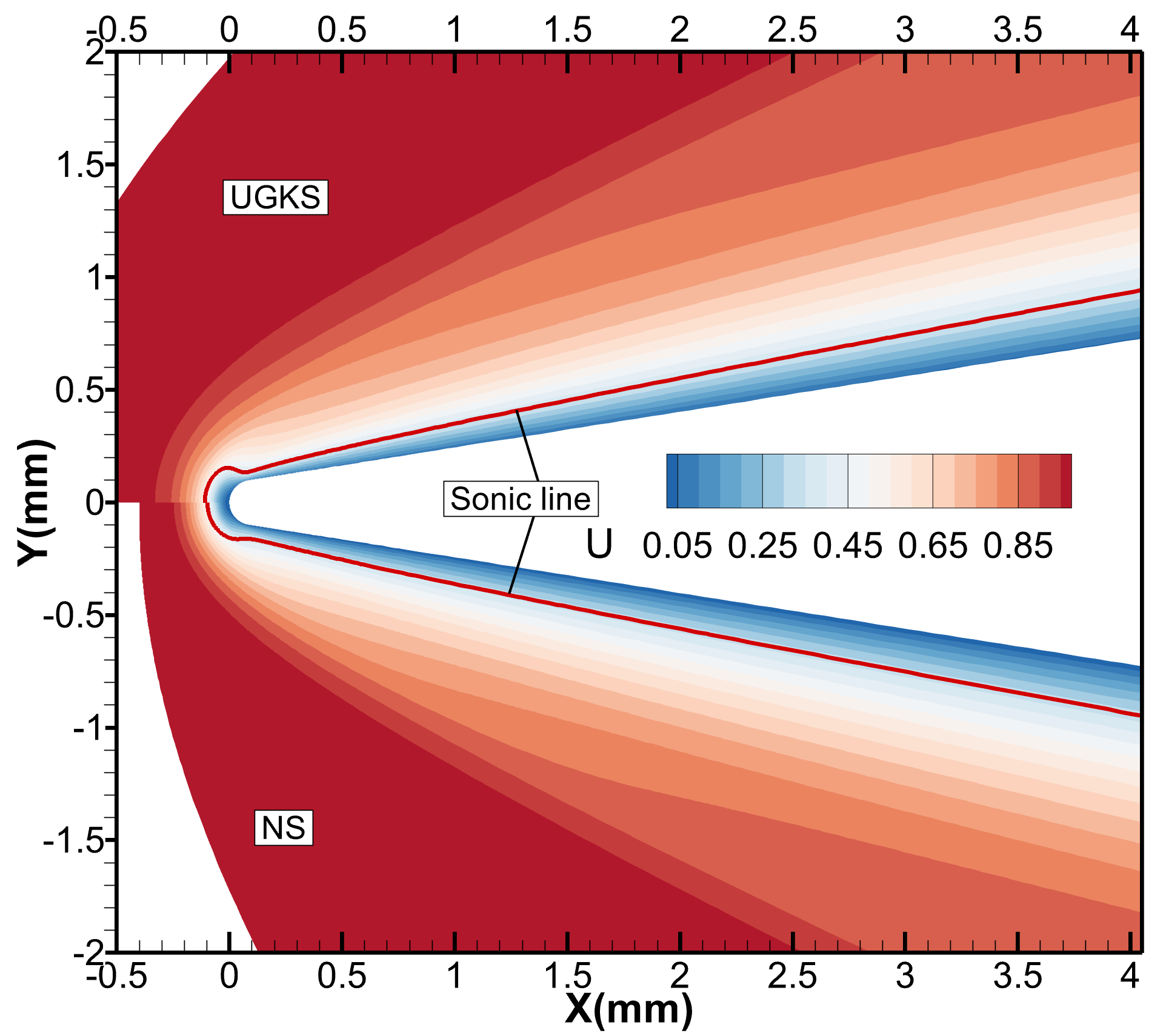}
    \caption{$U$}
    \label{fig:argon50km_vsNS_b}
  \end{subfigure}
  \hfill
  \begin{subfigure}[t]{0.32\textwidth}
    \includegraphics[width=\textwidth]{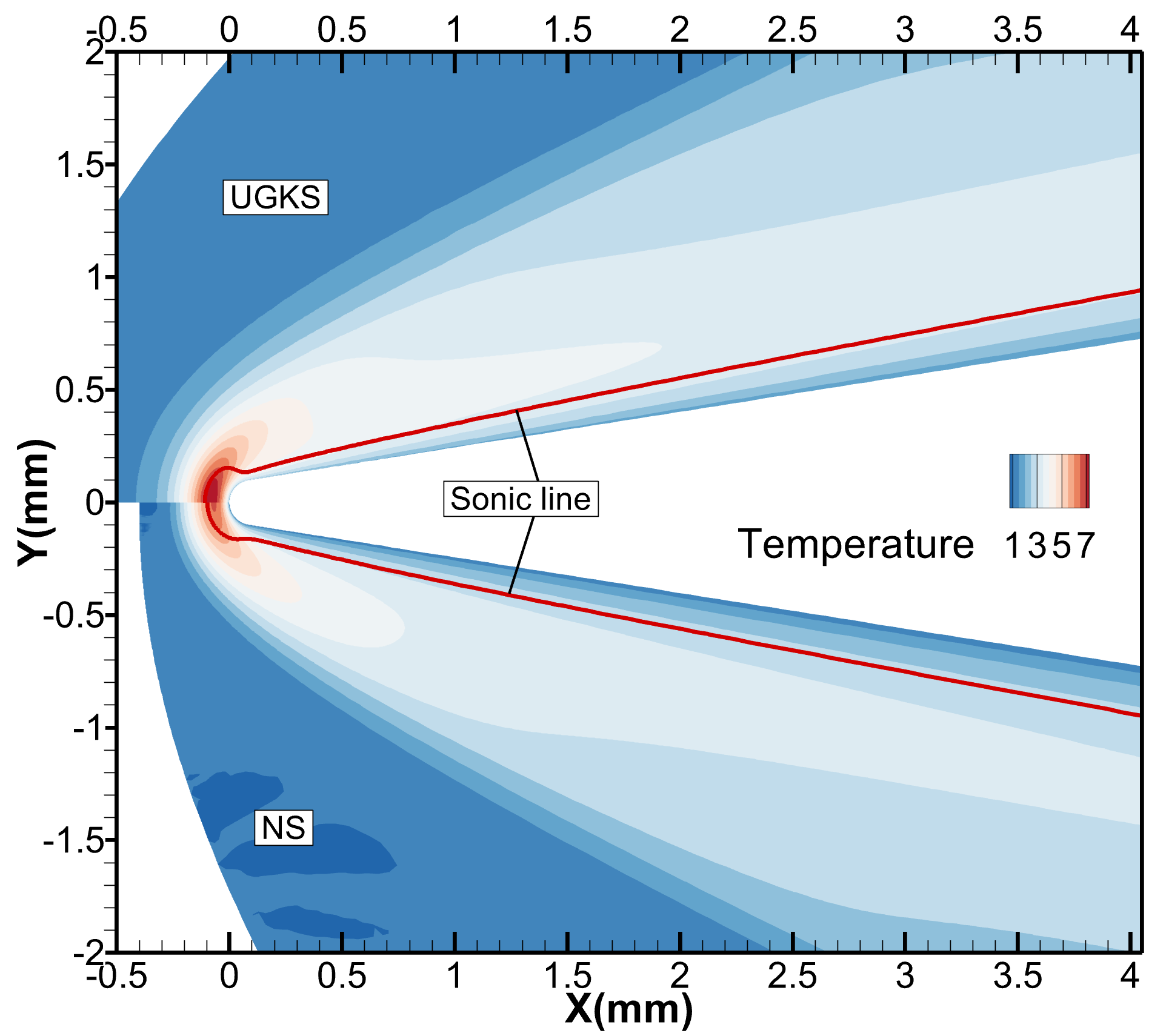}
    \caption{Temperature}
    \label{fig:argon50km_vsNS_c}
  \end{subfigure}
  \caption{Steady base flow for argon at 50\,km with \(Re_{R_n}=9.83\): (a)~density, (b)~streamwise velocity and (c)~temperature.
  Upper half: UGKS; lower half: NS.}
  \label{fig:argon50km_vsNS}
\end{figure}

\begin{figure}[!htbp]
  \centering
  \begin{subfigure}[t]{0.32\textwidth}
    \includegraphics[width=\textwidth]{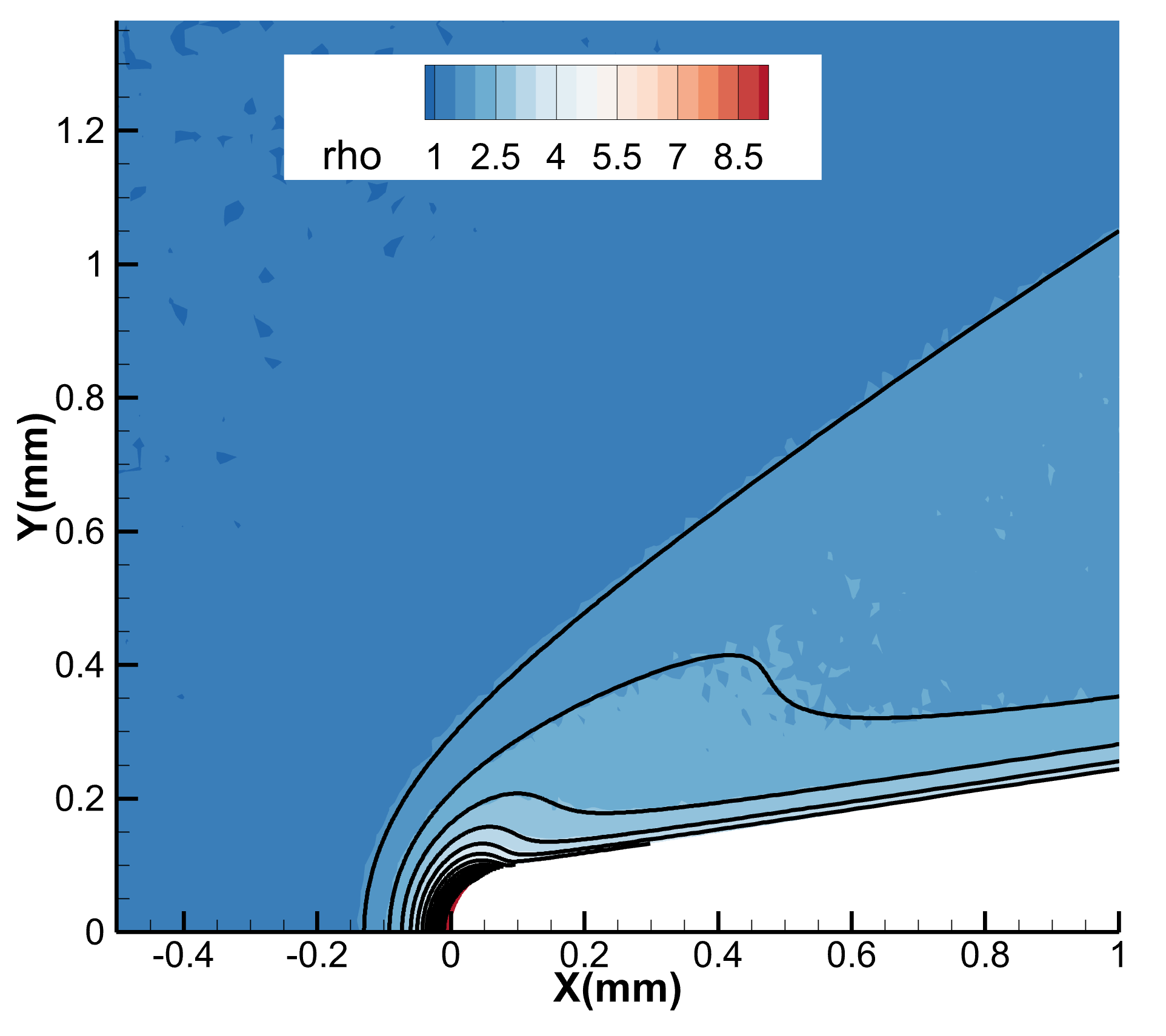}
    \caption{Density}
    \label{fig:argon50km_vsDSMC_a}
  \end{subfigure}
  \hfill
  \begin{subfigure}[t]{0.32\textwidth}
    \includegraphics[width=\textwidth]{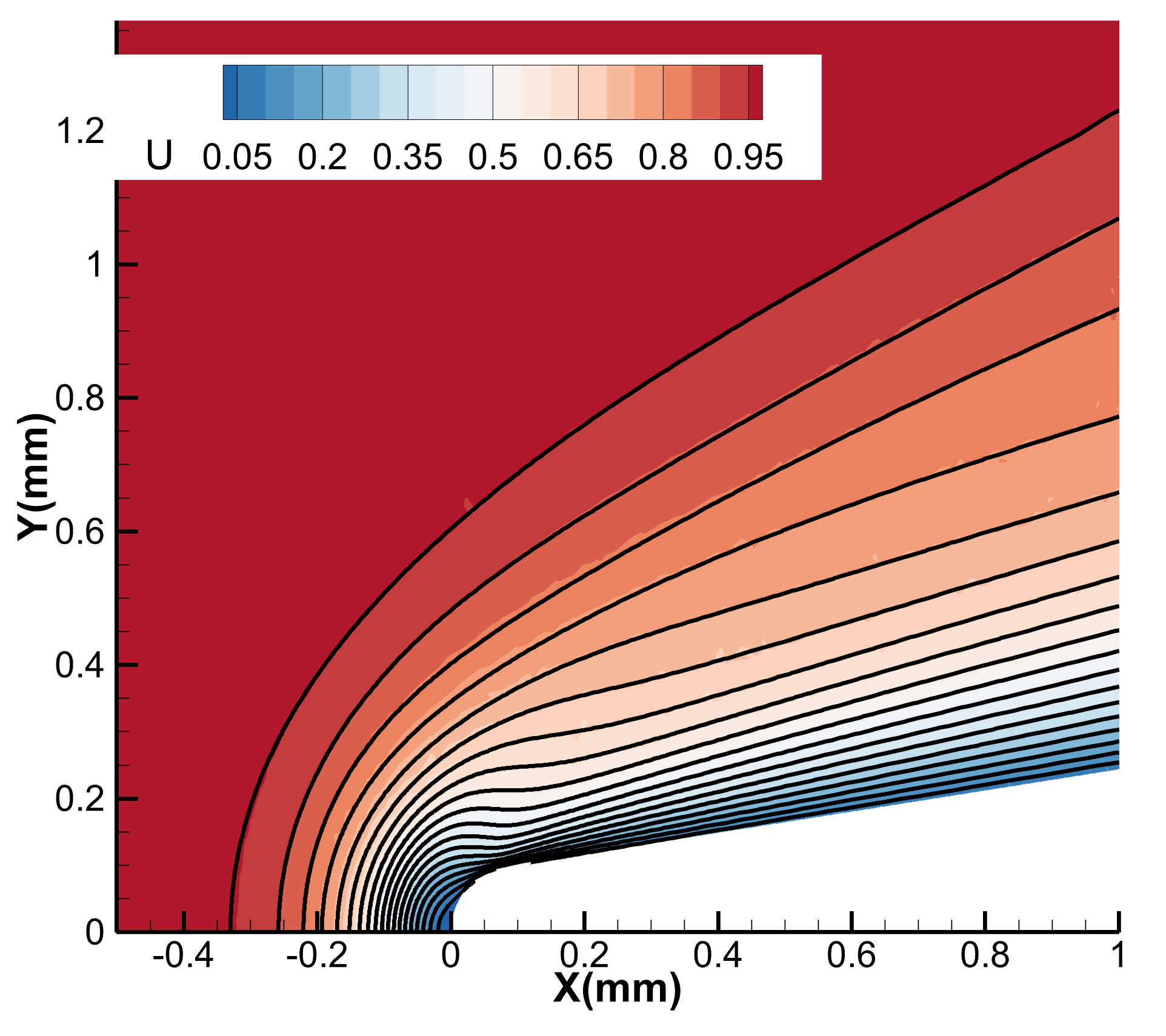}
    \caption{$U$}
    \label{fig:argon50km_vsDSMC_b}
  \end{subfigure}
  \hfill
  \begin{subfigure}[t]{0.32\textwidth}
    \includegraphics[width=\textwidth]{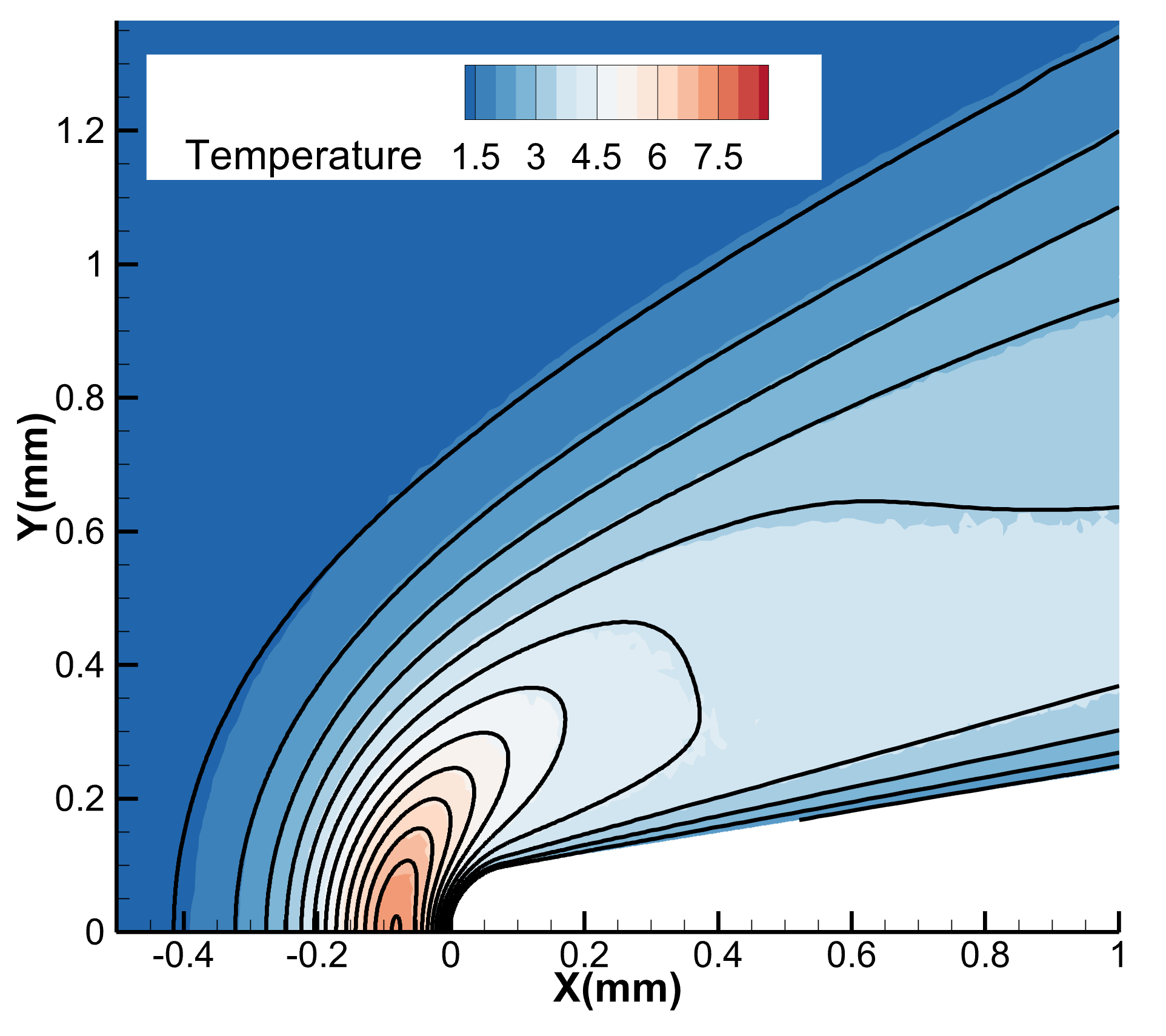}
    \caption{Temperature}
    \label{fig:argon50km_vsDSMC_c}
  \end{subfigure}
  \caption{Argon base flow at 50\,km with \(Re_{R_n}=9.83\) compared with DSMC (DS2V)~\cite{bird1994molecular}: (a)~density, (b)~streamwise velocity and (c)~temperature.
  Colour contours: DSMC; black isolines: UGKS.}
  \label{fig:argon50km_vsDSMC}
\end{figure}

\begin{figure}[!htbp]
  \centering
  \begin{subfigure}[t]{0.24\textwidth}
    \includegraphics[width=\textwidth]{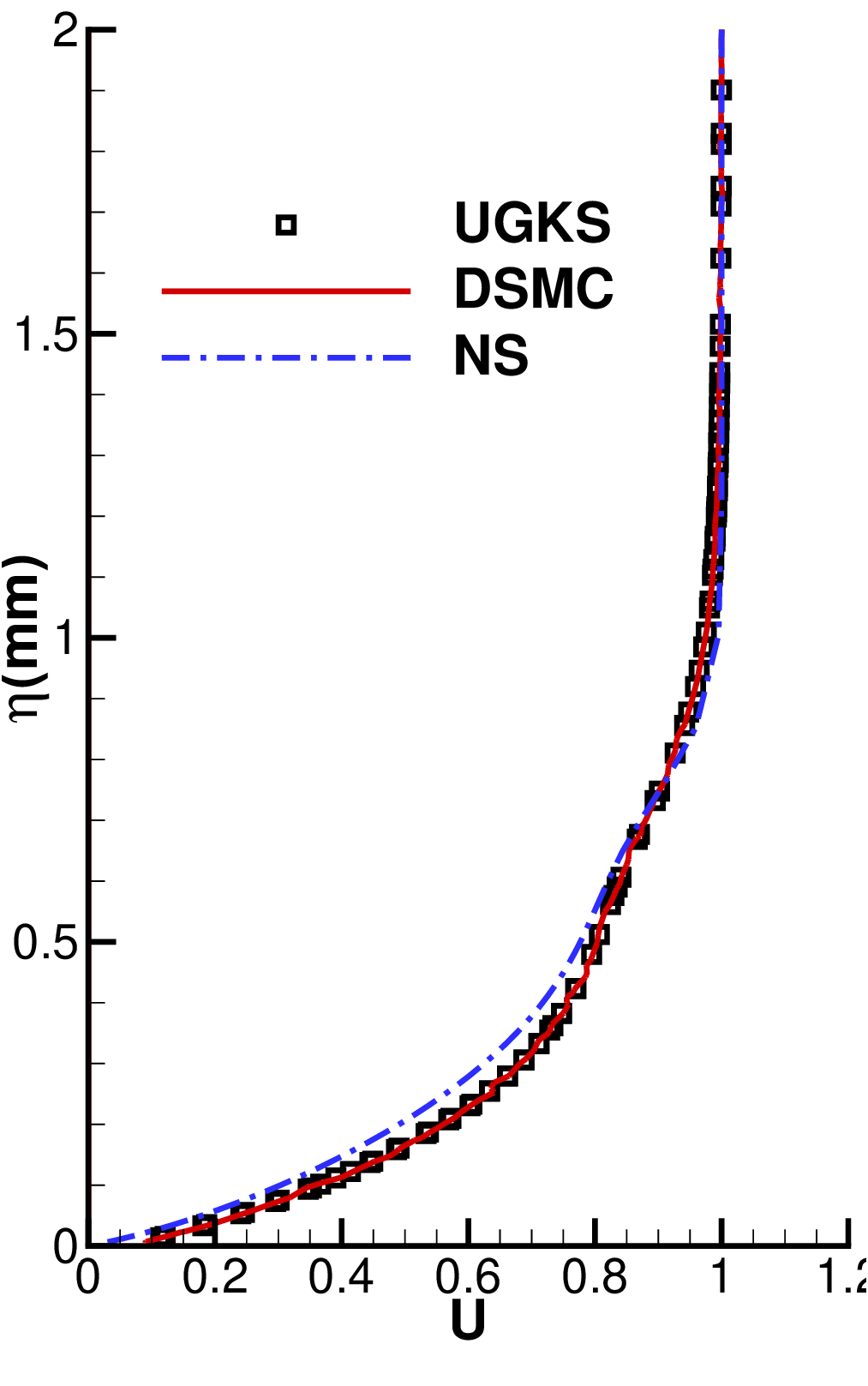}
    \caption{$U$, $\xi=1\,\mathrm{mm}$}
    \label{fig:argon50km_profile_a}
  \end{subfigure}
  \hfill
  \begin{subfigure}[t]{0.24\textwidth}
    \includegraphics[width=\textwidth]{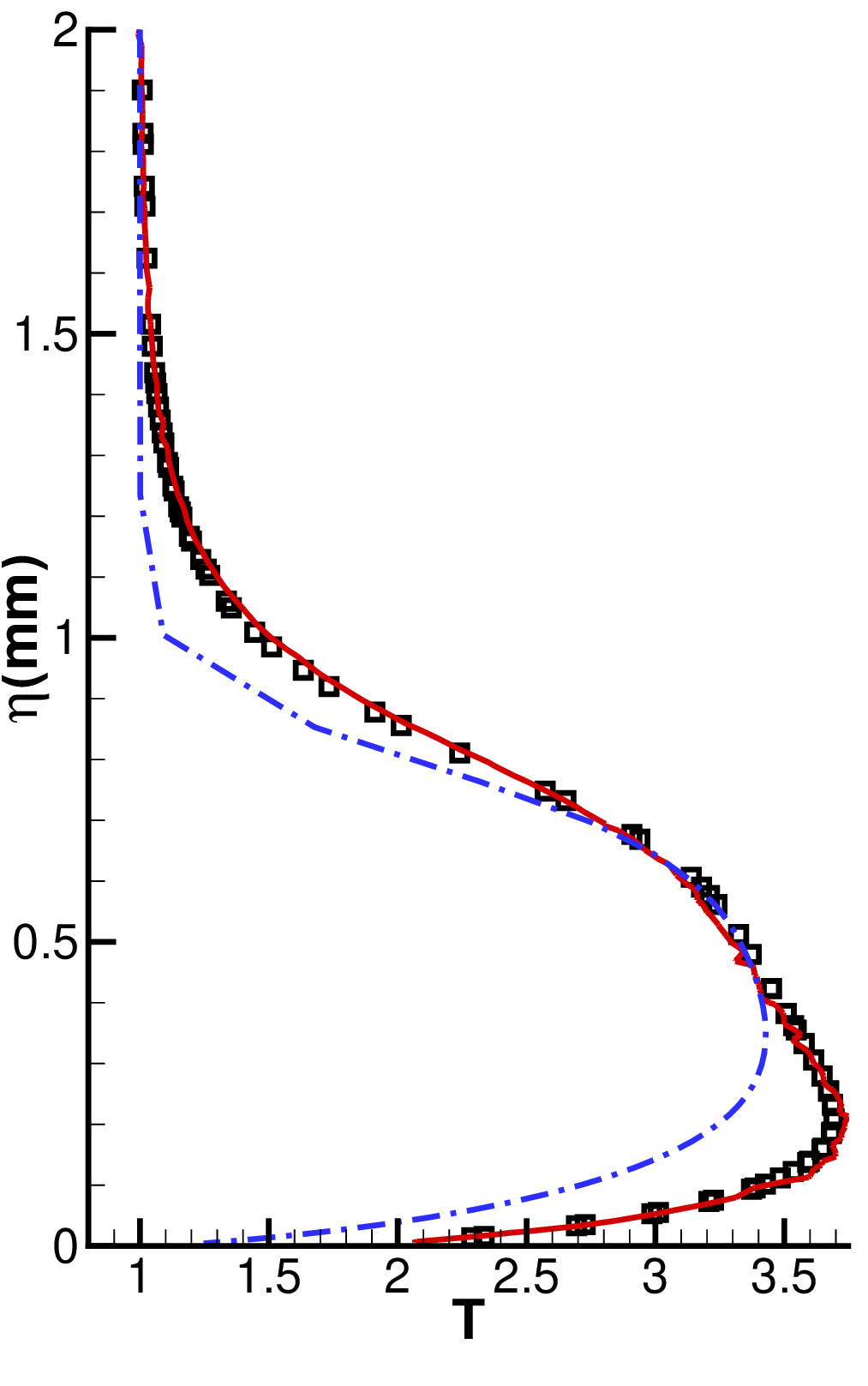}
    \caption{$T$, $\xi=1\,\mathrm{mm}$}
    \label{fig:argon50km_profile_b}
  \end{subfigure}
  \hfill
  \begin{subfigure}[t]{0.24\textwidth}
    \includegraphics[width=\textwidth]{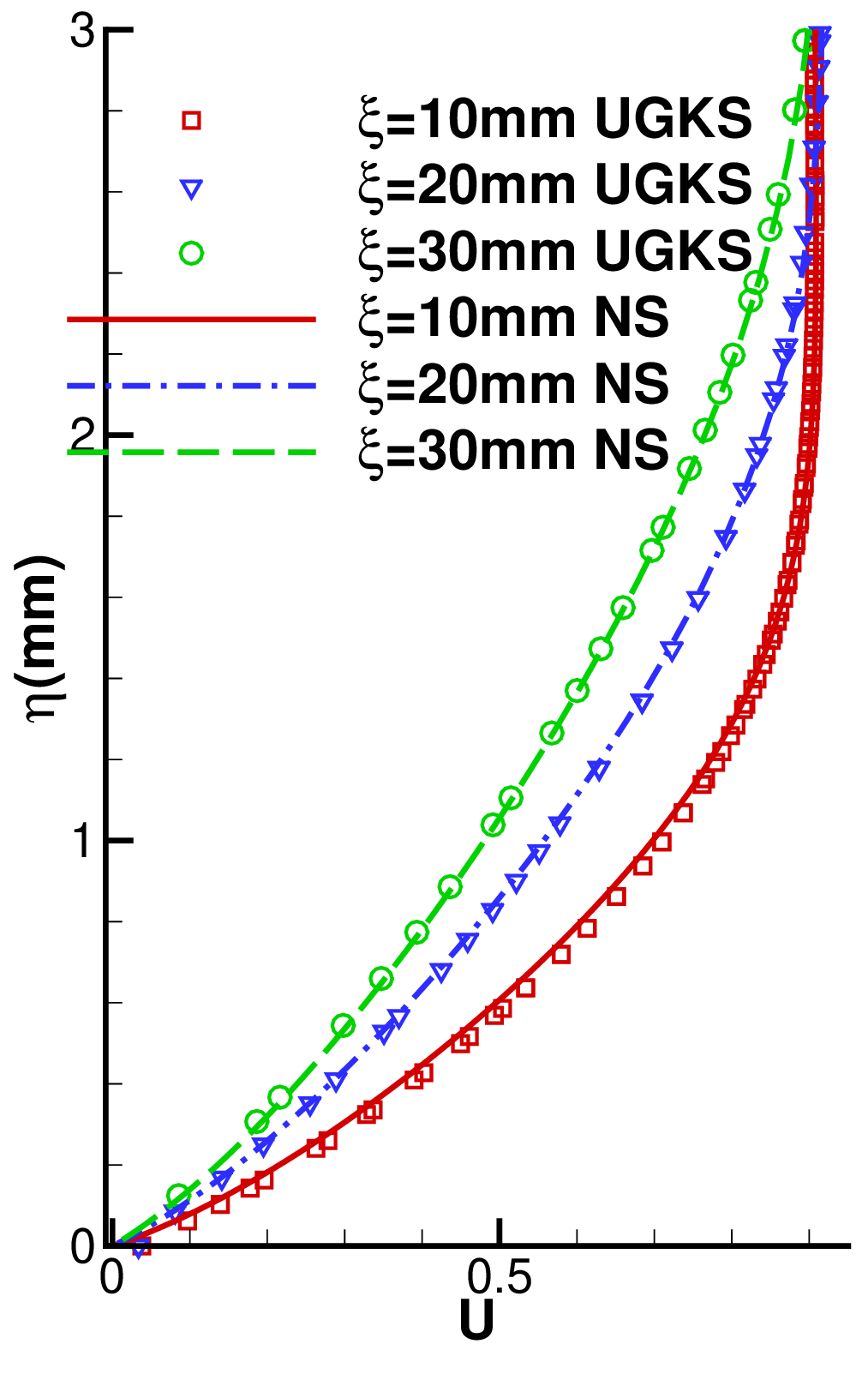}
    \caption{$U$, multi-station}
    \label{fig:argon50km_profile_c}
  \end{subfigure}
  \hfill
  \begin{subfigure}[t]{0.24\textwidth}
    \includegraphics[width=\textwidth]{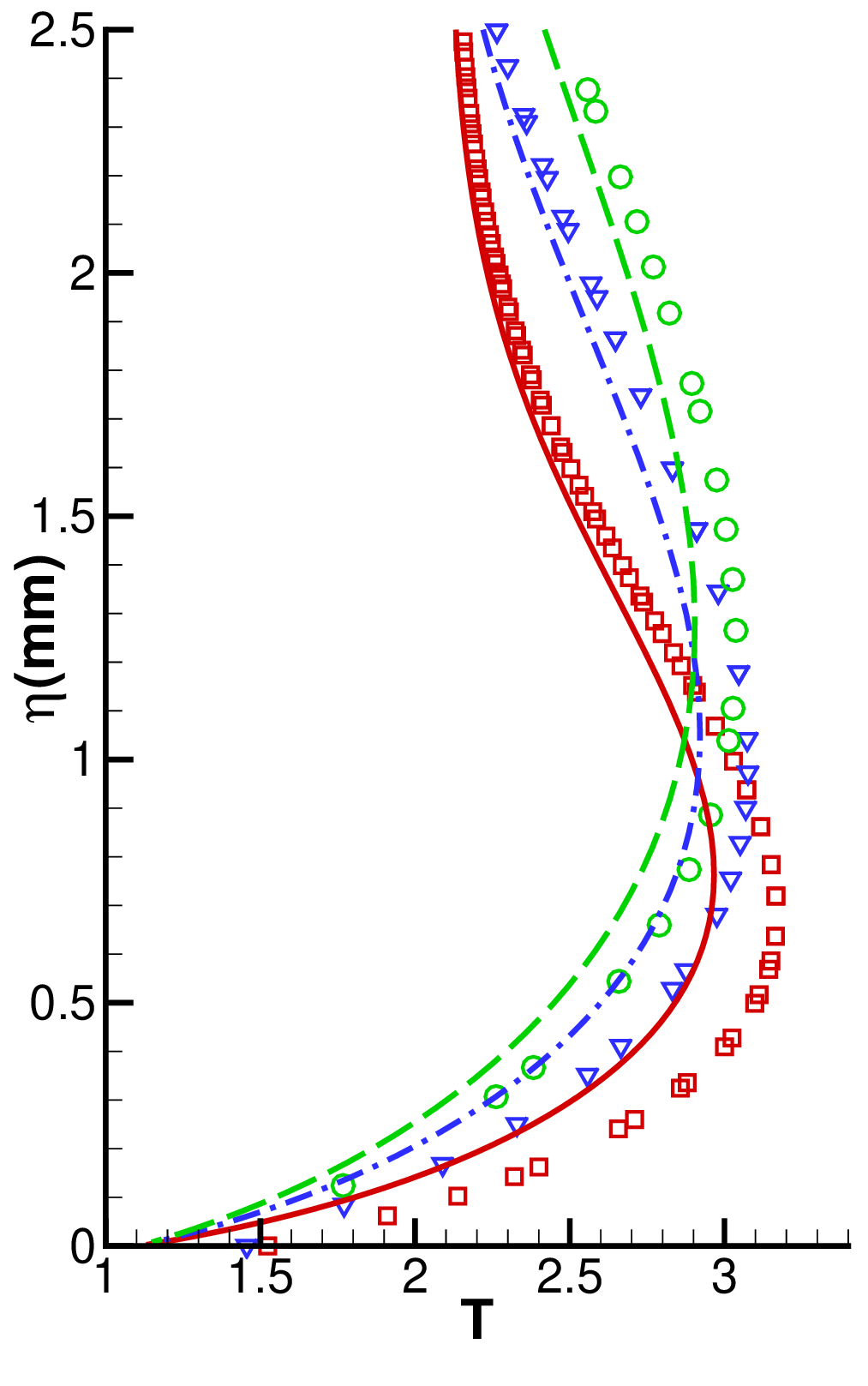}
    \caption{$T$, multi-station}
    \label{fig:argon50km_profile_d}
  \end{subfigure}
  \caption{Wall-normal base-flow profiles for argon at 50\,km with \(Re_{R_n}=9.83\):
  (a)--(b)~streamwise velocity and temperature at \(\xi=1\,\mathrm{mm}\);
  (c)--(d)~profiles at downstream stations.
  Slip and temperature jump are stronger than at \(40\,\mathrm{km}\).
  Farther downstream the velocity profiles inside the boundary layer approach each other, while the temperature discrepancy persists.}
  \label{fig:argon50km_profile}
\end{figure}

The resulting base flow is again compared with NS and with DSMC solutions from DS2V~\cite{bird1994molecular}.
Relative to the \(40\,\mathrm{km}\) case, the bow shock stands farther from the body and the shock layer is thicker, consistent with the lower freestream density.
Figure~\ref{fig:argon50km_vsNS} compares the UGKS density, streamwise velocity and temperature with the NS solution; Fig.~\ref{fig:argon50km_vsDSMC} shows the corresponding DSMC comparison.
The UGKS--NS discrepancy near the wall is amplified.
At \(\xi=1\,\mathrm{mm}\) (Figs.~\ref{fig:argon50km_profile_a} and~\ref{fig:argon50km_profile_b}), velocity slip and temperature jump are much stronger than at \(40\,\mathrm{km}\).
Farther downstream (Figs.~\ref{fig:argon50km_profile_c} and~\ref{fig:argon50km_profile_d}), inside the boundary layer the UGKS and NS velocity profiles approach each other, while the temperature difference remains large even at \(\xi=20\) and \(30\,\mathrm{mm}\).
These base-flow features provide the background for the disturbance evolution discussed below.
Agreement with DSMC remains close throughout the shock layer and the near-wall region.

\subsection{Disturbance field}
\begin{figure}[!htbp]
  \centering
  \begin{subfigure}[t]{0.40\textwidth}
    \includegraphics[width=\textwidth]{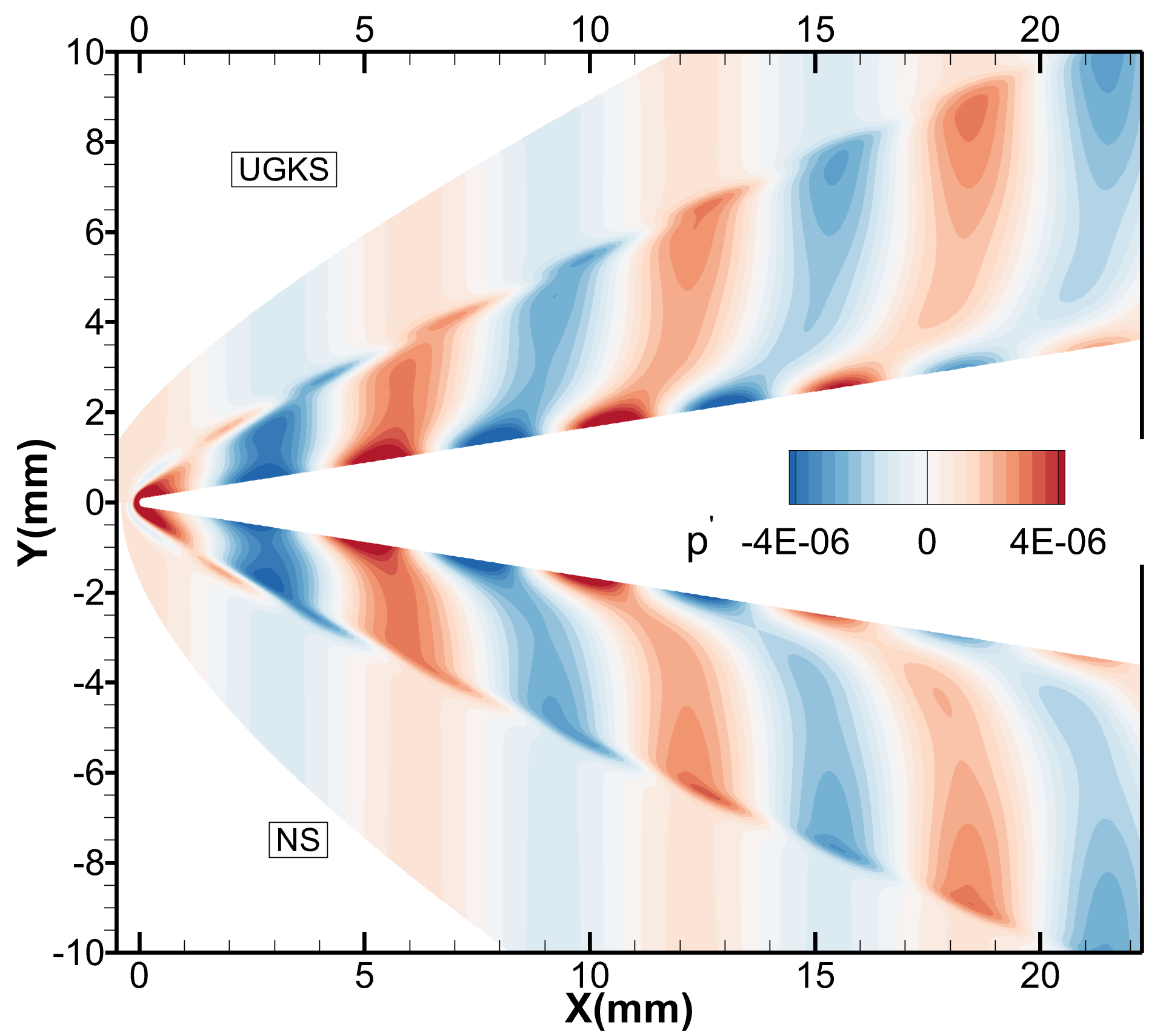}
    \caption{$\mathrm{Re}\{p'\}$, full field}
    \label{fig:argon50km_delta_a}
  \end{subfigure}
  \quad
  \begin{subfigure}[t]{0.40\textwidth}
    \includegraphics[width=\textwidth]{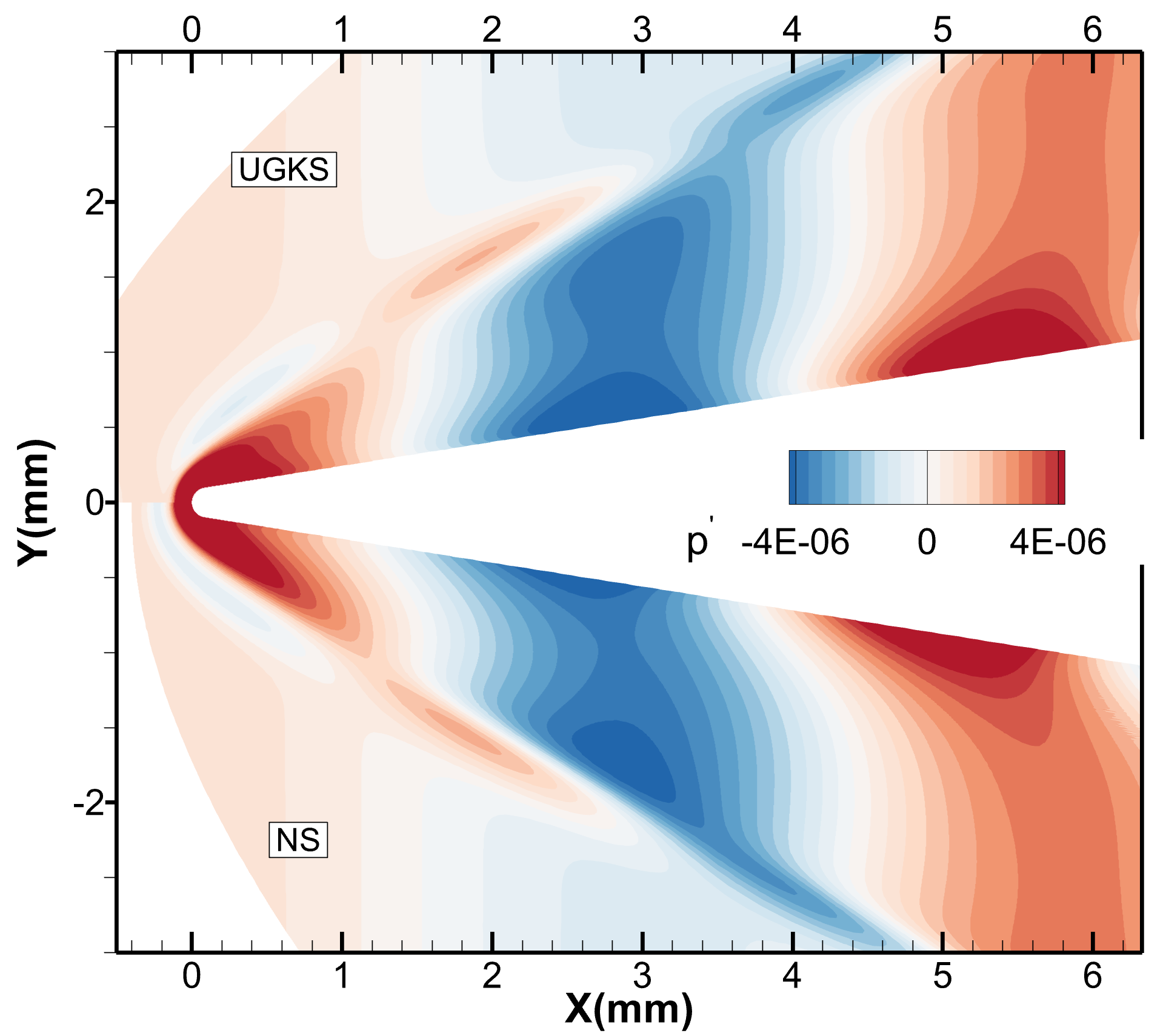}
    \caption{$\mathrm{Re}\{p'\}$, near nose}
    \label{fig:argon50km_delta_b}
  \end{subfigure}\\[0.6em]
  \begin{subfigure}[t]{0.40\textwidth}
    \includegraphics[width=\textwidth]{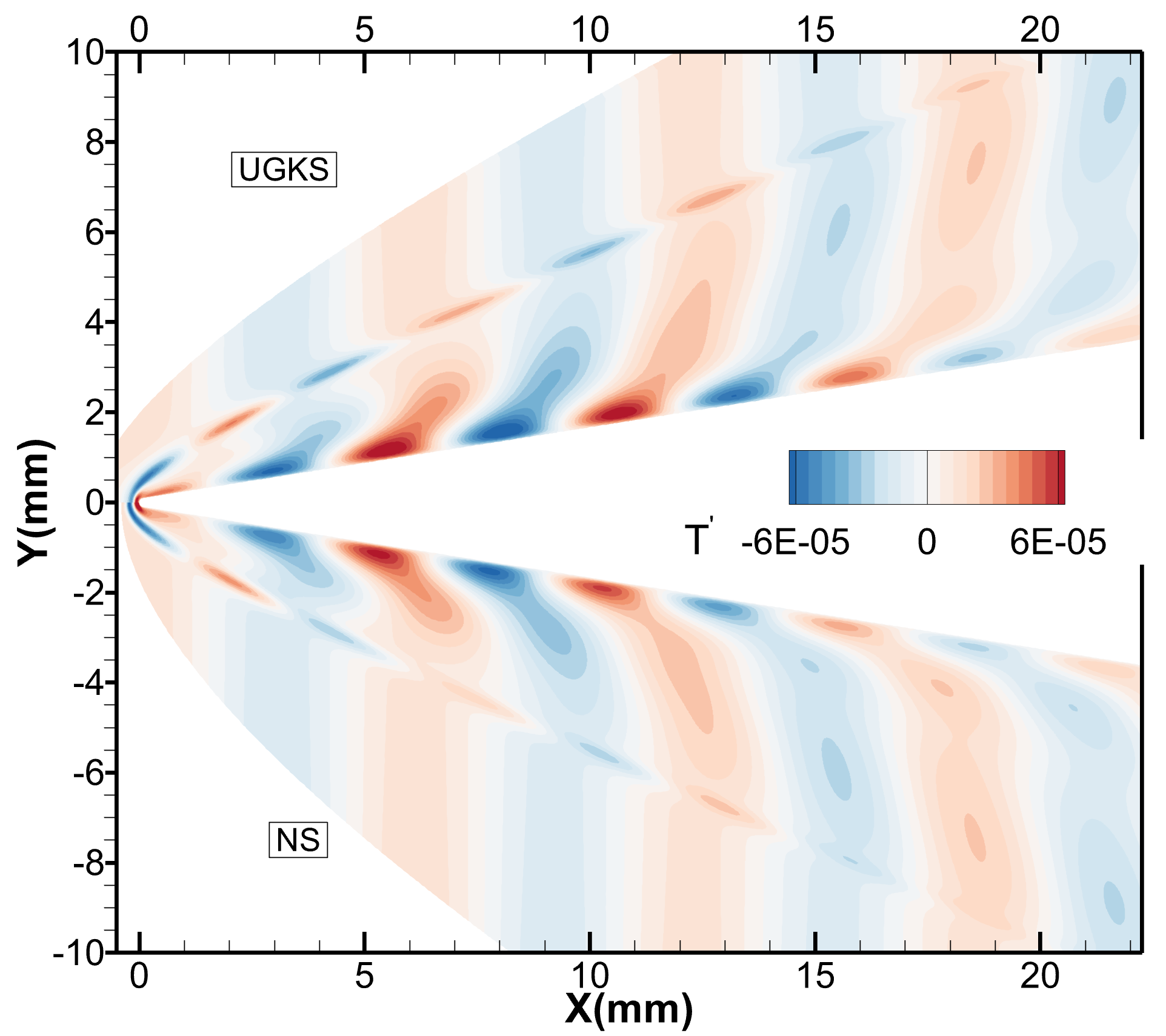}
    \caption{$\mathrm{Re}\{T'\}$, full field}
    \label{fig:argon50km_delta_c}
  \end{subfigure}
  \quad
  \begin{subfigure}[t]{0.40\textwidth}
    \includegraphics[width=\textwidth]{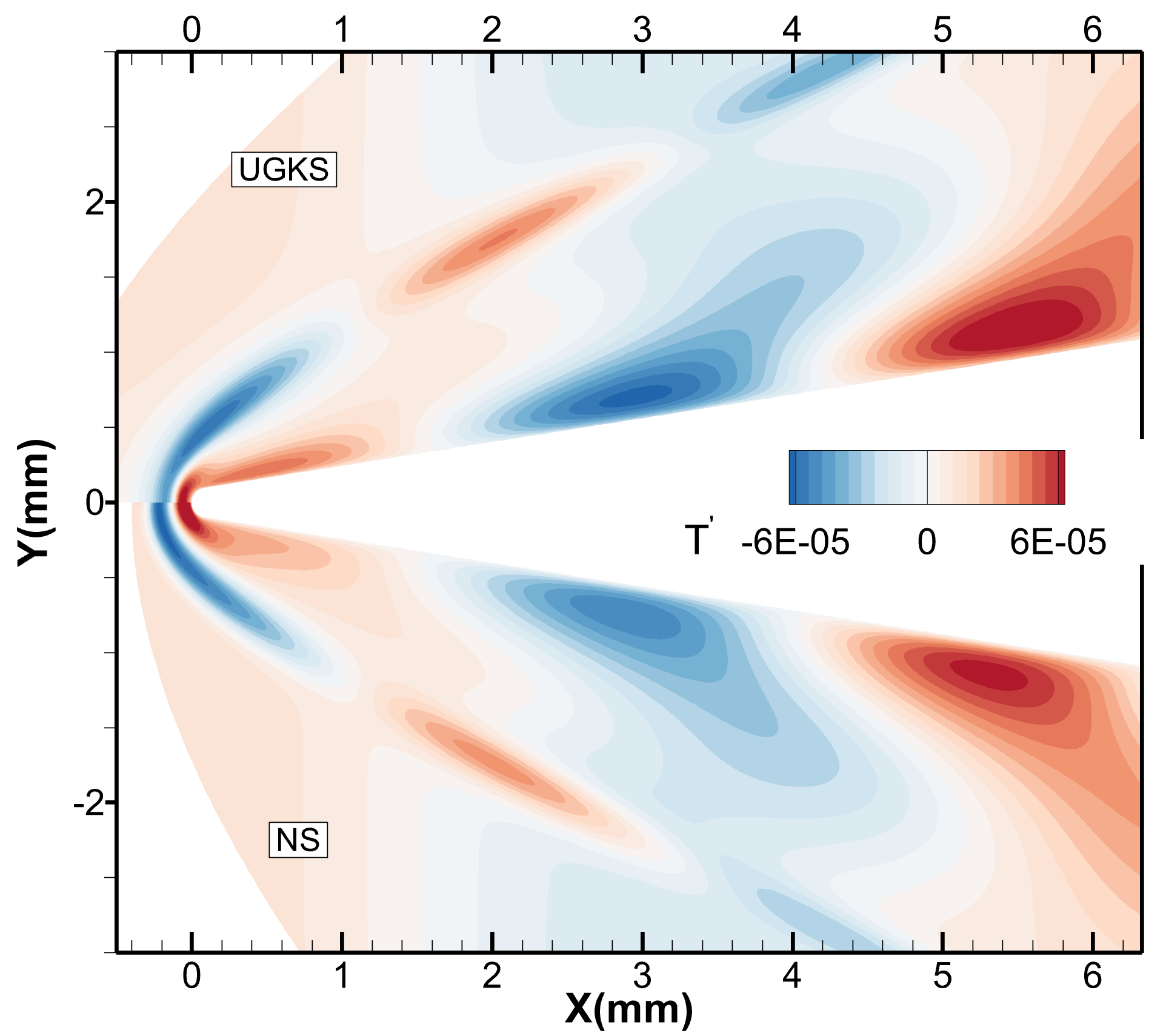}
    \caption{$\mathrm{Re}\{T'\}$, near nose}
    \label{fig:argon50km_delta_d}
  \end{subfigure}\\[0.6em]
  \begin{subfigure}[t]{0.40\textwidth}
    \includegraphics[width=\textwidth]{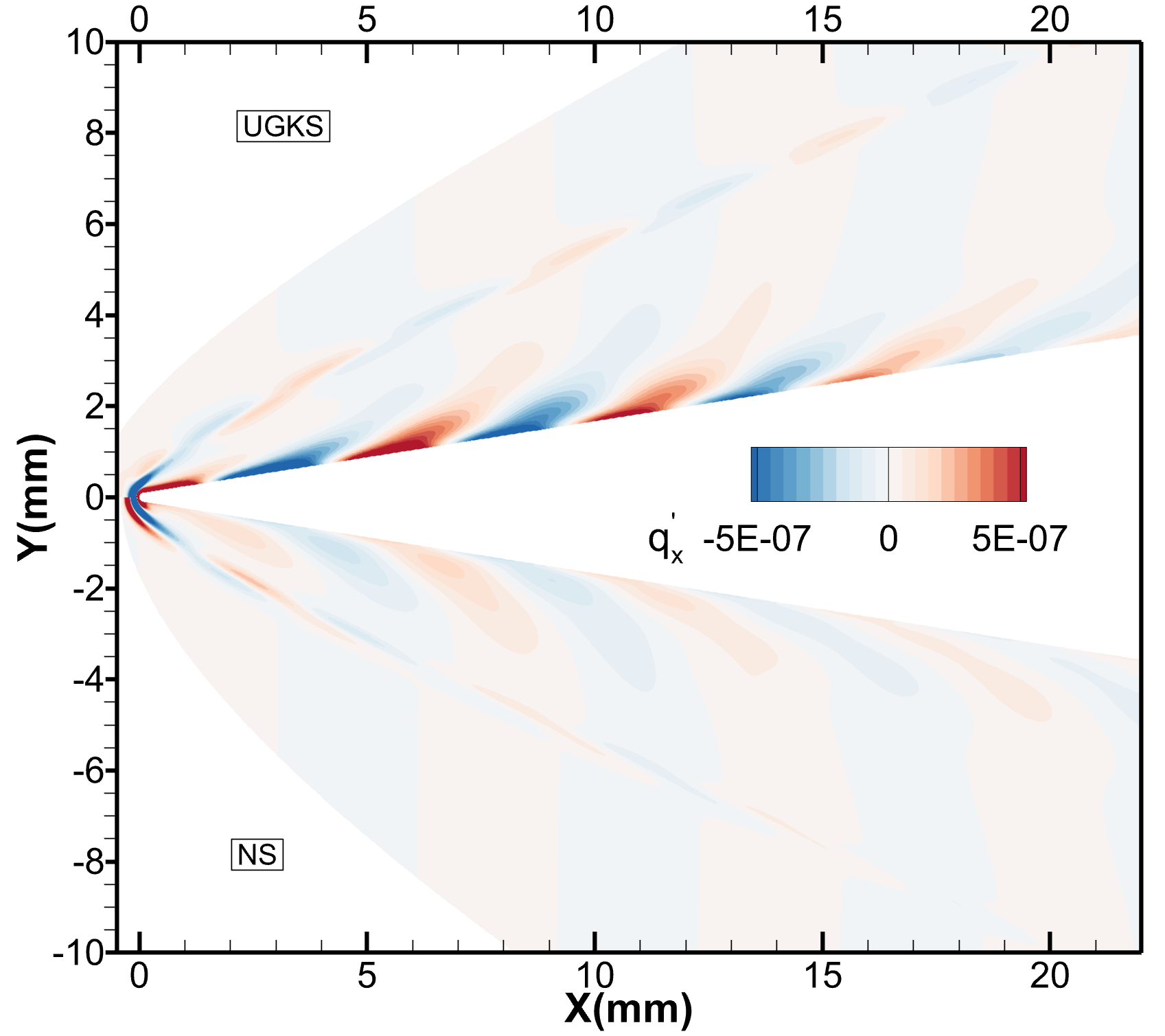}
    \caption{$\mathrm{Re}\{q'_x\}$, full field}
    \label{fig:argon50km_delta_e}
  \end{subfigure}
  \quad
  \begin{subfigure}[t]{0.40\textwidth}
    \includegraphics[width=\textwidth]{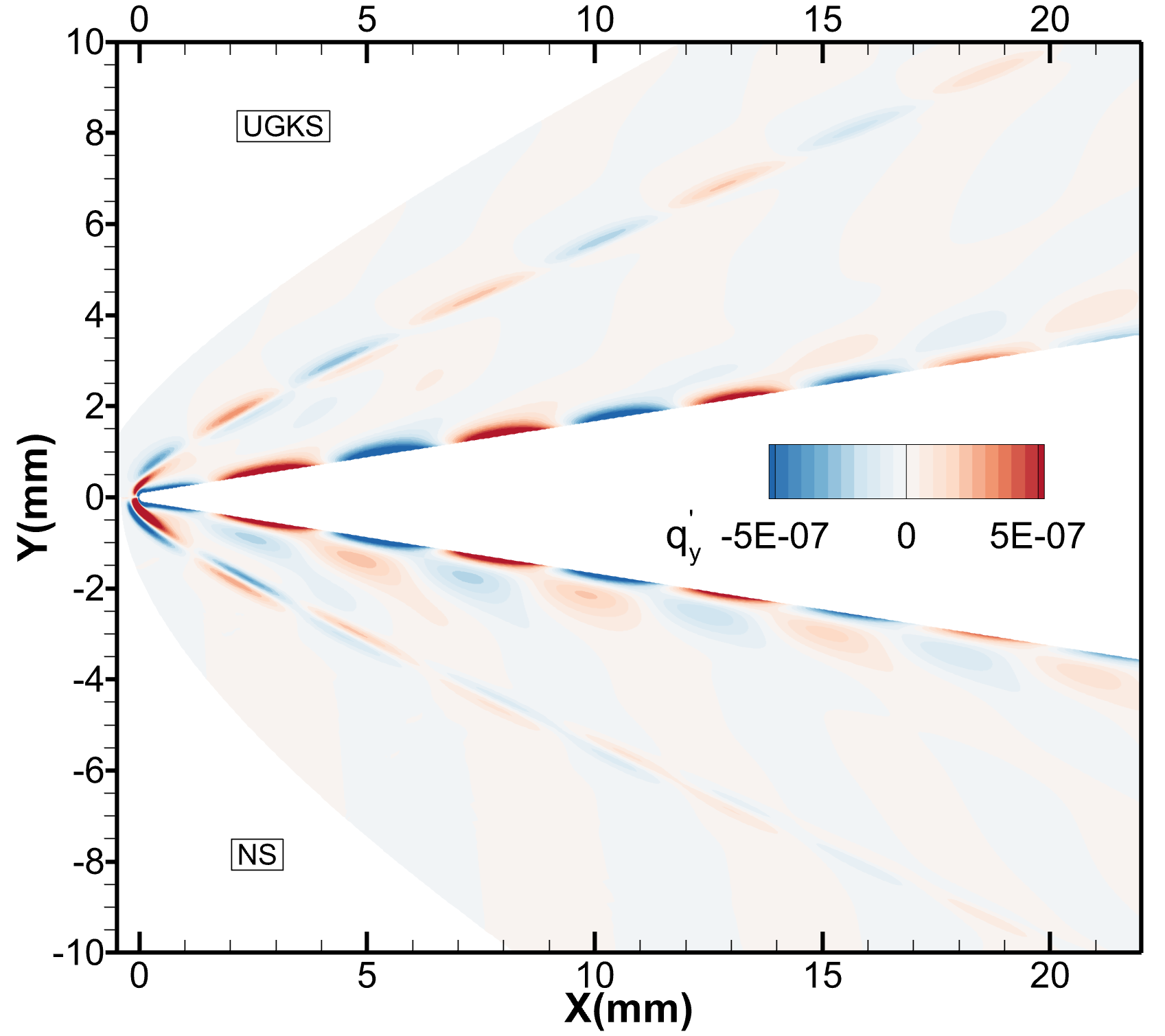}
    \caption{$\mathrm{Re}\{q'_y\}$, full field}
    \label{fig:argon50km_delta_f}
  \end{subfigure}
  \caption{Real parts of the pressure, temperature and heat-flux perturbations for freestream fast-acoustic incidence on the argon blunt wedge at 50\,km with \(Re_{R_n}=9.83\):
  (a)--(b)~$\mathrm{Re}\{p'\}$; (c)--(d)~$\mathrm{Re}\{T'\}$;
  (e)--(f)~full-field $\mathrm{Re}\{q'_x\}$ and $\mathrm{Re}\{q'_y\}$.
  Upper half: UGKS; lower half: Fourier-decomposed DNS in (a)--(d), and NS with Fourier's law in (e)--(f).}
  \label{fig:argon50km_delta}
\end{figure}

\begin{figure}[!htbp]
  \centering
  \begin{subfigure}[t]{0.32\textwidth}
    \includegraphics[width=\textwidth]{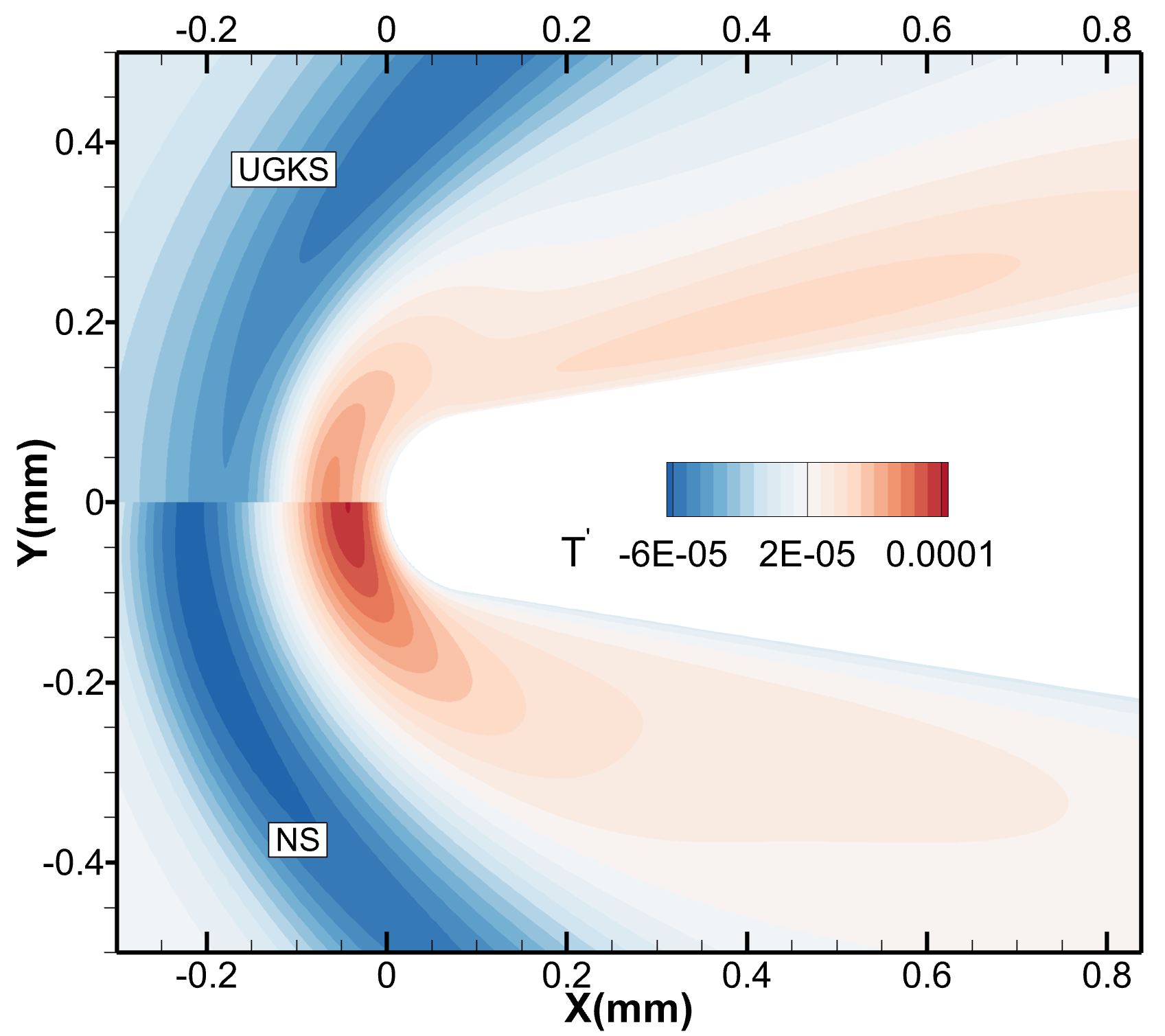}
    \caption{$\mathrm{Re}\{T'\}$}
    \label{fig:argon50km_qPrime_a}
  \end{subfigure}
  \hfill
  \begin{subfigure}[t]{0.32\textwidth}
    \includegraphics[width=\textwidth]{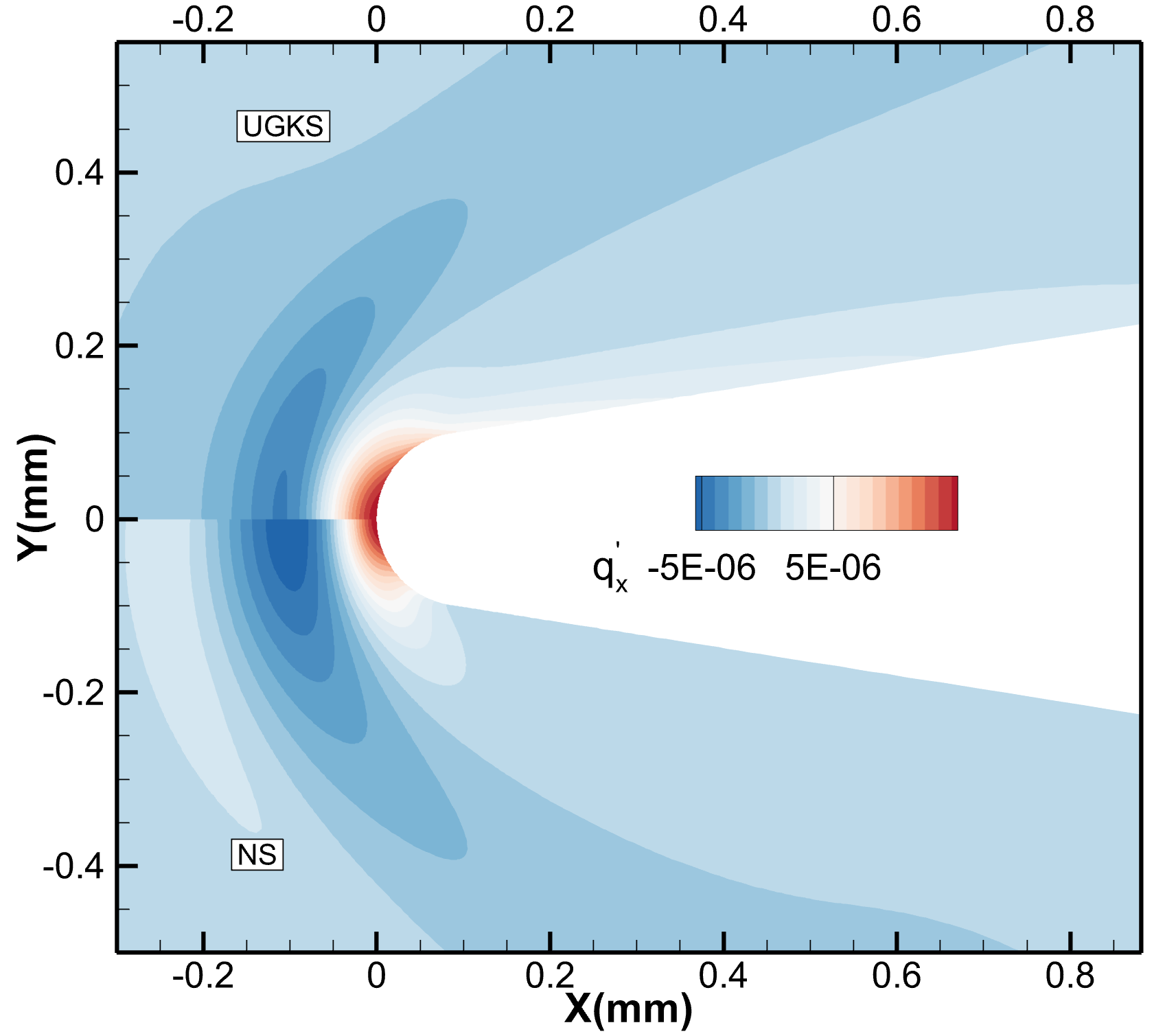}
    \caption{$\mathrm{Re}\{q'_x\}$}
    \label{fig:argon50km_qPrime_b}
  \end{subfigure}
  \hfill
  \begin{subfigure}[t]{0.32\textwidth}
    \includegraphics[width=\textwidth]{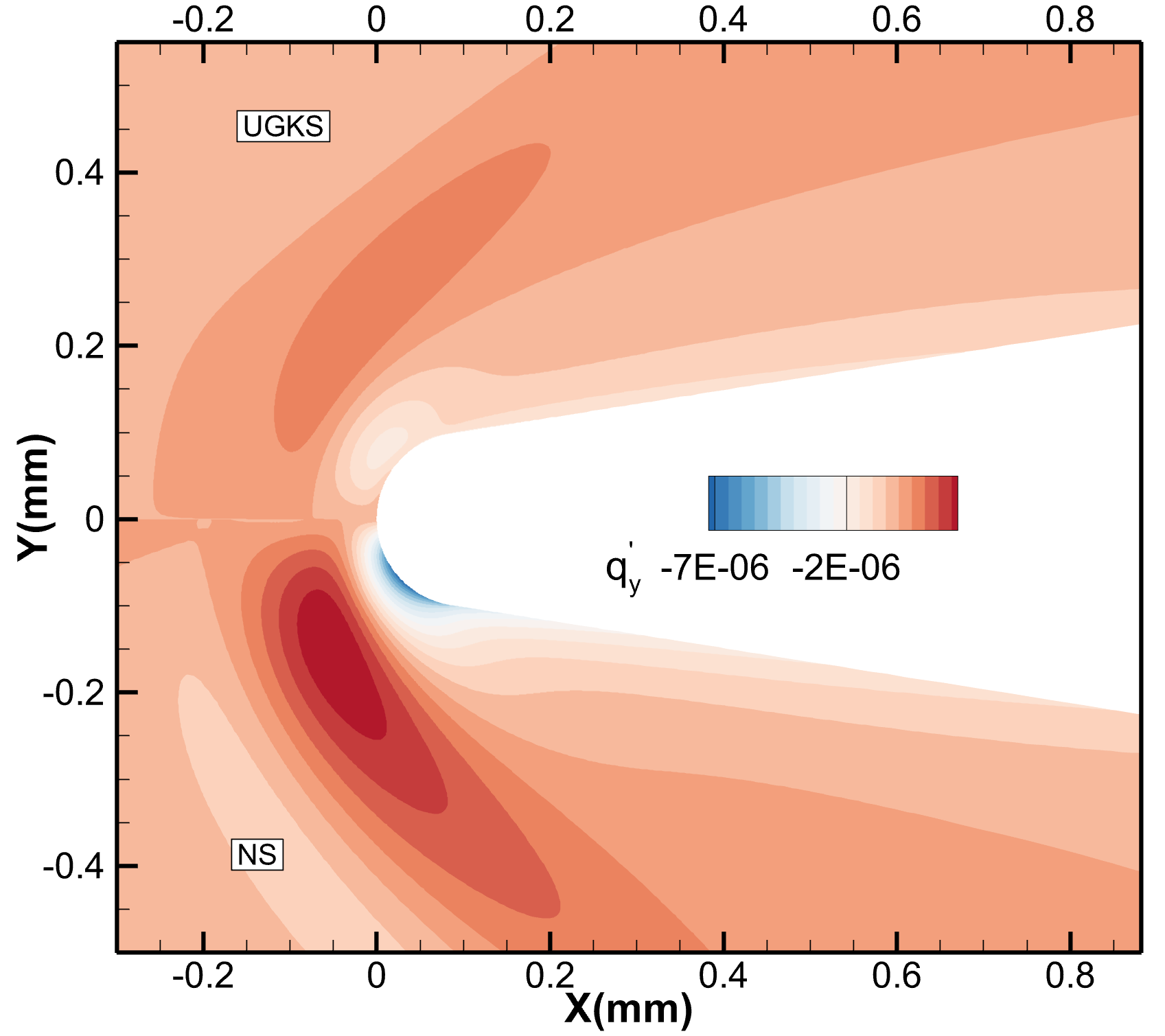}
    \caption{$\mathrm{Re}\{q'_y\}$}
    \label{fig:argon50km_qPrime_c}
  \end{subfigure}
  \caption{Near-nose real parts of the temperature and heat-flux perturbations for freestream fast-acoustic incidence at 50\,km with \(Re_{R_n}=9.83\):
  (a)~$\mathrm{Re}\{T'\}$, with a negative band along the shock front and a positive core toward the wall;
  (b)~$\mathrm{Re}\{q'_x\}$; (c)~$\mathrm{Re}\{q'_y\}$.
  Upper half: UGKS (heat flux from distribution-function moments);
  lower half: NS (Fourier's law).
  Waveform and wavelength agree; the wall-normal structure is compared in Fig.~\ref{fig:argon50km_mode}.}
  \label{fig:argon50km_qPrime}
\end{figure}

\begin{figure}[!htbp]
  \centering
  \begin{subfigure}[t]{0.4\textwidth}
    \includegraphics[width=\textwidth]{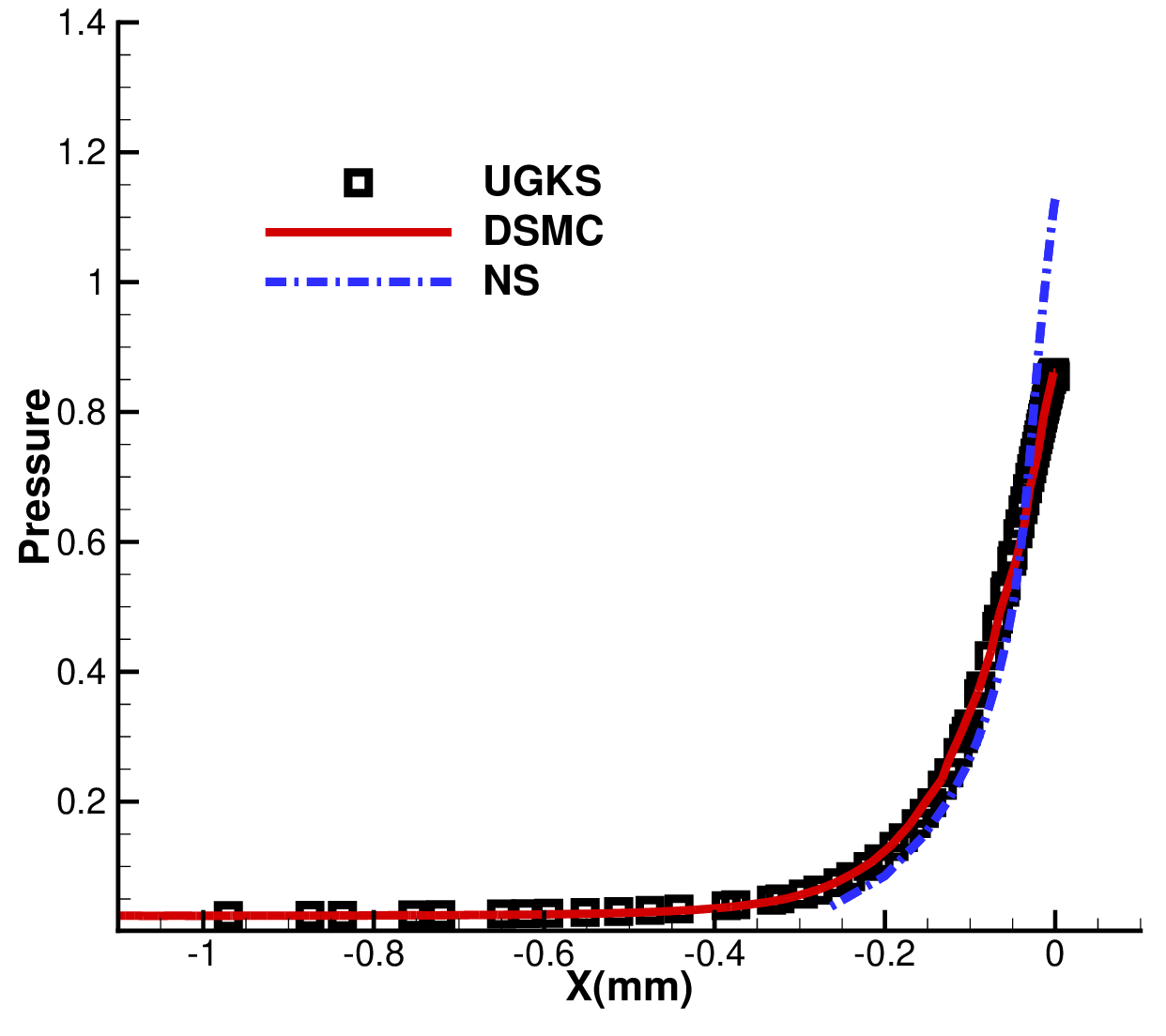}
    \caption{Base $p$}
    \label{fig:argon50km_stag_a}
  \end{subfigure}
  \quad
  \begin{subfigure}[t]{0.4\textwidth}
    \includegraphics[width=\textwidth]{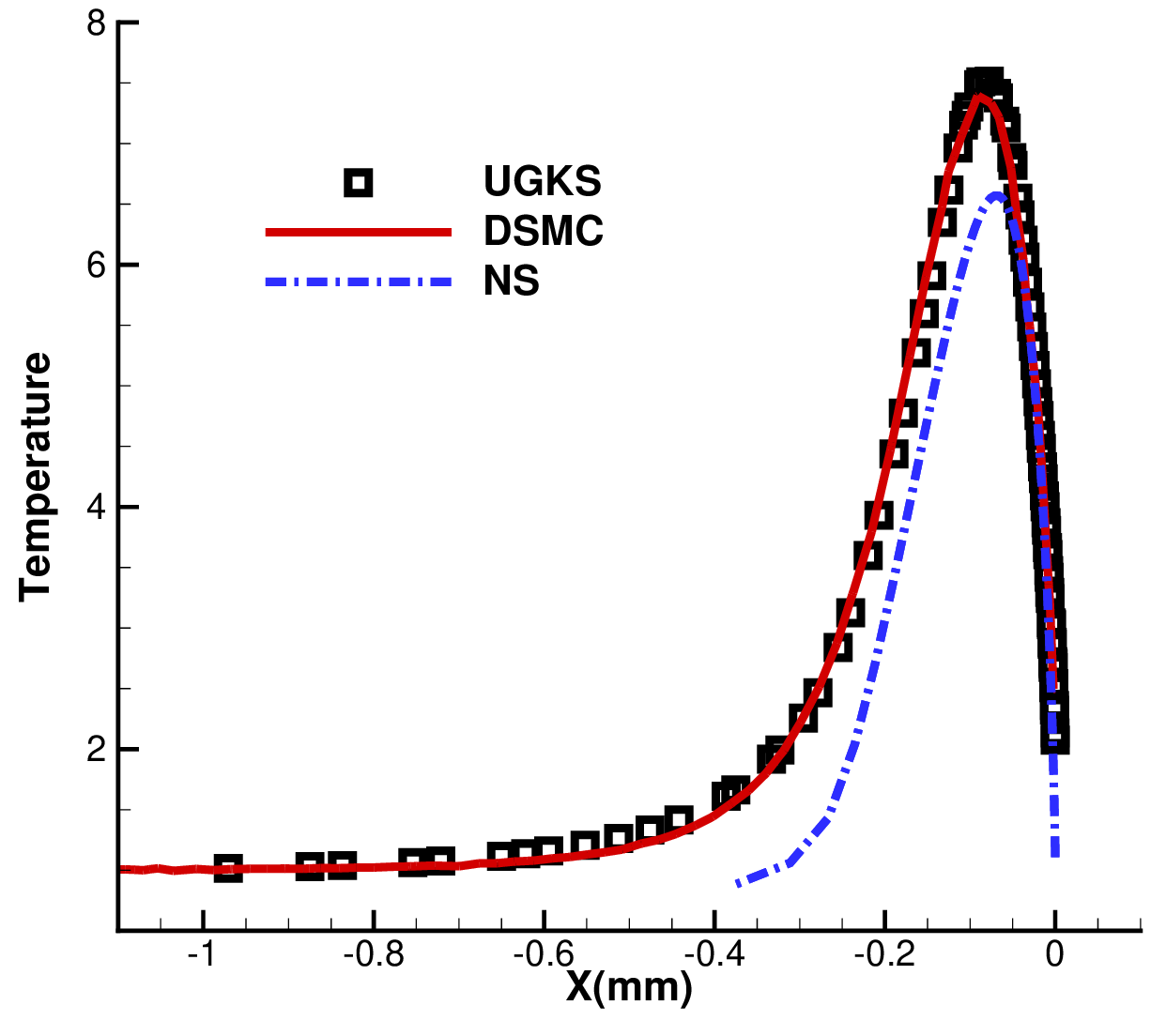}
    \caption{Base $T$}
    \label{fig:argon50km_stag_b}
  \end{subfigure}\\[0.6em]
  \begin{subfigure}[t]{0.4\textwidth}
    \includegraphics[width=\textwidth]{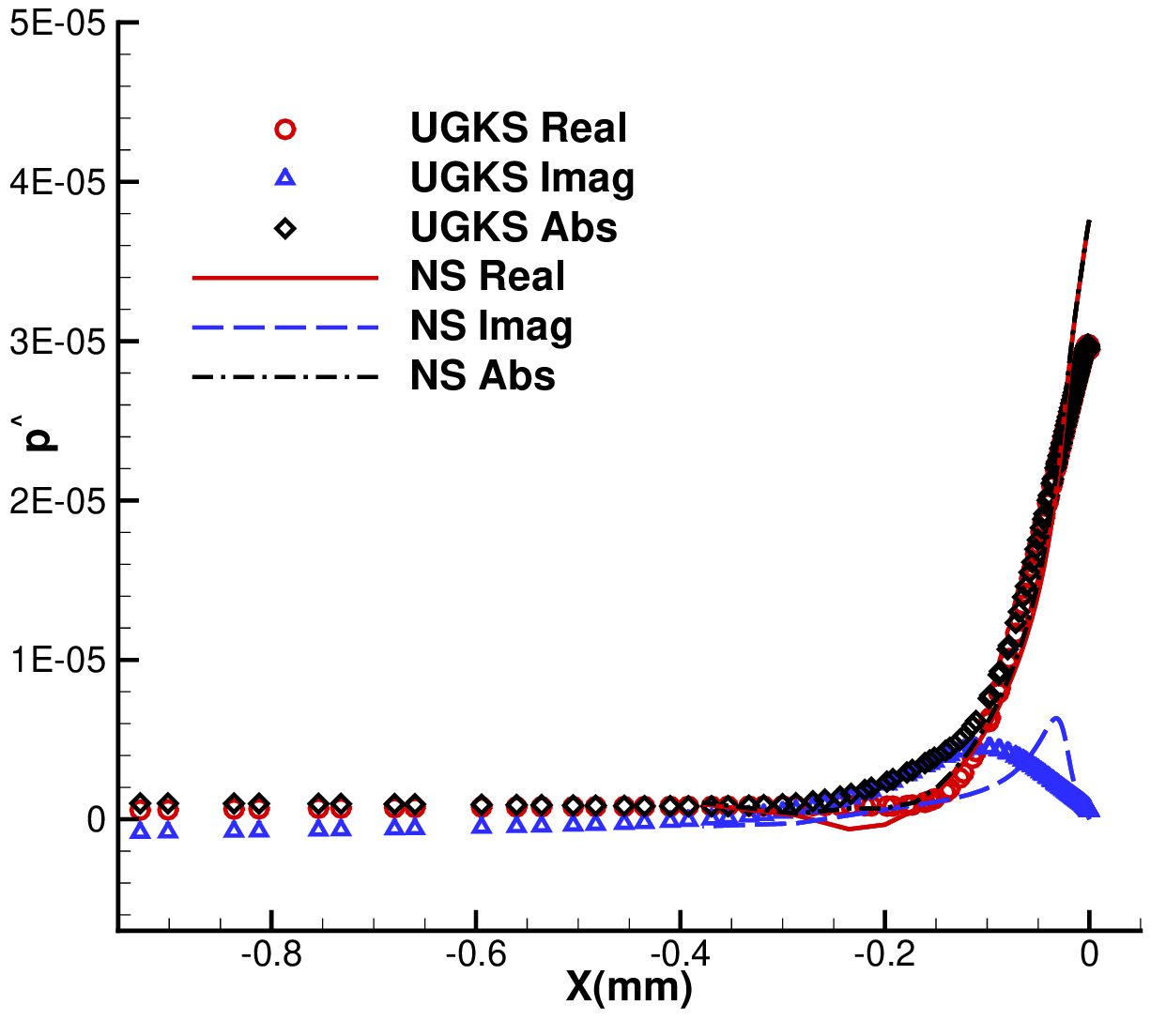}
    \caption{$p'$}
    \label{fig:argon50km_stag_c}
  \end{subfigure}
  \quad
  \begin{subfigure}[t]{0.4\textwidth}
    \includegraphics[width=\textwidth]{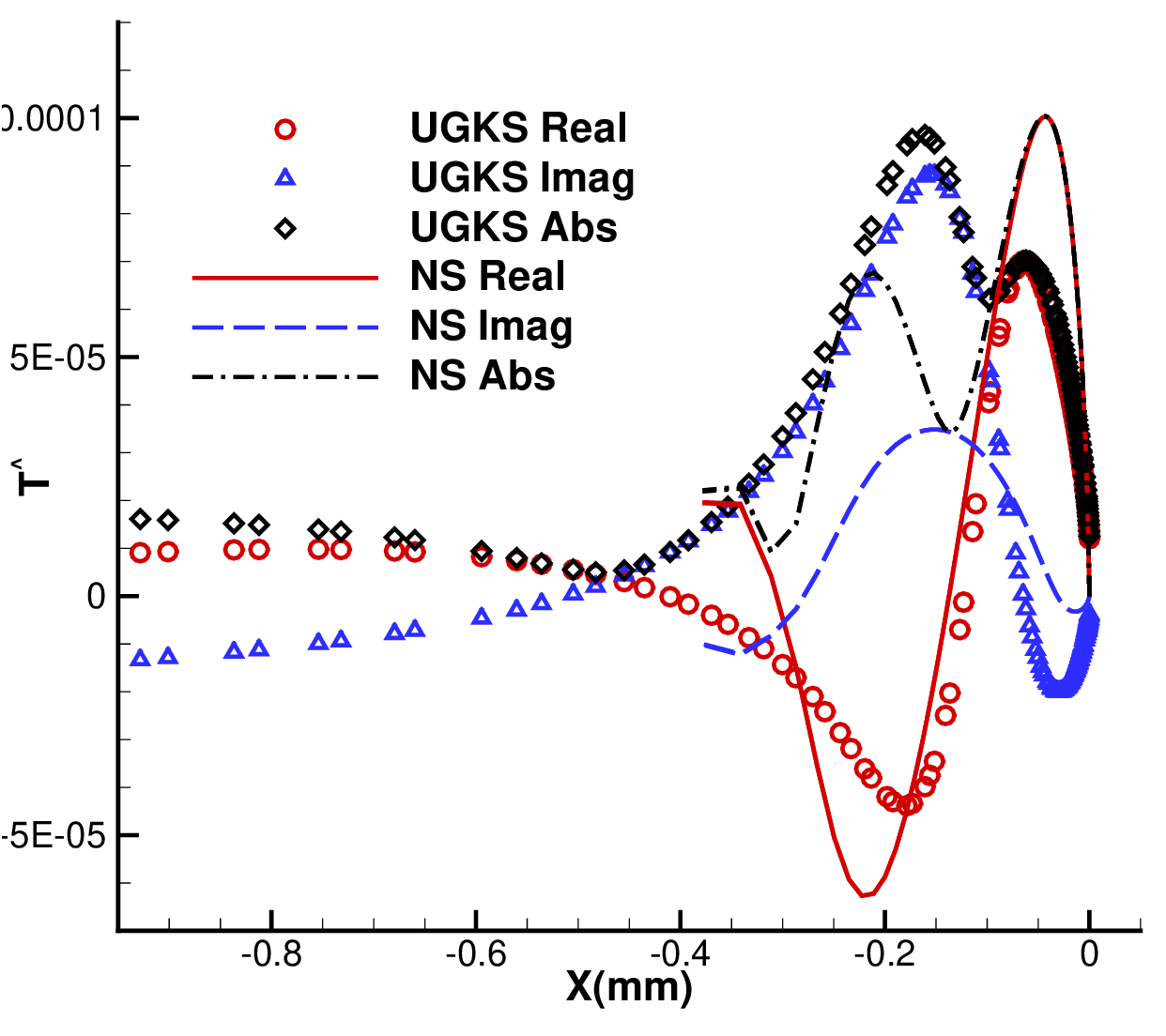}
    \caption{$T'$}
    \label{fig:argon50km_stag_d}
  \end{subfigure}
  \caption{Stagnation-line base flow and disturbance for argon at 50\,km with \(Re_{R_n}=9.83\)
  (\(X=0\) at the nose wall; negative \(X\) upstream):
  (a)~base pressure and (b)~base temperature, showing the UGKS/DSMC temperature precursor ahead of the NS front;
  (c)~pressure and (d)~temperature perturbations.}
  \label{fig:argon50km_stag}
\end{figure}

\begin{figure}[!htbp]
  \centering
  \begin{subfigure}[t]{0.25\textwidth}
    \includegraphics[width=\textwidth]{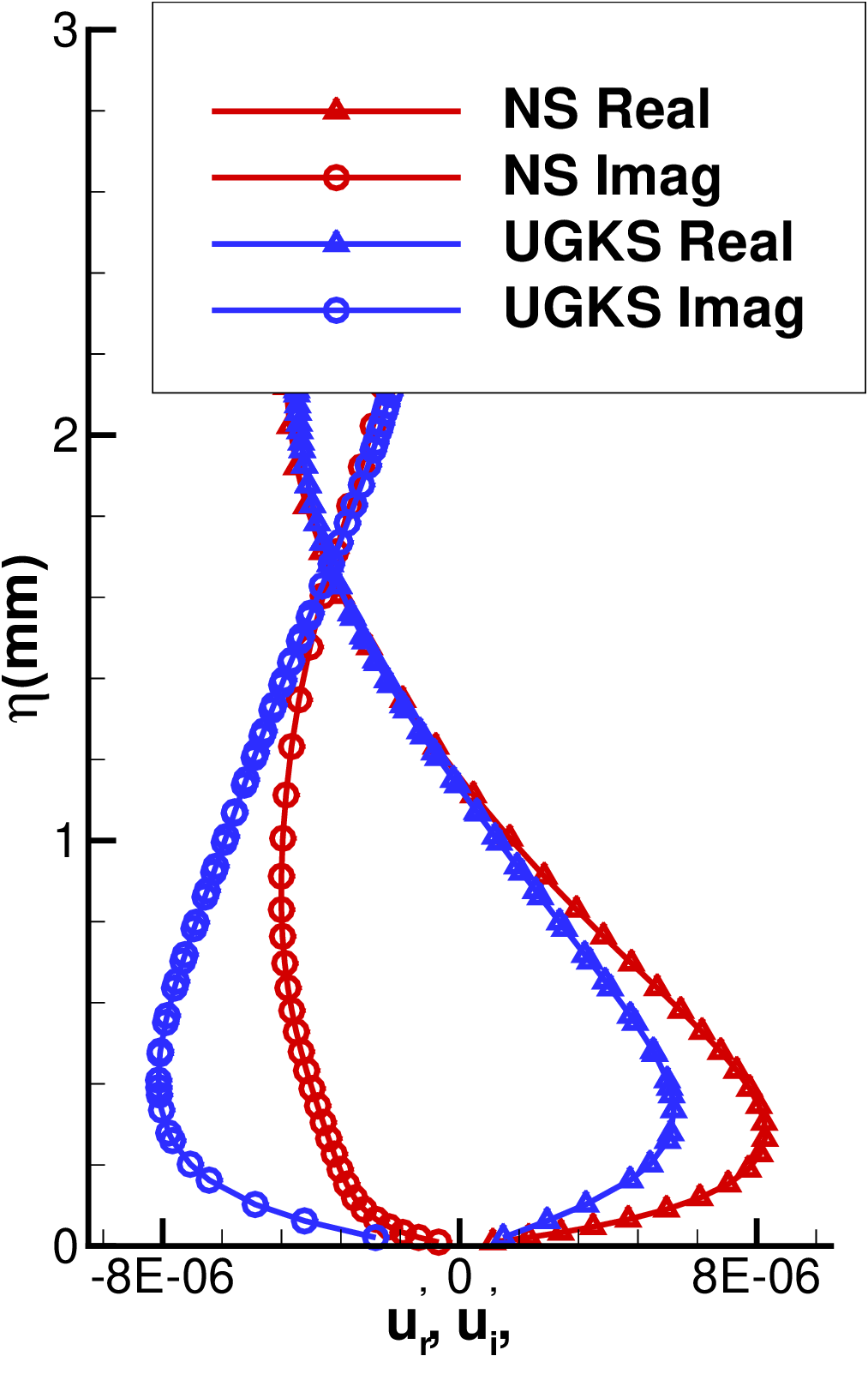}
    \caption{$U'$, $\xi=10\,\mathrm{mm}$}
    \label{fig:argon50km_mode_a}
  \end{subfigure}
  \begin{subfigure}[t]{0.25\textwidth}
    \includegraphics[width=\textwidth]{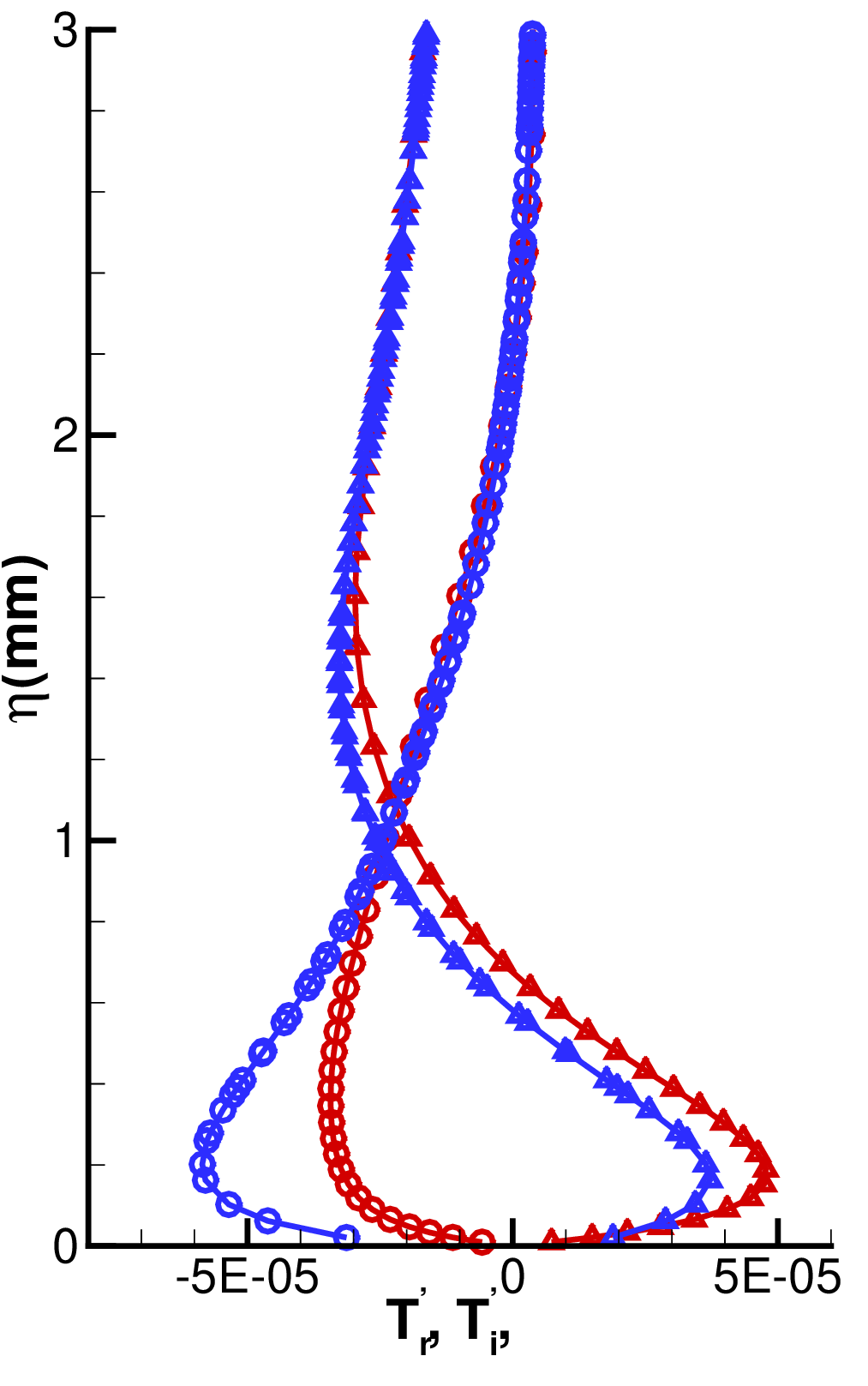}
    \caption{$T'$, $\xi=10\,\mathrm{mm}$}
    \label{fig:argon50km_mode_b}
  \end{subfigure}
  \begin{subfigure}[t]{0.25\textwidth}
    \includegraphics[width=\textwidth]{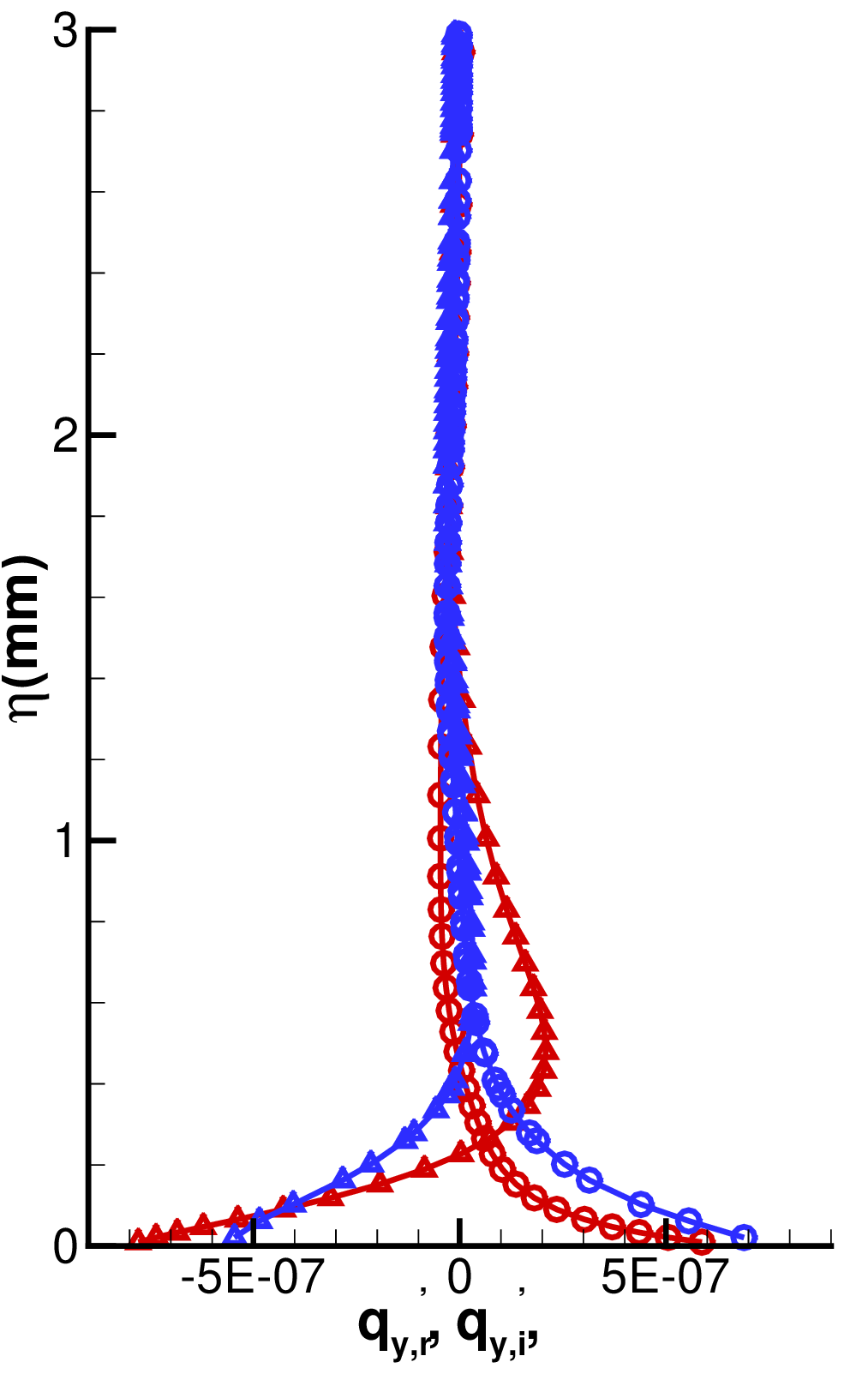}
    \caption{$q'_y$, $\xi=10\,\mathrm{mm}$}
    \label{fig:argon50km_mode_c}
  \end{subfigure}\\[0.6em]
  \begin{subfigure}[t]{0.25\textwidth}
    \includegraphics[width=\textwidth]{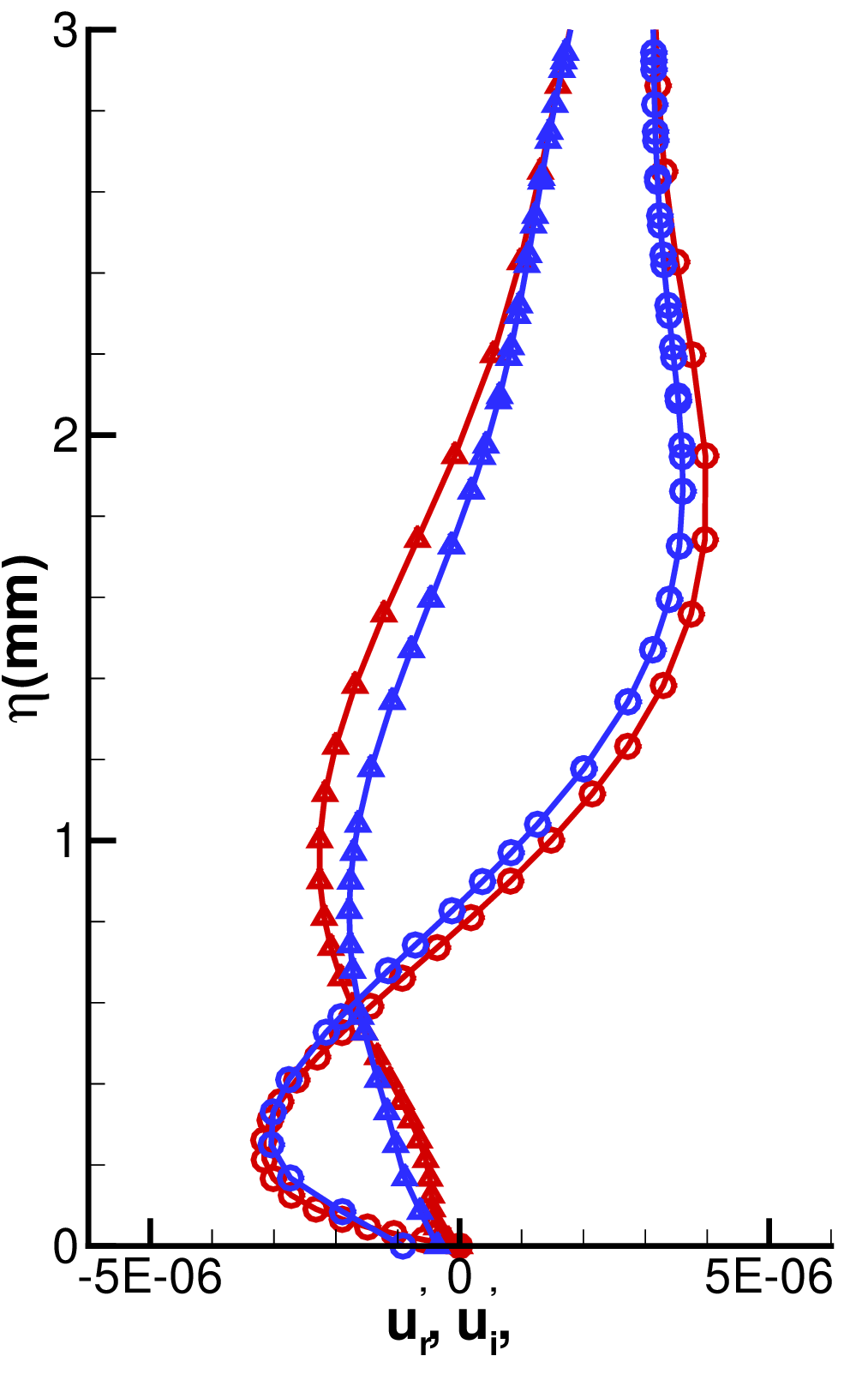}
    \caption{$U'$, $\xi=20\,\mathrm{mm}$}
    \label{fig:argon50km_mode_d}
  \end{subfigure}
  \begin{subfigure}[t]{0.25\textwidth}
    \includegraphics[width=\textwidth]{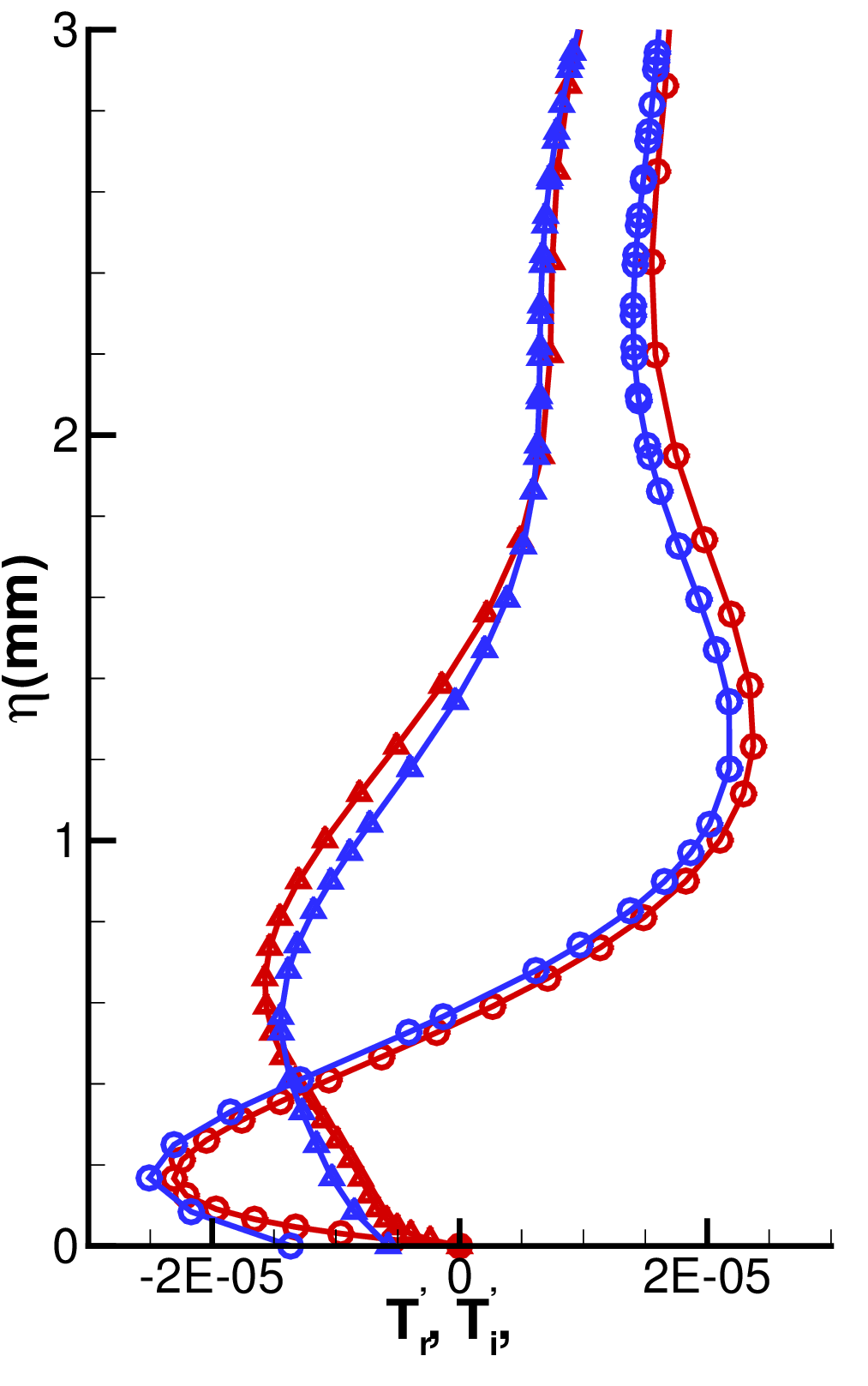}
    \caption{$T'$, $\xi=20\,\mathrm{mm}$}
    \label{fig:argon50km_mode_e}
  \end{subfigure}
  \begin{subfigure}[t]{0.25\textwidth}
    \includegraphics[width=\textwidth]{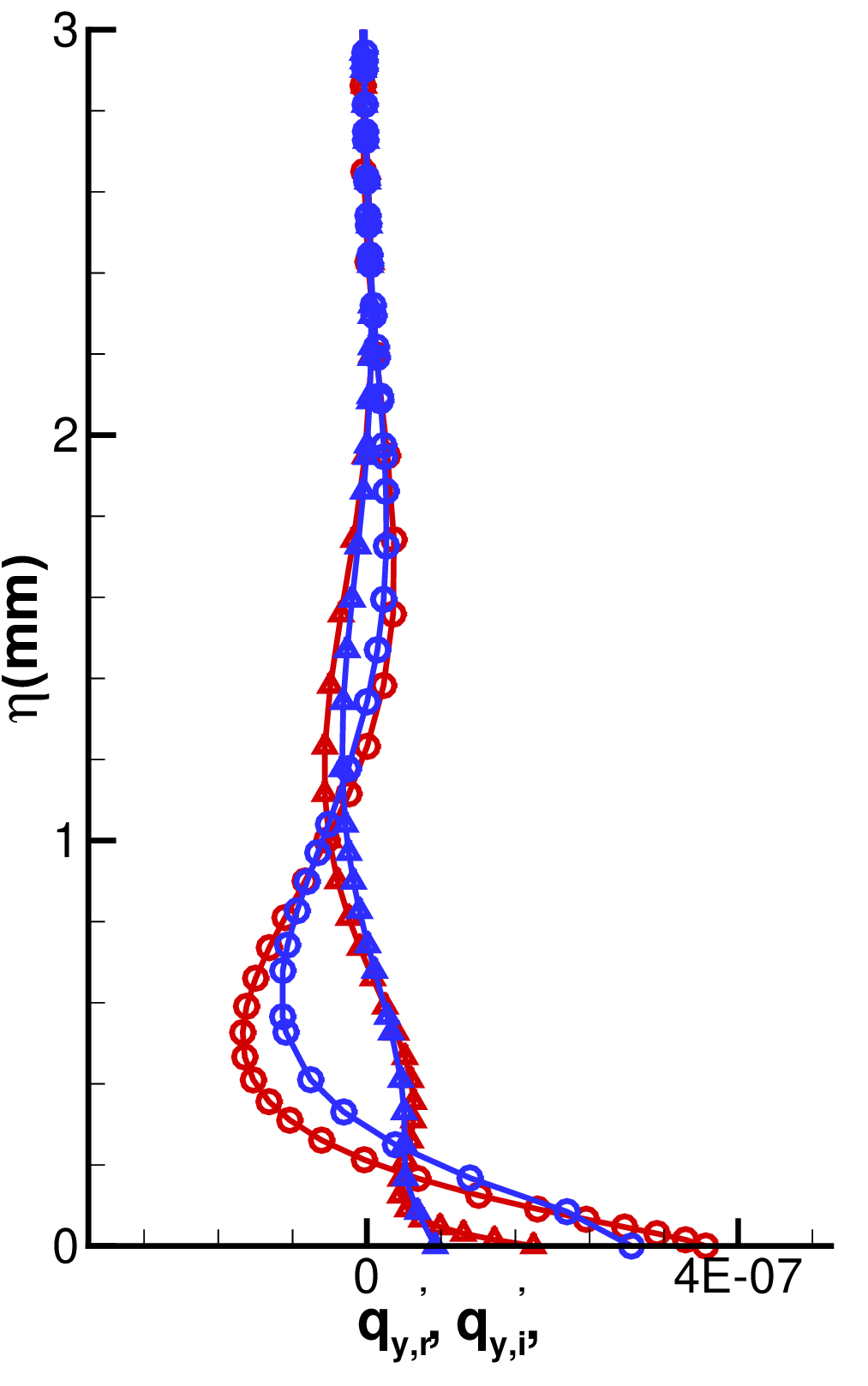}
    \caption{$q'_y$, $\xi=20\,\mathrm{mm}$}
    \label{fig:argon50km_mode_f}
  \end{subfigure}
  \caption{Wall-normal disturbance profiles for freestream fast-acoustic incidence at 50\,km with \(Re_{R_n}=9.83\):
  (a)~$U'$, (b)~$T'$ and (c)~$q'_y$ at \(\xi=10\,\mathrm{mm}\);
  (d)~$U'$, (e)~$T'$ and (f)~$q'_y$ at \(\xi=20\,\mathrm{mm}\).
  NS sets \(U'\) and \(T'\) to zero at the wall; UGKS slip and temperature jump leave both nonzero.
  At \(\xi=10\,\mathrm{mm}\), UGKS \(q'_y\) is a single peak at \(\eta=0\) and the NS peak stands off the wall.
  By \(\xi=20\,\mathrm{mm}\) the cut crosses the two bands of Fig.~\ref{fig:argon50km_delta_f}: a wall lobe and an outer lobe of opposite sign, deeper for NS.}
  \label{fig:argon50km_mode}
\end{figure}

The real parts of the pressure and temperature perturbations are compared with the continuum DNS in Fig.~\ref{fig:argon50km_delta}(a)--(d): full-field and near-nose views of \(\mathrm{Re}\{p'\}\) and \(\mathrm{Re}\{T'\}\) (UGKS upper half, DNS lower half).
Panels (e) and (f) of the same figure give the full-field heat-flux components \(\mathrm{Re}\{q'_x\}\) and \(\mathrm{Re}\{q'_y\}\) (UGKS upper half, NS lower half).
To show the two-dimensional receptivity pattern just after the detached shock, Fig.~\ref{fig:argon50km_qPrime} plots near-nose \(\mathrm{Re}\{T'\}\), \(\mathrm{Re}\{q'_x\}\) and \(\mathrm{Re}\{q'_y\}\) (UGKS upper half, NS lower half).
The two solvers define heat flux differently: NS evaluates the disturbed Fourier flux by a finite difference and removes the base-flow contribution,
\begin{equation*}
  \mathbf{q}'
  =
  -\left[
    \kappa(T+T')\nabla(T+T')
    -
    \kappa(T)\nabla T
  \right],
\end{equation*}
whereas UGKS obtains \(\mathbf{q}\) as a moment of the distribution function and evaluates the perturbation by linearization,
\begin{equation*}
  \delta\mathbf{q}_i
  = \sum_{k'} w_{k'}\,\frac{\partial \mathbf{q}_i}{\partial f_{i,k'}}\,\delta f_{i,k'}
  +\frac{\partial \mathbf{q}_i}{\partial \mathbf{W}_i}\,\delta\mathbf{W}_i,
\end{equation*}
as in Sec.~2.
In Fig.~\ref{fig:argon50km_qPrime_a}, \(\mathrm{Re}\{T'\}\) has a clear two-layer wave structure.
Along the upstream front of the main compression layer (near \(X\approx -0.15\,\mathrm{mm}\)), a continuous negative band follows the bow-shock arc, with a trough of about \(-6\times 10^{-5}\); this matches the negative extremum on the stagnation line in Fig.~\ref{fig:argon50km_stag_d}.
As the gas crosses the shock and stagnates toward the wall, strong compression turns the field into a deep red positive core of about \(1\times 10^{-4}\).
The same wavelength and the same wave pattern appear in the heat-flux components of Figs.~\ref{fig:argon50km_qPrime_b} and~\ref{fig:argon50km_qPrime_c}, and again over the full field in Figs.~\ref{fig:argon50km_delta_e} and~\ref{fig:argon50km_delta_f}.
Ahead of the stagnation point, \(\mathrm{Re}\{q'_x\}\) forms a tight positive--negative dipole, and \(\mathrm{Re}\{q'_y\}\) carries the disturbance off the stagnation line toward the shoulder.
Farther downstream the two halves of Fig.~\ref{fig:argon50km_delta_f} separate.
On the UGKS side the strong band of \(\mathrm{Re}\{q'_y\}\) stays on the wall, and the second band standing off the wall is weak.
On the NS side that off-wall band is stronger and stands clear of the surface, with the opposite sign to the wall band.
The wall-normal cuts in Fig.~\ref{fig:argon50km_mode} resolve the same contrast.


Figure~\ref{fig:argon50km_stag} examines the stagnation line, with \(X=0\) at the nose wall and negative \(X\) upstream.
The freestream approaches the wall in the positive-\(X\) direction.
At \(Re_{R_n}=9.83\), the bow-shock front starts near \(X\approx -0.4\,\mathrm{mm}\) and the macroscopic stand-off is about \(0.4\,\mathrm{mm}\).
In the base pressure and temperature (Figs.~\ref{fig:argon50km_stag_a} and~\ref{fig:argon50km_stag_b}), the most clear kinetic feature is a temperature precursor together with a thickened shock layer.
Temperature rises ahead of the pressure rise: UGKS and DSMC leave the freestream temperature already near \(X\approx -0.5\,\mathrm{mm}\), while pressure compression remains weak until \(X\gtrsim -0.3\,\mathrm{mm}\).
NS, by contrast, keeps a thin shock and a delayed temperature front that rises steeply only near \(X\approx -0.3\,\mathrm{mm}\).
After the strong compression, the base temperature peaks near \(X\approx -0.1\,\mathrm{mm}\) (about \(7.5\) for UGKS/DSMC), then drops sharply toward the cold isothermal wall for \(X\in[-0.1,0]\,\mathrm{mm}\).
The shock layer and the near-wall thermal layer therefore merge in a very narrow region, with essentially no inviscid post-shock pocket.
UGKS agrees closely with DSMC, whereas NS underpredicts the peak temperature (about \(6.5\)) and gives a slightly higher near-wall stagnation pressure.
This precursor comes from molecular non-equilibrium inside the strong shock.
At this low Reynolds number, hot molecules from the post-shock region can stream upstream against the freestream over a distance of order the mean free path and raise the thermal energy of the cold gas before macroscopic adiabatic compression is established.
Fourier heat conduction in NS is local and cannot capture this long-range molecular penetration, so the predicted shock front is too thin and too late.
The early temperature gradient also changes the receptivity environment seen by the incident acoustic wave: by \(X\approx -0.4\,\mathrm{mm}\), the local acoustic impedance and transport already differ from the freestream.
The pressure and temperature disturbances along the same line are shown in Figs.~\ref{fig:argon50km_stag_c} and~\ref{fig:argon50km_stag_d}.
As the flow stagnates toward the wall, \(|p'|\) and \(\mathrm{Re}\{p'\}\) grow almost monotonically and reach a wall maximum of about \(4.5\times 10^{-5}\).
In the strong-deceleration region \(X\in[-0.15,0]\,\mathrm{mm}\), the NS imaginary part of \(p'\) peaks higher than UGKS, indicating a continuum phase lag in the near-wall acoustic response.
The temperature disturbance \(T'\) is more multimodal: its amplitude peaks in the strong base-temperature rise near \(X\approx -0.15\,\mathrm{mm}\), while \(\mathrm{Re}\{T'\}\) forms a deep negative trough of about \(-7\times 10^{-5}\) near the compression front (\(X\approx -0.2\,\mathrm{mm}\)) and then flips sign before the wall.
The Imag parts of UGKS and NS separate clearly across the shock, reflecting non-Fourier thermal relaxation under the strong temperature gradient.
Thus, already on the stagnation line that feeds the wall response, the incident acoustic wave carries a kinetic phase and modal imprint from the non-equilibrium shock and near-wall thermal layer.

The wall-normal profiles are where the two wall conditions separate (Fig.~\ref{fig:argon50km_mode}).
The NS wall is no-slip and isothermal, so both \(U'\) and \(T'\) are zero at \(\eta=0\).
Diffuse reflection on the UGKS wall admits a slip velocity and a temperature jump, so neither disturbance vanishes there.
The wall-normal heat flux is the cut through that full-field pattern.
At \(\xi=10\,\mathrm{mm}\) the cut still meets a single peak: UGKS attached to the wall, and the NS peak standing off the surface (Fig.~\ref{fig:argon50km_mode_c}).
By \(\xi=20\,\mathrm{mm}\) the second, off-wall band of Fig.~\ref{fig:argon50km_delta_f} has formed, and the same cut is double-lobed, with one lobe on the wall and a lobe of opposite sign farther out (Fig.~\ref{fig:argon50km_mode_f}).
This outer lobe is produced downstream of the \(\xi=10\,\mathrm{mm}\) station.
It is deeper in the NS profile, where the off-wall band in the lower half of Fig.~\ref{fig:argon50km_delta_f} is the stronger one.

\FloatBarrier

\begin{figure}[!htbp]
  \centering
  \begin{subfigure}[t]{0.4\textwidth}
    \includegraphics[width=\linewidth]{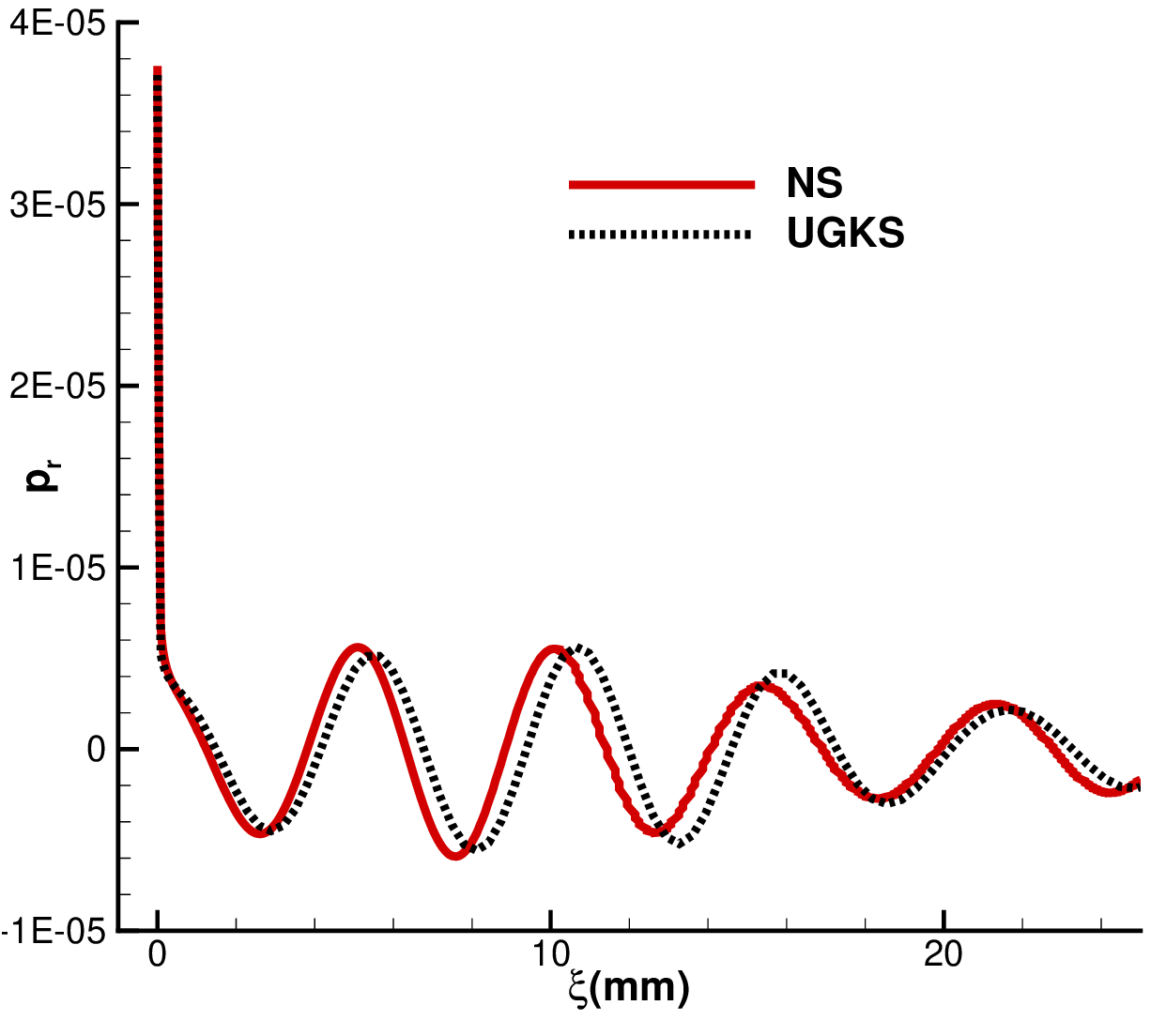}
    \caption{$\mathrm{Re}\{p'_w\}$}
    \label{fig:argon50km_wallDisp_a}
  \end{subfigure}
  \quad
  \begin{subfigure}[t]{0.4\textwidth}
    \includegraphics[width=\linewidth]{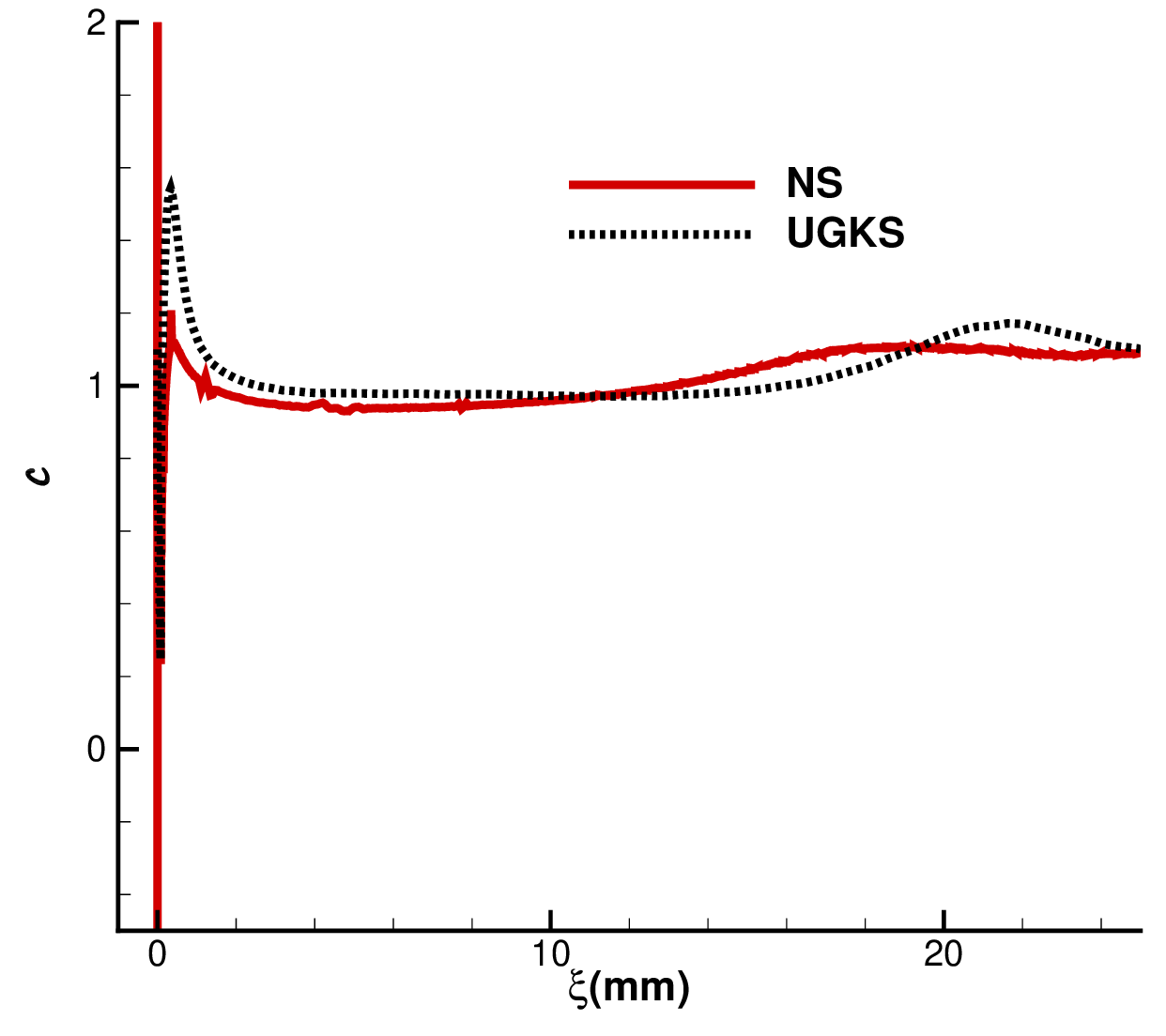}
    \caption{Phase speed}
    \label{fig:argon50km_wallDisp_b}
  \end{subfigure}\\[1.4em]
  \begin{subfigure}[t]{0.4\textwidth}
    \includegraphics[width=\linewidth]{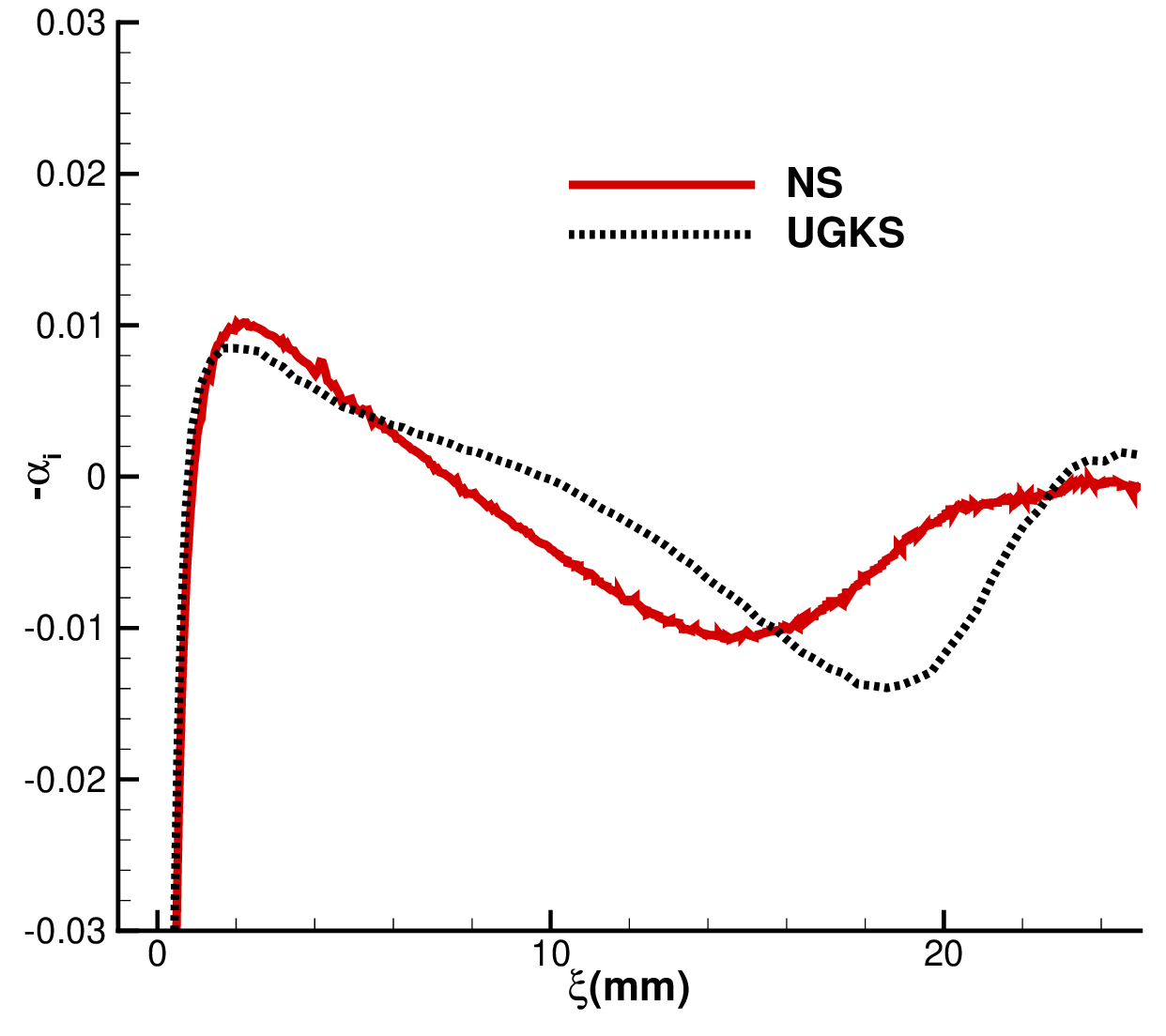}
    \caption{Growth rate}
    \label{fig:argon50km_wallDisp_c}
  \end{subfigure}
  \quad
  \begin{subfigure}[t]{0.4\textwidth}
    \includegraphics[width=\linewidth]{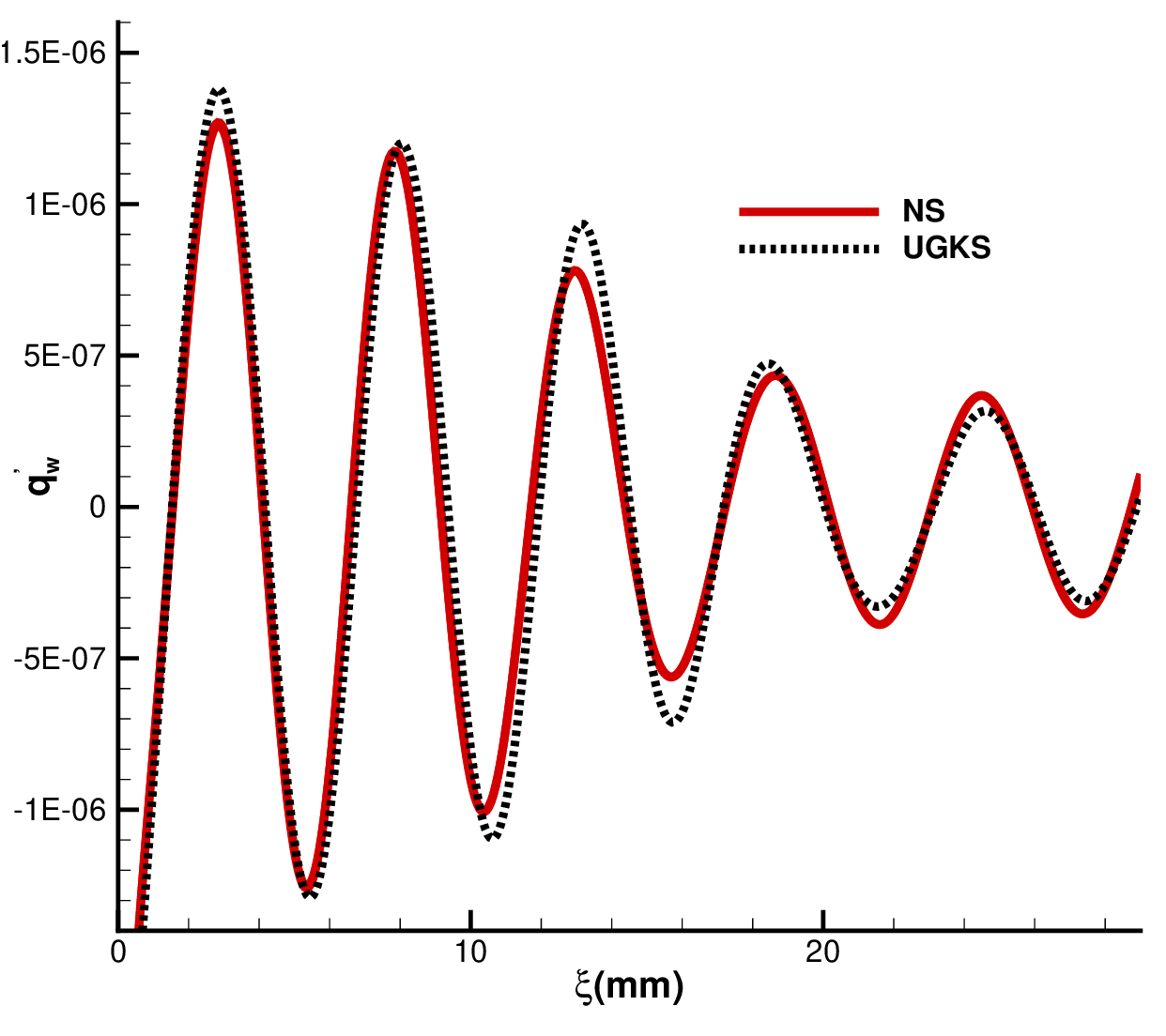}
    \caption{Wall heat flux}
    \label{fig:argon50km_wallDisp_d}
  \end{subfigure}
  \caption{Wall response for freestream fast-acoustic incidence at 50\,km with \(Re_{R_n}=9.83\):
  (a)~$\mathrm{Re}\{p'_w\}$;
  (b)~phase speed \(c_r\);
  (c)~spatial growth rate \(-\alpha_i\);
  (d)~wall heat-flux perturbation $q'_w$.
  Wall pressure and $q'_w$ keep a common phase.
  Slip delays the zero crossing of \(-\alpha_i\) from about \(7\,\mathrm{mm}\) for NS to about \(9\,\mathrm{mm}\) for UGKS.}
  \label{fig:argon50km_wallDisp}
\end{figure}

Figure~\ref{fig:argon50km_wallDisp} summarizes the wall response that follows from this near-wall structure: \(\mathrm{Re}\{p'_w\}\), the phase speed \(c_r\), the spatial growth rate \(-\alpha_i\), and the wall heat-flux perturbation \(q'_w\).
The wall-pressure traces share the same period and phase, and \(q'_w\) follows that same phase, with the UGKS crests only slightly higher near the first lobe and near \(\xi\approx 13\,\mathrm{mm}\).
These surface signals remain part of the macroscopic wave system of Fig.~\ref{fig:argon50km_delta}.
The growth rate records the slip already seen in the wall-normal profiles.
Diffuse reflection admits a tangential velocity at the wall and therefore lowers the nominal near-wall shear rate relative to the no-slip NS boundary.
The reduced shear weakens the viscous damping of the disturbance.
Both curves still peak near \(0.01\) at \(\xi\approx 2\text{--}3\,\mathrm{mm}\).
The zero crossing is delayed: UGKS stays amplifying until \(\xi\approx 9\,\mathrm{mm}\), while NS has entered the decaying region by \(\xi\approx 7\,\mathrm{mm}\) (Fig.~\ref{fig:argon50km_wallDisp_c}).
Slip thereby extends the streamwise interval over which the instability wave is amplified.
The stations \(\xi=10\) and \(20\,\mathrm{mm}\) used above therefore lie downstream of this reversal.
The double-lobed heat flux is the later of the two, the wall-normal cut through the off-wall band of Fig.~\ref{fig:argon50km_delta_f}.

\section{Conclusions}
\label{sec:conclusions}

A complex linearized unified gas-kinetic scheme (CLUGKS) has been established to investigate hypersonic flow stability and receptivity from continuum to rarefied regimes.
By expanding the discrete UGKS residual under a harmonic ansatz with complex disturbance amplitudes, the unsteady perturbation problem is recast into a steady elliptic formulation and solved using LU-SGS accelerated pseudo-time marching.
This strategy circumvents high-dimensional kinetic eigenvalue spectrum calculations while inherently resolving kinetic shock profiles, diffuse wall-slip dynamics, and non-Fourier heat-flux transport directly from distribution moments.

The solver was verified against three canonical benchmarks: an acoustic standing wave over an isothermal diffuse wall, a near-critical Mach\,2 shock--disturbance interaction verified against spatial linear interaction analysis, and blunt-wedge freestream acoustic receptivity compared against DNS.
These verifications confirm the scheme's multiscale capability to capture acoustic propagation, shock distortion, and wall-mode coupling within a unified framework.

Subsequent investigations of argon flows over a blunt wedge reveal how receptivity characteristics transition with altitude.
At an equivalent altitude of \(40\,\mathrm{km}\) (\(\mathit{Re}_{R_n}=39\)), non-equilibrium effects are localized within the bow shock and a thin surface layer.
The baseline UGKS profile matches DSMC solutions and differs from Navier--Stokes predictions primarily through near-wall slip and thermal jump.
Under fast-acoustic excitation, UGKS and NS yield comparable wall-pressure wavelengths and phases; however, kinetic effects mitigate early wall-mode attenuation, allowing the spatial growth rate in UGKS to remain positive further downstream.
Although the present computational domain covers only the early stage of modal growth, such incremental discrepancies in spatial amplification rate will compound over extended linear regimes, generating substantial downstream amplitude deviations.

At \(50\,\mathrm{km}\) (\(\mathit{Re}_{R_n}=9.83\)), intensified rarefaction widens the disparity between kinetic and continuum models.
Whereas classical continuum boundary conditions enforce \(U'=0\) and \(T'=0\), kinetic diffuse reflection permits finite fluctuating slip and temperature jump at the surface.
At \(\xi=10\,\mathrm{mm}\), the UGKS wall-normal heat-flux disturbance exhibits a solitary peak at the surface, contrasting with the detached peak predicted by NS.
Further downstream (\(\xi=20\,\mathrm{mm}\)), \(\mathrm{Re}\{q'_y\}\) evolves into a distinct double-lobed profile, with NS exhibiting a significantly deeper outer lobe.
Concurrently, tangential wall slip relieves near-surface shear, shifting the neutral stability location ($-\alpha_i = 0$) downstream from \(\approx 7\,\mathrm{mm}\) to \(\approx 9\,\mathrm{mm}\) and expanding the linear growth zone.

In summary, CLUGKS offers a robust kinetic paradigm for leading-edge receptivity where continuum formulations lose validity.
While standard Navier--Stokes models can approximate primary wall-pressure patterns at moderate altitudes, wall-normal heat-flux disturbances and non-Fourier transport mechanisms increasingly diverge and mandate kinetic treatment as altitude ascends.

Future investigations will extend this framework beyond the present Mach\,5 monatomic argon configuration.
Elevated Mach numbers will test the intensification of kinetic shock precursors and non-Fourier heat flux under stronger shock heating.
Additionally, modulating wall temperature will reveal how thermal boundary conditions shape Knudsen-layer disturbance interactions, while integrating polyatomic internal energy relaxation via UGKS extensions will yield a comprehensive view of how shock strength, wall heating, and thermochemical non-equilibrium jointly dictate hypersonic receptivity.

\section*{Acknowledgements}
Thanks Cao Junzhe for providing the DSMC data and helpful discussions.
This work was supported by the National Natural Science Foundation of China (Grant Nos.~92371107, 12572253, 11872270 and U24B2006), and the Hong Kong Research Grants Council (Grant No.~16208324).

\bibliographystyle{elsarticle-num}
\bibliography{sample}

@article{xu2015direct,  
  title={Direct modeling for computational fluid dynamics: {Construction} and application of unified gas-kinetic schemes},
  author={Xu, Kun},
  year={2015},
  publisher={World Scientific}
}

@article{xu2010unified,
  title={A unified gas-kinetic scheme for continuum and rarefied flows},
  author={Xu, Kun and Huang, Juan-Chen},
  journal={Journal of Computational Physics},
  volume={229},
  number={20},
  pages={7747--7764},
  year={2010},
  publisher={Elsevier}
}

@book{bird1994molecular,
  title={Molecular gas dynamics and the direct simulation of gas flows},
  author={Bird, Graeme A},
  year={1994},
  publisher={Oxford university press}
}

@article{zhu2019ugkwp,
  title={Unified gas-kinetic wave-particle methods. {II}. Multiscale simulation on unstructured mesh},
  author={Zhu, Yajun and Liu, Chang and Zhong, Chengwen and Xu, Kun},
  journal={Physics of Fluids},
  volume={31},
  number={6},
  year={2019},
  publisher={AIP Publishing}
}

@article{liu2020ugkwp,
  title={Unified gas-kinetic wave-particle methods {I}: Continuum and rarefied gas flow},
  author={Liu, Chang and Zhu, Yajun and Xu, Kun},
  journal={Journal of Computational Physics},
  volume={401},
  pages={108977},
  year={2020},
  publisher={Elsevier}
}

@article{wei2024adaptive,
  title={Adaptive unified gas-kinetic scheme for diatomic gases with rotational and vibrational nonequilibrium},
  author={Wei, Yufeng and Long, Wenpei and Xu, Kun},
  journal={Computer Physics Communications},
  volume={305},
  pages={109324},
  year={2024},
  publisher={Elsevier}
}

@article{zhu2016implicit,
  title={Implicit unified gas-kinetic scheme for steady state solutions in all flow regimes},
  author={Zhu, Yajun and Zhong, Chengwen and Xu, Kun},
  journal={Journal of Computational Physics},
  volume={315},
  pages={16--38},
  year={2016},
  publisher={Elsevier}
}

@article{zhu2019implicit,
  title={An implicit unified gas-kinetic scheme for unsteady flow in all {Knudsen} regimes},
  author={Zhu, Yajun and Zhong, Chengwen and Xu, Kun},
  journal={Journal of Computational Physics},
  volume={386},
  pages={190--217},
  year={2019},
  publisher={Elsevier}
}

@article{he2026revisiting,
  author  = {He, Jianhong and Su, Caihong},
  title   = {Revisiting the interaction of linear waves with an oblique shock: a spatial analysis and the inverse problem},
  journal = {Journal of Fluid Mechanics},
  year    = {2026},
  volume  = {1039},
  pages   = {A47},
}

@article{alsmeyer1976,
  author  = {Alsmeyer, H.},
  title   = {Density profiles in argon and nitrogen shock waves measured by the absorption of an electron beam},
  journal = {Journal of Fluid Mechanics},
  year    = {1976},
  volume  = {74},
  number  = {3},
  pages   = {497--513},
}

@article{li2026receptivity,
  author  = {Li, Chao and Su, Caihong and Han, Yufeng},
  title   = {Hypersonic Boundary-Layer Receptivity over a Small Bluntness Wedge to Freestream Disturbances},
  journal = {AIAA Journal},
  year    = {2026},
}

@misc{zhang2026asymptoticpreservingadjointunifiedgas,
  title         = {An asymptotic-preserving adjoint unified gas kinetic scheme for sensitivity analysis},
  author        = {Zhang, Yue and Cao, Junzhe and Long, Wenpei and Xu, Kun},
  year          = {2026},
  eprint        = {2608.03236},
  archivePrefix = {arXiv},
  primaryClass  = {math.NA},
}

@article{ou2024rarefaction,
  author  = {Ou, Jun and Wang, Cong and Chen, Jian},
  title   = {Rarefaction effects on hypersonic boundary-layer stability},
  journal = {Acta Mechanica Sinica},
  volume  = {40},
  number  = {12},
  pages   = {123184},
  year    = {2024},
}

@article{klothakis2022linear,
  author  = {Klothakis, Alexandros and Quintanilha, Jr., H{\'e}lio and Sawant, Saurabh S. and Protopapadakis, Eftychios and Theofilis, Vassilis and Levin, Deborah A.},
  title   = {Linear stability analysis of hypersonic boundary layers computed by a kinetic approach: {A} semi-infinite flat plate at {$M_\infty = 4.5\text{--}9$}},
  journal = {Theoretical and Computational Fluid Dynamics},
  volume  = {36},
  number  = {1},
  pages   = {117--139},
  year    = {2022},
}

@article{lee2019flow,
  author  = {Lee, Cunbiao and Jiang, Xi},
  title   = {Flow structures in transitional and turbulent boundary layers},
  journal = {Physics of Fluids},
  volume  = {31},
  number  = {11},
  pages   = {111301},
  year    = {2019},
}

@article{tu2021progress,
  author  = {Tu, Guohua and Chen, Jianda and Yuan, Xianxu and Yang, Qiang and Duan, Meng and Yang, Qing and Duan, Ying and Chen, Xi and Wan, Bing and Xiang, Xingyuan},
  title   = {Progress in flight tests of hypersonic boundary layer transition},
  journal = {Acta Mechanica Sinica},
  volume  = {37},
  number  = {11},
  pages   = {1589--1609},
  year    = {2021},
}

@article{zhong2012direct,
  author  = {Zhong, Xiaolin and Wang, Xiaowen},
  title   = {Direct Numerical Simulation on the Receptivity, Instability, and Transition of Hypersonic Boundary Layers},
  journal = {Annual Review of Fluid Mechanics},
  volume  = {44},
  number  = {1},
  pages   = {527--561},
  year    = {2012},
}

@article{goldstein1989boundary,
  author  = {Goldstein, Marvin E. and Hultgren, Lennart S.},
  title   = {Boundary-Layer Receptivity to Long-Wave Free-Stream Disturbances},
  journal = {Annual Review of Fluid Mechanics},
  volume  = {21},
  number  = {1},
  pages   = {137--166},
  year    = {1989},
}

@article{saric2002boundary,
  author  = {Saric, William S. and Reed, Helen L. and Kerschen, Edward J.},
  title   = {Boundary-Layer Receptivity to Freestream Disturbances},
  journal = {Annual Review of Fluid Mechanics},
  volume  = {34},
  number  = {1},
  pages   = {291--319},
  year    = {2002},
}

@article{fedorov2011transition,
  author  = {Fedorov, Alexander},
  title   = {Transition and Stability of High-Speed Boundary Layers},
  journal = {Annual Review of Fluid Mechanics},
  volume  = {43},
  number  = {1},
  pages   = {79--95},
  year    = {2011},
}

@article{zou2022linear,
  author  = {Zou, Sen and Zhong, Chengwen and Bi, Lin and Yuan, Xianxu and Tang, Zhigong},
  title   = {A new linear stability analysis approach for microchannel flow based on the {Boltzmann Bhatnagar--Gross--Krook} equation},
  journal = {Physics of Fluids},
  volume  = {34},
  number  = {12},
  pages   = {124114},
  year    = {2022},
}

@article{wu2024novel,
  title={A novel linear stability method based on {Shakhov Bhatnagar--Gross--Krook} equations for {Rayleigh--B{\'e}nard} convection in rarefied gases},
  author={Wu, Chao and Bi, Lin and Zhao, Jin and Tang, Zhigong and Yuan, Xianxu and Wen, Dongsheng},
  journal={Physics of Fluids},
  volume={36},
  number={12},
  year={2024},
  publisher={AIP Publishing}
}

@article{zou2024normal,
  author  = {Zou, Sen and Bi, Lin and Zhong, Chengwen and Yuan, Xianxu and Tang, Zhigong},
  title   = {A normal-mode approach for high-speed rarefied plane {Couette} flow},
  journal = {Physics of Fluids},
  volume  = {36},
  number  = {1},
  pages   = {014121},
  year    = {2024},
}

@article{wu2024efficient,
title = {Efficient solutions of eigenvalue problems in rarefied gas flows},
journal = {Journal of Computational Physics},
volume = {549},
pages = {114607},
year = {2026},
issn = {0021-9991},
author = {Lei Wu}
}

@article{broadwell1964study,
  author  = {Broadwell, James E.},
  title   = {Study of rarefied shear flow by the discrete velocity method},
  journal = {Journal of Fluid Mechanics},
  volume  = {19},
  number  = {3},
  pages   = {401--414},
  year    = {1964},
}

@article{chen2020three,
  author  = {Chen, Yuhan and Zhu, Yajun and Xu, Kun},
  title   = {A three-dimensional unified gas-kinetic wave-particle solver for flow computation in all regimes},
  journal = {Physics of Fluids},
  volume  = {32},
  number  = {9},
  pages   = {096108},
  year    = {2020},
}

@article{long2024nonequilibrium,
  author  = {Long, Wenpei and Wei, Yipeng and Xu, Kun},
  title   = {Nonequilibrium flow simulations using unified gas-kinetic wave-particle method},
  journal = {AIAA Journal},
  volume  = {62},
  number  = {4},
  pages   = {1411--1433},
  year    = {2024},
}

@article{liu2014unified,
  author  = {Liu, Sha and Yu, Peng and Xu, Kun and Zhong, Chengwen},
  title   = {Unified gas kinetic scheme for diatomic molecular simulations in all flow regimes},
  journal = {Journal of Computational Physics},
  volume  = {259},
  pages   = {96--113},
  year    = {2014},
}

@article{wei2026unified,
  author  = {Wei, Yipeng and Cao, Junzhe and Xu, Kun},
  title   = {Unified gas-kinetic scheme for reactive flow with multi-scale transport and chemical non-equilibrium},
  journal = {Journal of Computational Physics},
  volume  = {546},
  pages   = {114514},
  year    = {2026},
}

@misc{long2026applicability,
  author        = {Long, Wenpei and Cao, Junzhe and Zhang, Yue and Xu, Kun},
  title         = {On the Applicability of the Gas-Kinetic Scheme with Kinetic Boundary Conditions for Near-Continuum Hypersonic Flows},
  year          = {2026},
  eprint        = {2605.23303v1},
  archivePrefix = {arXiv},
  primaryClass  = {physics.flu-dyn}
}

@techreport{lees1946investigation,
  author      = {Lees, Lester and Lin, Chia Chiao},
  title       = {Investigation of the stability of the laminar boundary layer in a compressible fluid},
  institution = {National Advisory Committee for Aeronautics},
  number      = {NACA Technical Note No. 1115},
  year        = {1946}
}

@techreport{mack1984boundary,
  author      = {Mack, Leslie M.},
  title       = {Boundary-layer linear stability theory},
  institution = {AGARD},
  number      = {AGARD-R-709},
  year        = {1984}
}

@article{bertolotti1991analysis,
  author  = {Bertolotti, Fabio P. and Herbert, Thorwald H.},
  title   = {Analysis of the linear stability of compressible boundary layers using the {PSE}},
  journal = {Theoretical and Computational Fluid Dynamics},
  volume  = {3},
  number  = {2},
  pages   = {117--124},
  year    = {1991},
}

@article{mack1975linear,
  author  = {Mack, Leslie M.},
  title   = {Linear stability theory and the problem of supersonic boundary-layer transition},
  journal = {AIAA Journal},
  volume  = {13},
  number  = {3},
  pages   = {278--289},
  year    = {1975},
}

@article{mack1990inviscid,
  author  = {Mack, Leslie M.},
  title   = {On the inviscid acoustic-mode instability of supersonic shear flows. {Part} 1. {Two}-dimensional waves},
  journal = {Theoretical and Computational Fluid Dynamics},
  volume  = {2},
  number  = {2},
  pages   = {97--123},
  year    = {1990},
}

@inproceedings{zhong2000receptivity,
  author    = {Zhong, Xiaolin},
  title     = {Receptivity of hypersonic boundary layers to freestream disturbances},
  booktitle = {38th Aerospace Sciences Meeting and Exhibit},
  year      = {2000},
  note      = {AIAA Paper 2000-0531},
}

@inproceedings{balakumar2015receptivity,
  author    = {Balakumar, Ponnampalam},
  title     = {Receptivity of hypersonic boundary layers to acoustic and vortical disturbances},
  booktitle = {45th AIAA Fluid Dynamics Conference},
  year      = {2015},
  note      = {AIAA Paper 2015-2473},
}

@article{liang2010effects,
  author  = {Liang, Xiao and Li, Xiaowei and Fu, Dexun and Ma, Yanwen},
  title   = {Effects of wall temperature on boundary layer stability over a blunt cone at {Mach} 7.99},
  journal = {Computers \& Fluids},
  volume  = {39},
  number  = {2},
  pages   = {359--370},
  year    = {2010},
}

@article{malik1991real,
  author  = {Malik, Mujeeb R. and Anderson, E. Clay},
  title   = {Real gas effects on hypersonic boundary-layer stability},
  journal = {Physics of Fluids A: Fluid Dynamics},
  volume  = {3},
  number  = {5},
  pages   = {803--821},
  year    = {1991},
}

@article{wan2020receptivity,
  author  = {Wan, Bingbing and Su, Caihong and Chen, Jian},
  title   = {Receptivity of a hypersonic blunt cone: role of disturbances in entropy layer},
  journal = {AIAA Journal},
  volume  = {58},
  number  = {9},
  pages   = {4047--4054},
  year    = {2020},
}

@article{chen2021receptivity,
  author  = {Chen, Yi and Tu, Guohua and Wan, Bingbing and Su, Caihong and Yuan, Xianxu and Chen, Jian},
  title   = {Receptivity of a hypersonic flow over a blunt wedge to a slow acoustic wave},
  journal = {Physics of Fluids},
  volume  = {33},
  number  = {8},
  pages   = {084107},
  year    = {2021},
}

@article{wan2023effects,
  author  = {Wan, Bingbing and Chen, Jian and Tu, Guohua and Xiang, Xing and Yuan, Xianxu and Duan, Meng},
  title   = {Effects of nose bluntness on entropy-layer stabilities over cones and wedges},
  journal = {Acta Mechanica Sinica},
  volume  = {39},
  number  = {12},
  pages   = {122176},
  year    = {2023},
}

@techreport{us1976atmosphere,
  author      = {{NOAA} and {NASA} and {USAF}},
  title       = {{U.S.} Standard Atmosphere, 1976},
  institution = {U.S. Government Printing Office},
  address     = {Washington, D.C.},
  year        = {1976},
}

@article{wang2024slip,
  author  = {Wang, Chenyue and Ou, Jihui and Chen, Jie},
  title   = {Slip effects on the receptivity of supersonic flat-plate boundary layer to freestream acoustic waves},
  journal = {Physics of Fluids},
  volume  = {36},
  number  = {9},
  pages   = {094125},
  year    = {2024},
}

@article{ou2019dins,
  author  = {Ou, Jihui and Chen, Jie},
  title   = {{DSMC} data-improved numerical simulation of hypersonic flow past a flat plate in near-continuum regime},
  journal = {Computers \& Fluids},
  volume  = {194},
  pages   = {104308},
  year    = {2019},
}

@article{ou2020nonlinear,
  author  = {Ou, Jihui and Chen, Jie},
  title   = {Nonlinear transport of rarefied {Couette} flows from low-speed to high-speed},
  journal = {Physics of Fluids},
  volume  = {32},
  number  = {11},
  pages   = {112021},
  year    = {2020},
}

@article{wang2025bluntcone,
  author  = {Wang, Chenyue and Ou, Jihui and Chen, Jie},
  title   = {Rarefaction effects on hypersonic boundary-layer stability over a blunt cone at varying degrees of wall cooling and nose bluntness in near-continuum regime},
  journal = {Physical Review Fluids},
  year    = {2025},
}

@article{zeng2026,
  author  = {Zeng, Chengming and Li, Qibing and Jiang, Dongjun},
  title   = {Rarefaction effects on hypersonic boundary-layer transition over straight cone},
  journal = {Physics of Fluids},
  volume  = {38},
  number  = {7},
  pages   = {078121},
  year    = {2026},
  month   = {07},
  issn    = {1070-6631},
}
\end{document}